\documentclass[a4paper,11pt]{article}
\pdfoutput=1

\usepackage{jcappub}

\usepackage{amssymb,amsmath,bm}
\usepackage{color}
\usepackage{slashed}
\usepackage{graphicx}
\usepackage{amsfonts}
\usepackage{lscape}
\usepackage{amsthm}
\usepackage{booktabs}
\usepackage{array}
\usepackage{tabulary}
\usepackage{hyperref}
\usepackage{cleveref}
\usepackage{xspace}

\usepackage[small,bf]{caption}
\newcommand{\eps}{\epsilon}

\newcommand{\be}{\begin{equation}}
\newcommand{\ee}{\end{equation}}

\newcommand{\CEvNS}{CE$\nu$NS\xspace}

\newcommand{\CaWO}{\mathrm{CaWO}_4}
\newcommand{\sapphire}{\mathrm{Al}_2\mathrm{O}_3}

\title{Prospects for exploring New Physics in Coherent Elastic Neutrino-Nucleus Scattering}

\author[a]{Julien Billard,}
\author[b]{Joseph Johnston,}
\author[c,d]{Bradley J. Kavanagh}
\abstract{Coherent Elastic Neutrino-Nucleus Scattering (\CEvNS) is a Standard Model process that, although predicted for decades, has only been detected recently by the COHERENT collaboration. Now that \CEvNS has been discovered, it provides a new probe for physics beyond the Standard Model. We study the potential to probe New Physics with \CEvNS through the use of low temperature bolometers at a reactor source. We consider contributions to \CEvNS due to a neutrino magnetic moment (NMM), Non-Standard Interactions (NSI) that may or may not change flavor, and simplified models containing a massive scalar or vector mediator. Targets consisting of Ge, Zn, Si, CaWO$_4$, and Al$_2$O$_3$ are examined. We show that by reaching a percentage-level precision measurement on the \CEvNS energy spectrum down to $\mathcal{O}(10)$ eV, forthcoming experiments will improve by two orders of magnitude both the \CEvNS-based NMM limit and the  search for new massive mediators. Additionally, we demonstrate that such dedicated low-threshold \CEvNS experiments will lead to unprecedented constraints on NSI parameters (particularly when multiple targets are combined) which will have major implications for the global neutrino physics program.}

\affiliation[a]{Univ Lyon, Universite Claude Bernard Lyon 1, CNRS/IN2P3, Institut de Physique nucleaire de Lyon, 4 rue Enrico Fermi, F-69622 France}
\affiliation[b]{Laboratory for Nuclear Science, Massachusetts Institute of Technology, Cambridge, MA, USA}
\affiliation[c]{GRAPPA, University of Amsterdam, Science Park 904, 1098 XH Amsterdam, The Netherlands}
\affiliation[d]{Laboratoire de Physique Th\'eorique et Hautes Energies, CNRS, UMR 7589, 4 Place Jussieu, F-75252, Paris, France}

\emailAdd{j.billard@ipnl.in2p3.fr}
\emailAdd{jpj13@mit.edu}
\emailAdd{b.j.kavanagh@uva.nl}

\begin{document} 
\maketitle
\flushbottom

\section{Introduction}
\label{sec:intro}
Coherent Elastic Neutrino-Nucleus Scattering (\CEvNS) is the scattering of neutrinos off nuclei through the Standard Model Weak neutral current \cite{Freedman:1973yd}. The interaction of neutrinos and quarks through $Z$-boson exchange ultimately gives rise to a coherent interaction between neutrinos and the nucleus as a whole \cite{Drukier:1983gj}, with the resulting scattering rate scaling roughly as $A^2$ for a nucleus containing $A$ nucleons. For sufficiently large target nuclei and a sufficiently large neutrino flux, this process should be observable. For a long time, however, \CEvNS escaped detection due to the difficulty of measuring the low energy ($\sim$~sub-keV) nuclear recoils produced by the neutrino-nucleus scattering events \cite{Formaggio:2013kya}. In spite of this challenge, \CEvNS was successfully observed for the first time in 2017 by the COHERENT collaboration \cite{Akimov:2017ade}, using a 14.6-kg CsI[Na] scintillating detector with a 4.2 keV energy threshold at the Spallation Neutron Source (SNS). 

This first detection of \CEvNS was achieved with a statistical significance of $6.7\sigma$\footnote{Strictly speaking, the absence of any \CEvNS events in the COHERENT data was ruled out at $6.7\sigma$.}, and the measured rate matched the Standard Model prediction within $1\sigma$ uncertainty. Future results will allow the Standard Model to be tested with increasing precision, including low energy measurements of the Weak angle $\sin^2\theta_W$ \cite{Lindner:2016wff}. Other applications include measurements of nuclear structure \cite{Patton:2013nwa} as well as nuclear reactor monitoring \cite{Hagmann:2004uv}. The study of \CEvNS is of particular interest for direct Dark Matter search experiments, for which \CEvNS scattering from solar neutrinos represents an almost irreducible background \cite{Gutlein:2010tq,Billard:2013qya,Dutta:2017nht}. Confirmation of \CEvNS improves our understanding of this background \cite{AristizabalSierra:2017joc}, while precision neutrino flux measurements in future multi-ton-scale liquid Xenon detectors (using the \CEvNS signal itself) may help to mitigate it \cite{Baudis:2013qla,Cerdeno:2016sfi}.

The successful observation of \CEvNS also opens up a novel probe of New Physics in the neutrino sector, which is the focus of this work. We explore the sensitivity of future reactor neutrino scattering experiments to new interactions and deviations from the predictions of the Standard Model. Such new interactions include:
\begin{itemize}
\item \textbf{Anomalously large neutrino magnetic moments} - Massive Dirac neutrinos should have a negligibly small magnetic moment, but loop-contributions from New Physics could raise this to a detectable level, adding a new feature to the nuclear recoil spectrum at low energy (see e.g.~Ref.~\cite{Studenikin:2016ykv} for a review).
\item \textbf{Non-Standard Interactions (NSI) of neutrinos} - Couplings between neutrinos and quarks may differ from those predicted by the Standard Model, affecting the overall normalisation of the \CEvNS rate.  The NSI framework \cite{Davidson:2003ha,Barranco:2005yy} allows us to constrain neutrino interactions without resorting to a particular mediator or coupling structure. Limits on NSI can also be recast as constraints on heavy \textbf{leptoquarks} \cite{Barranco:2007tz}.
\item \textbf{Couplings to new massive scalar and vector mediators} -  We can consider adding new mediators to the Standard Model, within the framework of Simplified Models, which are widely used in interpreting direct dark matter searches \cite{Abdallah:2015ter}. This approach requires us to focus on a specific mediator (either a new vector or scalar boson) but allows us to explore constraints coming from modifications to the shape of the \CEvNS recoil spectrum.
\item \textbf{Active-to-sterile neutrino oscillations} - In principle, measurements of \CEvNS could be used to constrain oscillations to sterile neutrinos \cite{Anderson:2012pn}. For the typical sterile neutrino masses required to explain the LSND anomaly \cite{Aguilar:2001ty} ($\Delta m_{41}^2 \sim 1 \,\,\mathrm{eV}^2$) and for typical neutrino energies from the CHOOZ reactor ($E_\nu \sim 5 \,\,\mathrm{MeV}$), the wavelength of active-sterile oscillations is $\lambda \sim 5 \,\,\mathrm{m}$. With detectors positioned at different baselines, it may be possible to detect these oscillations. Here, we assume a single fixed location for the detectors, in which case such active-to-sterile oscillations could appear only as a \%-level change in the total \CEvNS event rate \cite{Dentler:2018sju}. This would be hard to disentangle from signal systematics (such as uncertainties on the reactor neutrino flux) and we therefore do not consider this possibility here.
\end{itemize}
The COHERENT collaboration have reported the first limits on non-standard neutrino-nucleus interactions (NSI) from \CEvNS  \cite{Akimov:2017ade} and further constraints are expected from the on-going COHERENT program \cite{Akimov:2018ghi}. The COHERENT-2017 observation has also been analysed by a number of groups \cite{Liao:2017uzy,Coloma:2017ncl,Ge:2017mcq,Dent:2017mpr,Kosmas:2017tsq,Denton:2018xmq,Abdullah:2018ykz} in the context of other New Physics scenarios, including those outlined above. Beyond COHERENT, a number of studies have looked at the New Physics sensitivity of \CEvNS searches. For example, Ref.~\cite{Lindner:2016wff} explored limits on NSI couplings and new massive vector mediators for a simplified Germanium detector with a threshold down to 100 eV. Reference~\cite{Dent:2016wcr} looked at the sensitivity of low-threshold Germanium and Silicon detectors to new light mediators.

Here, we perform a broad survey of the New Physics prospects of future \CEvNS experiments, considering a wide range of experimental scenarios and taking into account realistic backgrounds and systematic uncertainties. We aim to further develop the ideas of Ref.~\cite{Dent:2016wcr}, demonstrating the importance of very low detector thresholds ($< 100 \,\mathrm{eV}$) for probing the parameter space of New Physics, especially in the case of new light mediators. We also show the impact of using multiple detector targets to break degeneracies between the various neutrino-quark couplings. We derive and compare the New Physics constraints from the COHERENT experiment (see Appendix~\ref{app:COHERENT})\footnote{During the preparation of this work, a number of other papers appeared \cite{Liao:2017uzy,Coloma:2017ncl,Ge:2017mcq,Dent:2017mpr,Kosmas:2017tsq,Denton:2018xmq,Abdullah:2018ykz} in which New Physics constraints are derived from the COHERENT results, similar to those presented here. We have checked that our constraints are broadly consistent with those presented elsewhere.} as well as comparing with complementary constraints from other probes such as fixed-target experiments and the LHC. This allows us to make recommendations for the optimal experimental setups and target ensembles for probing a variety of New Physics signatures. 

We begin by describing our assumed detector properties and statistical approach in Sec.~\ref{sec:experiment}, followed by the projected discovery reach of these detectors for a vanilla \CEvNS signal in Sec.~\ref{sec:discovery}. We explore the prospects for constraining neutrino magnetic moments in Sec.~\ref{sec:magnetic}, non-standard neutrino-nucleus interactions in Sec.~\ref{sec:NSI} and new vector and scalar mediators coupled to neutrinos in Sec.~\ref{sec:simplifiedmodels}. Finally, we discuss our conclusions and the implications of this work in Sec.~\ref{sec:conclusions}.



\section{Experimental considerations}
\label{sec:experiment}

In this section, we summarise on-going and planned \CEvNS measurement projects (Sec.~\ref{sec:projects}). We then describe our assumptions for detector targets, experimental sites and backgrounds (Sec.~\ref{sec:detector}). Finally, we discuss the statistical procedure we employ for obtaining projected limits on New Physics from \CEvNS measurements (Sec.~\ref{sec:statistics}).

\subsection{Ongoing or planned \texorpdfstring{\CEvNS}{CENNS} projects}
\label{sec:projects}
We give hereafter a brief description of some of the planned or ongoing cryogenic experiments that aim at probing the low-energy range (sub-100 eV) of the \CEvNS signal:
\begin{itemize}
\item MINER is a planned cryogenic bolometer-based experiment that will be located at the 1 MW thermal power research reactor at the Mitchell Institute in Texas. The experiment will use Si and Ge bolometers with a total target mass of 10 kg and a projected energy threshold of 200 eV with background discrimination~\cite{Agnolet:2016zir}.
\item NuCLEUS is a planned few-gram-scale cryogenic bolometer-based experiment that will use a combination of CaWO$_{4}$ and Al$_{2}$O$_{3}$ detectors with no intrinsic particle identification. The NuCLEUS strategy is to focus primarily on lowering the energy threshold with gram-scale detectors~\cite{Strauss:2017cam}, and to get an indirect background rejection power thanks to surrounding veto detectors~\cite{Strauss:2017cuu}.

\item Ricochet is a planned kg-scale cryogenic bolometer-based experiment that will use a combination of Zn-superconducting metal and Ge-semiconductor detectors, with tens of gram-scale individual detectors. The Ricochet strategy is to focus on both lowering the energy threshold to 10-50 eV, and maintaining significant background rejection capabilities down to this energy threshold~\cite{Billard:2016giu}.
\end{itemize}
Finally, despite its slightly higher energy threshold, it is also worth mentioning the ongoing CONUS experiment which uses Ge-semiconductor detectors. It is located at the 3.9 GW thermal power Brokdorf nuclear power plant in Germany. With a total target mass of $\sim4$ kg and a threshold of around 300 eV, CONUS began taking data in April 2018 \cite{CONUSwebpage} and preliminary results show $2.4\sigma$ evidence for the observation of \CEvNS \cite{CONUStalk2}.
\subsection{Signal and Background modeling}
\label{sec:detector}

Based on the various target materials envisioned by the ongoing and planned low-threshold cryogenic experiments, in the following we consider five possible target materials : Zinc, Germanium, Silicon, CaWO$_4$ and Al$_2$O$_3$ (sapphire). In addition, we consider two possible phases of operation (Phase 1 and Phase 2). The energy thresholds, target payloads
\footnote{Here, we use the term `payload' to refer to the total mass of active target material in the detector. This sometimes allows us to avoid potential ambiguity between the total detector mass and the mass of individual target nuclei.}
, and expected background rejection powers for the two phases are summarised in Tab.~\ref{tab:experiments}. When calculating the discovery significance for \CEvNS and limits on the neutrino magnetic moment, we study the dependence on the exposure, varying the experimental runtime from 0 to 1000 days. In this case, we use the target payload listed under Phase 1 (only varying the energy threshold between the two phases). For constraints on NSI and new mediator models, we fix the total runtime at 365 days and assume the target payloads given in Tab~\ref{tab:experiments} for Phases 1 and 2. Eventually, we will assume background rejection capabilities for each of these targets as follows:
\begin{itemize}
\item Ge and Si: are semiconductor materials that can perform electronic/nuclear recoil discrimination down to the energy threshold providing the ionization baseline resolution is good enough~\cite{Phipps:2016gdx}.
\item Zn: is a superconducting metal for which discrimination between electronic and nuclear recoils could potentially be performed thanks to pulse shape discrimination due to the difference in the thermalization time profiles, as discussed in~\cite{Billard:2016giu}.
\item CaWO$_4$ and Al$_2$O$_3$: are two scintillating crystals that are planned to be used surrounded by highly-performing vetoing detectors that should allow significant neutron and gamma mitigation~\cite{Strauss:2017cuu}.
\end{itemize}
Eventually, we will consider a gamma background rejection power of 1000 for each of these targets with an additional neutron background mitigation of 10 for the CaWO$_4$ and Al$_2$O$_3$. We assume the background rejection and signal efficiencies to be flat over the entire energy range of interest as proper measurements and simulations would be required to use more realistic efficiency curves, which is beyond the scope of this paper.

Regarding the signal and background modeling, here we explore \CEvNS signals for detectors located at the Chooz nuclear reactor complex in North-East France. The Chooz complex hosts two 4.25 GW thermal power reactors, providing a high-luminosity reactor neutrino source allowing us to study \CEvNS even from distances of a few hundred meters away from the nuclear reactor cores. For the neutrino spectrum, we follow the calculations of Refs.~\cite{Huber:2011wv,Mueller:2011nm} and use isotopic abundances in the Chooz reactor as given in Ref.~\cite{Ardellier:2006mn}. This neutrino flux calculation matches that used in previous studies of \CEvNS at Chooz \cite{Billard:2016giu}. In the same spirit as what was done in Ref.~\cite{Strauss:2017cuu}, we will consider two potential sites:
\begin{itemize}
\item The Near Site (NS): is the near hall of the Double Chooz experiment, located 400 meters away from the two cores. This site has an overburden of about 110 m.w.e (meter water equivalent) significantly reducing cosmogenic induced neutrons and other backgrounds~\cite{Billard:2016giu}.
\item The Very Near Site (VNS): is a potential site located within the power plant perimeter, which is currently being investigated for the placement of reactor neutrino experiments \cite{Vivier}. For the sake of illustration we will consider a distance from the two cores of 80 meters with an overbudden of 10 m.w.e as suggested in~\cite{Vivier}. Compared to the NS, the neutrino flux is increased by a factor of 25, accompanied by an intrinsic larger background which will have to be carefully characterized.
\end{itemize}

\begin{table}[t]
\begin{center}
  \begin{tabulary}{0.9\textwidth}{L|LL|LL|LL}
       &   \multicolumn{2}{c|}{Phase 1} &     \multicolumn{2}{c}{Phase 2}  &  \multicolumn{2}{c}{Background reduction}       \\
     Target  &  $E_\mathrm{th}$ [eV]  & Mass [g]   &  $E_\mathrm{th}$ [eV]  & Mass [g]  &   gamma & neutron                               \\
    \hline
    \hline
    Zn & 50 & 500 & 10 & 5000 & 1000 & 1\\
    Ge & 50 & 500 & 10 & 5000 & 1000 & 1\\
    Si & 50 & 500 & 10 & 5000 & 1000 & 1\\
    CaWO$_4$ & 20 & 6.84 & 7 & 68.4 & 1000 & 10 \\
    Al$_2$O$_3$ & 20 & 4.41  & 4 &44.1 & 1000 & 10 \\
  \end{tabulary}  
\end{center}
\caption{ \textbf{Summary of detector parameters.} We show the threshold energy $E_\mathrm{th}$ and payload mass for detectors in Phase 1 and Phase 2. Note that when considering the Discovery Significance (Sec.~\ref{sec:discovery}) and Magnetic Moment constraints (Sec.~\ref{sec:magnetic}), we keep the masses fixed to those listed under Phase 1 and iterate the runtime between 0 and 1000 days for both Phases 1 and 2. For projections on NSI (Sec.~\ref{sec:NSI}) and new mediators (Sec.~\ref{sec:simplifiedmodels}), we consider a ten-fold increase in the payload mass between Phase 1 and Phase 2 (as shown in this table) and keep the runtime fixed at 365 days. The last columns correspond to the background mitigation factor for gamma and neutron backgrounds.}
\label{tab:experiments}
\end{table}

\begin{figure}[t!]
\centering
\includegraphics[width=0.85\textwidth]{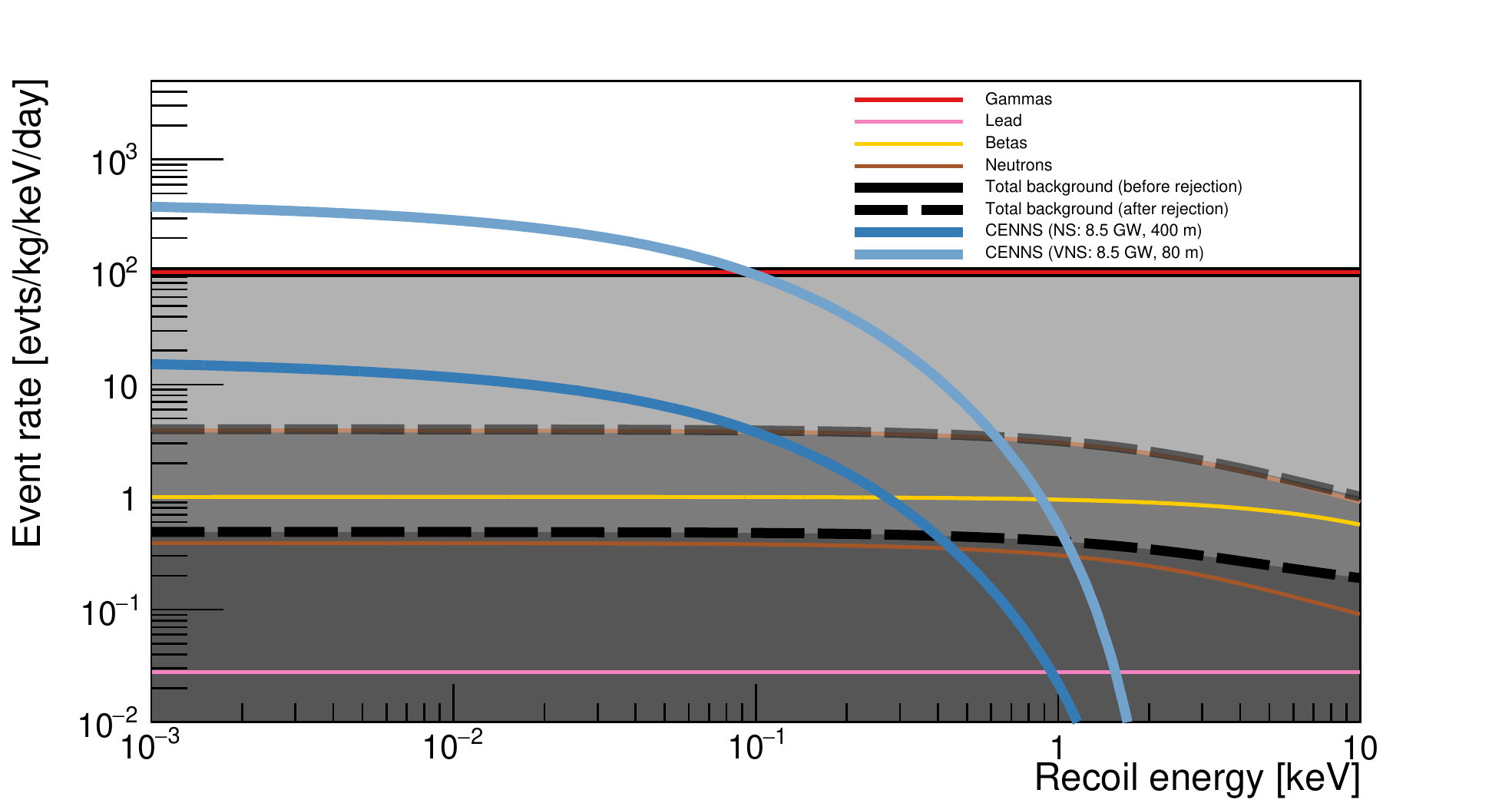}
\caption{Expected event rate as a function of the recoil energy for a Ge target detector installed in the Near Site situated 400 away from two high power reactors of 4.25 GW. The standard \CEvNS event rate as expected from the NS and VNS are shown as the dark and light blue solid lines. The leading background contributions (see text) are also shown: gammas (red), surface events from $^{206}$Pb (pink) and $^{210}$Pb-related betas (yellow), and neutrons (brown). The black solid and long-dashed lines represent the total background before and after electronic recoil discrimination (assumed to be a factor of 1000), for both the VNS (upper, light gray dashed) and the NS (lower, dark gray dashed). For illustration purposes, the background rejection capabilities expected from the various targets have been extrapolated down to 1 eV and considered as independent of energy.}
\label{fig:backgrounds}
\end{figure}

We also consider the following backgrounds:
\begin{itemize}
\item \textbf{Compton Background:}
High energy gammas, that will produce electronic recoils from Compton scattering, are produced by natural radioactivity of the surrounding materials, mostly from U/Th/K contaminations. In this study, we will conservatively assume a gamma background level of 100 event/kg/day/keV at both the NS and VNS experimental sites, shown as the red solid line in Fig.~\ref{fig:backgrounds}, which is of the order of what is observed in several low-background surface and shallow site experiments (see Ref.~\cite{Strauss:2017cuu} and references therein). For targets other than germanium, the Compton background is rescaled following $Z/Z_{Ge}$, where $Z$ is the atomic number.
\item \textbf{Neutron Background:}
High energy neutrons are the ultimate background to \CEvNS searches as they will also produce nuclear recoils. They can therefore only be statistically subtracted if their energy spectrum is well known. They arise from surrounding material radioactivity and from the cosmogenic activation of the surrounding cavern, which is $\sim$120 and $\sim$10 meter-water-equivalent deep for the NS and VNS respectively. The neutron-induced background expected at NS (brown) shown in Fig.~\ref{fig:backgrounds} has been carefully estimated from Monte Carlo simulations which included a moderate water neutron shield~\cite{Billard:2016giu}. Note that no reactor-correlated backgrounds 400 (NS) or 80 (VNS) meters away from the reactors are expected. Therefore, the neutron background will be additionally subtracted from the time variation of the nuclear reactor power~\cite{Billard:2016giu}. The average operating profile at the Chooz nuclear power plant is that 40\% of the time one of the two reactors (but not both) is off. The neutron background is assumed to be similar between all considered targets, but 10 times larger at the VNS than the NS as roughly expected from the smaller overburden.
\item \textbf{Surface event backgrounds:} these arise from $^{222}$Rn surface contamination of the detectors and of the surrounding materials, which then decays to the long-lived $^{210}$Pb radioactive isotope. Over the course of the experiment, the $^{210}$Pb will undergo beta-decays to $^{210}$Bi and then to $^{210}$Po, and will eventually alpha-decay to $^{206}$Pb. During this radioactive cascade, several types of beta-electrons, Auger-electrons, gammas, X-rays, and an alpha particle are emitted. These particles generate surface events that may fall into the sub-100 eV window. The surface event contributions from the $^{206}$Pb recoils (pink) and betas (yellow) shown in Fig.~\ref{fig:backgrounds} come from the EDELWEISS background characterization~\cite{Hehn:2016nll}. As one can conclude from Fig.~\ref{fig:backgrounds}, because all the targets/experiments we consider are assumed to have electron recoil discrimination capabilities, this third class of background will be negligible compared to the remaining gammas and neutrons. They will hence be neglected in the following.
\end{itemize}

\subsection{Statistical procedure}
\label{sec:statistics}

In order to calculate the projected discovery reach and exclusion limits, we use a likelihood-based approach. For a given mock data set $\mathbf{D}$ (the number of observed events  $\{N_\mathrm{obs}^{(i)}\}$ over a number of bins), the likelihood is given by


\begin{align}
\mathcal{L}(\mathbf{D}| \boldsymbol{\theta}, \boldsymbol{\psi}) = \mathcal{L}(\boldsymbol{\psi}) \times \prod_{i}^{N_\mathrm{bins}} P\left(N_\mathrm{obs}^{(i)} \right| \left.N_\mathrm{sig}^{(i)}(\boldsymbol{\theta}) + N_\mathrm{bg}^{(i)}(\boldsymbol{\psi})\right)\,.
\end{align}
Here, $P(k|\lambda)$ is the Poisson probability of observing $k$ events for $\lambda$ expected events. The number of signal events in the $i$th bin $N_\mathrm{sig}^{(i)}(\boldsymbol{\theta})$ will depend on the parameters of interest $\boldsymbol{\theta}$, while the background normalisation is controlled by a set of nuisance parameters $\boldsymbol{\psi}$. The uncertainties on these parameters are encoded in the likelihood $\mathcal{L}(\boldsymbol{\psi})$. The uncertainty on the normalisation of each background is assumed to be 10\% \cite{Billard:2016giu}. It is worth noticing that for Zn, Si and Ge targets aiming at discriminating electronic and nuclear recoils on an event-by-event basis, a two-dimensional likelihood function could be used, as in Ref.~\cite{Arnaud:2017usi}. This would ultimately improve the statistical background subtraction over the one-dimensional method used hereafter, but requires more detailed detector modeling  which is beyond the scope of this work.

Remaining completely agnostic about the low energy value of $\sin^2\theta_W$ drastically weakens constraints on New Physics. Without a prior on $\sin^2\theta_W$, it is difficult to know whether a modification of the \CEvNS rate comes from a modification of this SM parameter or from some New Physics contribution. Of course, the low energy value of $\sin^2\theta_W$ is not completely unknown; atomic parity violation measurements \cite{Bouchiat:1983uf,Porsev:2009pr} constrain $\sin^2\theta_W$ at the 1\% level (for momentum transfers $q \sim 2.4\,\,\mathrm{MeV}$). There are also uncertainties associated with maintaining a fixed threshold and uncertainties in the reactor flux. All of these uncertainties are included in a $\pm$5\% systematic on the \CEvNS signal and on all new physics signals. This uncertainty is incorporated into the model by multiplying the signals by a gaussian envelope.

For a given value of the parameters of interest $\boldsymbol{\theta}$ (which could be the neutrino magnetic moment, or the mass and coupling of a new vector or scalar mediator), we generate a number of mock data sets $\mathbf{D}$ and calculate the projected 90\% confidence level upper limits for each, using the asymptotic properties of the likelihood ratio test \cite{Wilks:1938dza,Cowan:2010js}. Limits were obtained with numerical minimisation, using the scipy package \cite{Jones:2001} with the L-BFGS algorithm. We then report the median upper limit, as well as the 95\% quantiles. 10000 mock data sets were generated for projected discovery significance, and 1000 were generated for all other bounds.

For bounds on other Non-Standard Interactions, the likelihood function is calculated for the ``Asimov data set'' \cite{Cowan:2010js}, where the number of counts in each bin is set equal to the expected counts without any Non-Standard Interactions. We then use the asymptotic properties of the likelihood ratio test to extract the 95\% confidence level allowed regions.

\section{Projected discovery significance for \texorpdfstring{\CEvNS}{CENNS}}
\label{sec:discovery}

In this section, we present the projected discovery significance for a \CEvNS signal, assuming only a Standard Model contribution to the rate. The \CEvNS cross section for a nucleus $N$ with atomic mass $A$ and atomic number $Z$ is given by \cite{Billard:2013qya}:
\begin{align}
\label{eq:CEvNS}
\frac{\mathrm{d}\sigma_{\nu-N}}{\mathrm{d}E_R} = \frac{G_F^2}{4\pi} Q_W^2 m_N \left(1 - \frac{m_N E_R}{2 E_\nu^2}\right) F^2(E_R)\,,
\end{align}
where the weak nuclear hypercharge is given by:
\begin{align}
\label{eq:weakcharge}
Q_W = (A-Z) - (1-4 \sin^2\theta_W)Z\,.
\end{align}
We assume a Helm form factor $F^2(E_R)$ \cite{Helm:1956zz}, which encodes the nuclear structure and leads to a loss of coherence at large recoil energies. Equation~\eqref{eq:CEvNS} gives the cross section for a fixed neutrino energy $E_\nu$, which must be at least $E_{\nu, \mathrm{min}} = \sqrt{m_N E_R/2}$ from kinematic considerations. The recoil rate per unit target mass with a given target nucleus can be obtained by convolving with the neutrino flux $\Phi(E_\nu)$:
\begin{align}
\label{eq:CEvNSfull}
\frac{\mathrm{d}R}{\mathrm{d}E_R} = \frac{1}{m_N} \int_{E_\nu > E_{\nu, \mathrm{min}}} \Phi(E_\nu) \frac{\mathrm{d}\sigma_{\nu-N}}{\mathrm{d}E_R}\, \mathrm{d}E_\nu\,.
\end{align}

\begin{figure}[t!]
\centering
\includegraphics[width=0.49\textwidth]{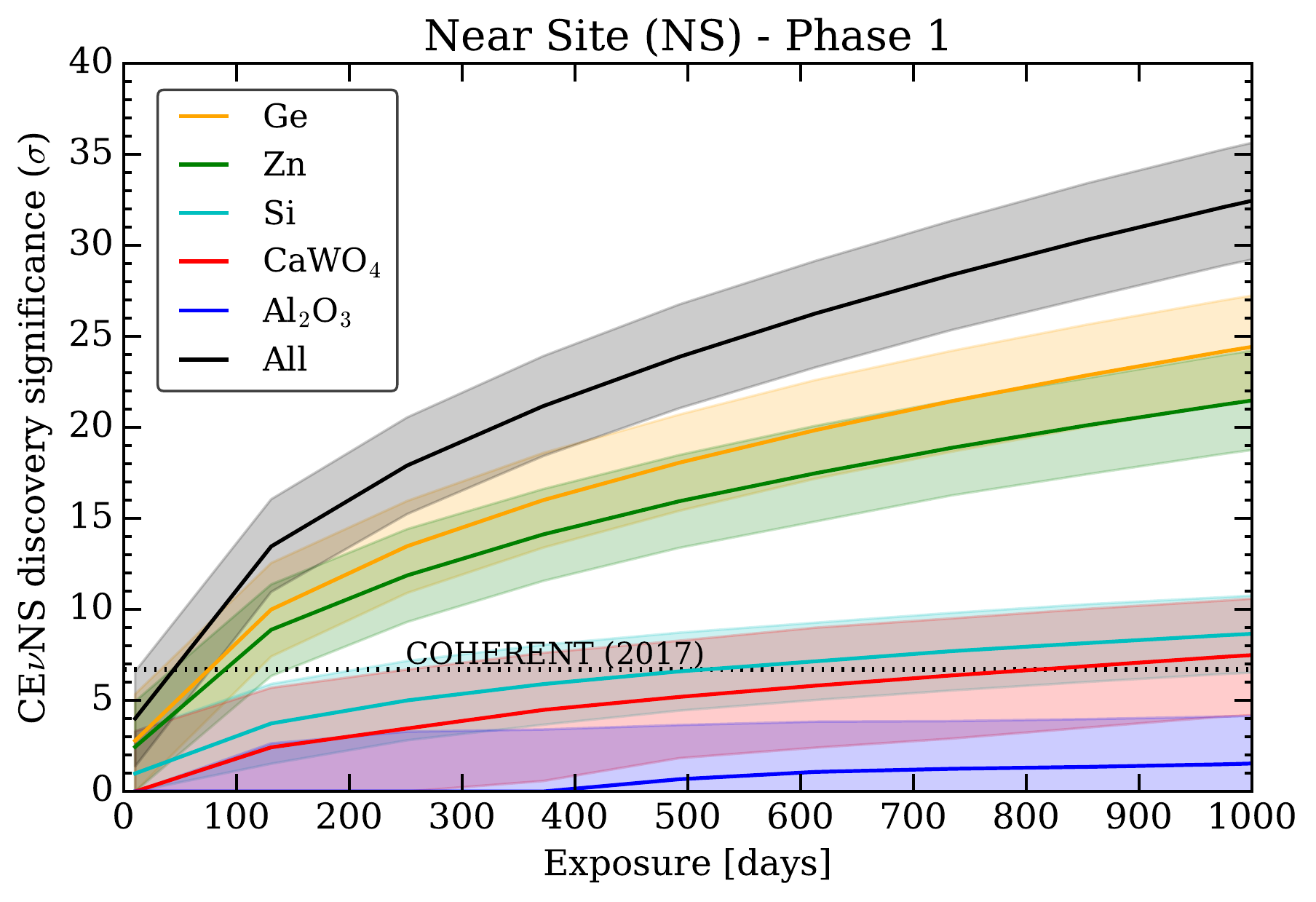}
\includegraphics[width=0.49\textwidth]{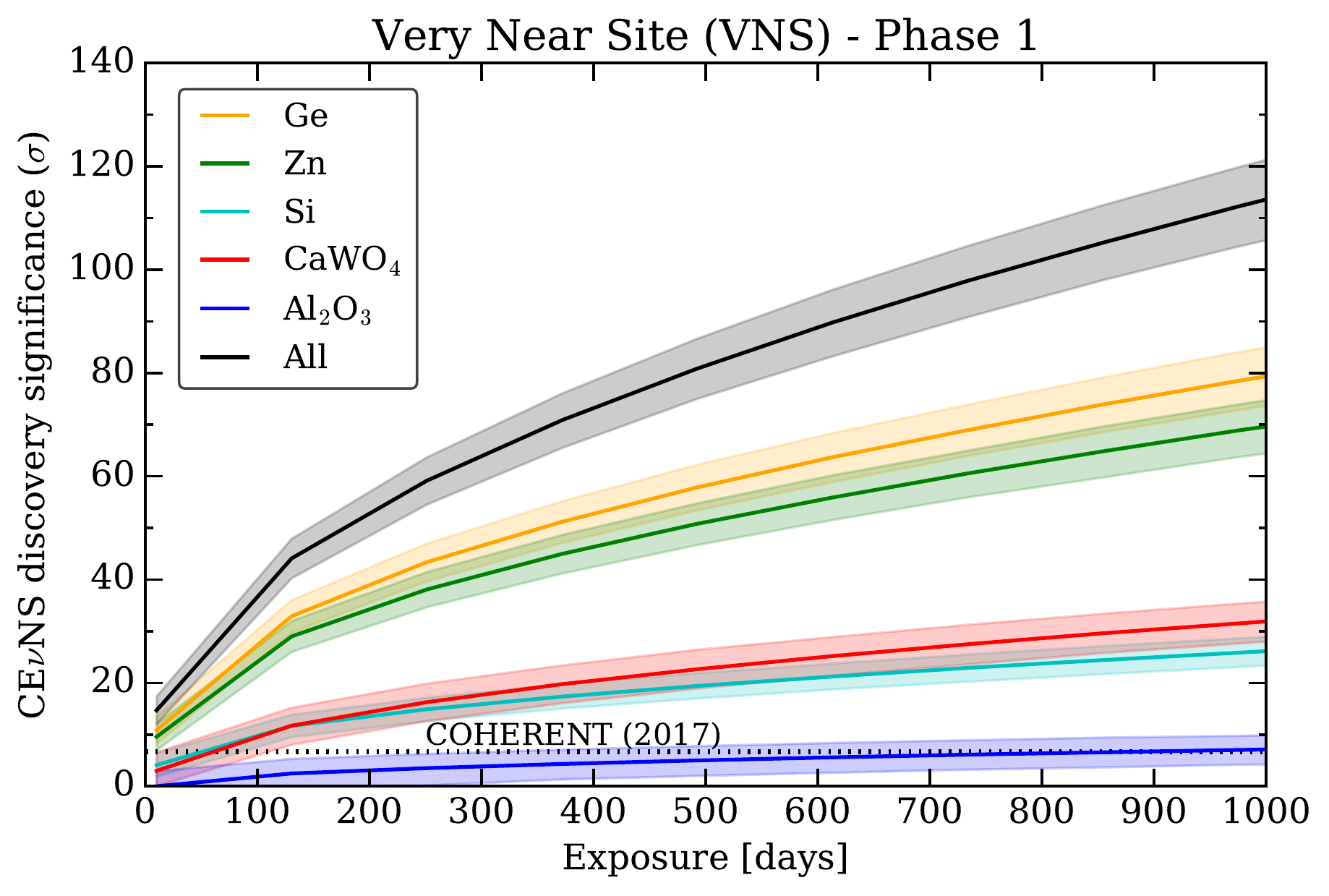}
\caption{\textbf{\CEvNS discovery significance as a function of exposure.} Median significance (and 95\% confidence level bands) for the discovery of \CEvNS using a number of detector targets: Ge (orange), Zn (green), Si (cyan), CaWO$_4$ (red) and Al$_2$O$_3$ (blue). In black, we show the discovery significance when all 5 targets are used simultaneously. Details of detector properties, backgrounds and systematic uncertainties are given in Sec.~\ref{sec:detector}. The significance of the COHERENT discovery of \CEvNS \cite{Akimov:2017ade} is shown as a dotted line at $6.7\sigma$. \textbf{Left:} Detectors located at the Near Site. \textbf{Right:} Detectors located at the Very Near Site.}
\label{fig:Discovery}
\end{figure}

Figure~\ref{fig:Discovery} shows the projected discovery significance using a range of Phase 1 detector setups. The left panel shows projections for the Near Site, while the right shows projections for the Very Near Site. For a given detector mass and fixed exposure, we note that the discovery significance is higher using a Ge target, compared with a Zn or Si target. As shown in Eq.~\eqref{eq:weakcharge}, the weak nuclear hypercharge scales roughly with the number of neutrons in the nucleus. The Ge nucleus has more neutrons than Zn or Si ($\sim41$ compared to $\sim35$ and $14$ respectively), leading to a larger \CEvNS signal and therefore a greater significance. For the $\CaWO$ and $\sapphire$ targets (which are planned to be used by the Nucleus experiment), the discovery significance is substantially smaller because the payloads we consider are much smaller (by a factor $\sim 100$) than for Ge, Zn and Si.

At the Near Site (left panel), it should be possible to reach the COHERENT-2017 significance of $6.7\sigma$ in around 70 days of exposure time using Ge or Zn targets alone. The black solid line shows the significance when all five targets are combined, which would surpass the COHERENT significance in roughly 30 days. The Very Near Site (right panel) gives even better prospects for the observation of a \CEvNS signal, owing primarily to the larger neutrino flux. In this case, a $5\sigma$ observation is possible with all targets but $\sapphire$ in less than 100 days.

\section{Neutrino magnetic moment}
\label{sec:magnetic}

In minimal extensions of the Standard Model, a massive Dirac neutrino may acquire a diagonal (i.e.~flavour-conserving) magnetic moment through radiative corrections, with a magnitude \cite{Fujikawa1980}:
\begin{align}
\mu_\nu \approx 3.2 \times 10^{-19} \, \left[ \frac{m_\nu}{1\,\,\mathrm{eV}}\right] \mu_B\,,
\end{align}
which, as we will see, is many orders of magnitude below current experimental limits. While New Physics contributions can lift the neutrino magnetic moment (NMM) \cite{Shrock:1974nd,Lee:1977tib,Shrock:1982sc}, these may also lead to unacceptably large contributions to the neutrino masses \cite{Pal:1991pm,Balantekin:2006sw}. If the Dirac NMM is generated at an energy scale $\Lambda \sim 1 \,\mathrm{TeV}$, it may be possible to obtain an NMM as high as $\mu_\nu \sim 10^{-15}\,\mu_B$, without exceeding current bounds on neutrino masses \cite{Bell:2005kz,Bell:2006wi}. 

For Majorana neutrinos, the relevant NMMs are instead \textit{transition} magnetic moments, for which the neutrino flavour changes in the interaction. In this case, the correction to the neutrino mass is typically suppressed with respect to the NMM (due to the different flavour symmetries involved) \cite{Bell:2006wi} and so much larger NMMs can be achieved (see e.g.~Refs.~\cite{Georgi:1990se,Davidson:2005cs,Bell:2006wi,Xing:2012gd}) without exceeding the neutrino mass bounds. The authors of Ref.~\cite{Lindner:2017uvt} consider a number of concrete realisations with large Majorana NMMs and find that it is possible to obtain a transition NMM of $\mu_{\nu_e \nu_\mu} \sim 10^{-12}\,\mu_B$ or larger with New Physics at the TeV scale. Note that because the flavour of the scattered neutrino is not measured in lab-based experiments, constraints on diagonal and transition NMMs are typically equal. 

Experimental constraints on $\mu_\nu$ are typically much weaker than the theoretical expectations for a Dirac neutrino, while current and near-future experiments may be sensitive to anomalously large Majorana NMMs. A measurement of a large NMM would therefore hint at the Majorana nature of the neutrino, as well as providing evidence New Physics beyond the Weak Scale.  A compilation of current constraints can be found in Ref.~\cite{Olive:2016xmw} but we highlight some of the strongest here:

\begin{itemize}
\item Large neutrino magnetic moments would increase the rate of energy loss in red giant stars, leading to an increase in the core mass at which helium burning becomes relevant \cite{Raffelt:1990pj,Raffelt:1999gv}.  Observations of red giants can then be used to place constraints, giving $\mu_\nu < 2.2 \times 10^{-12} \,\,\mu_B$ (90\% CL) \cite{Arceo-Diaz:2015pva}. 

\item Measurements of the electron recoil spectra due to solar neutrinos may also be used to constrain an anomalously large neutrino magnetic moment \cite{Grifols:2004yn,Liu:2004ny,Arpesella:2008mt}. Using 1292 live days of solar neutrino data, the BOREXINO experiment obtain $\mu_\nu < 2.8 \times 10^{-11} \,\,\mu_B$ (90\% CL) \cite{Borexino:2017fbd}. Note that oscillations should modify the flavour composition of solar neutrinos, meaning that measurements using solar and reactor neutrinos may differ.

\item The strongest constraint on the neutrino magnetic moment of reactor $\overline{\nu}_e$ comes from the GEMMA experiment, which measures $\overline{\nu}_e$-$e$ scattering. The resulting limit is $\mu_\nu < 2.9 \times 10^{-11} \,\,\mu_B$ (90\% CL) \cite{Beda:2013mta}.
\end{itemize}

\begin{figure*}[t!]
\centering
\includegraphics[width=0.49\textwidth]{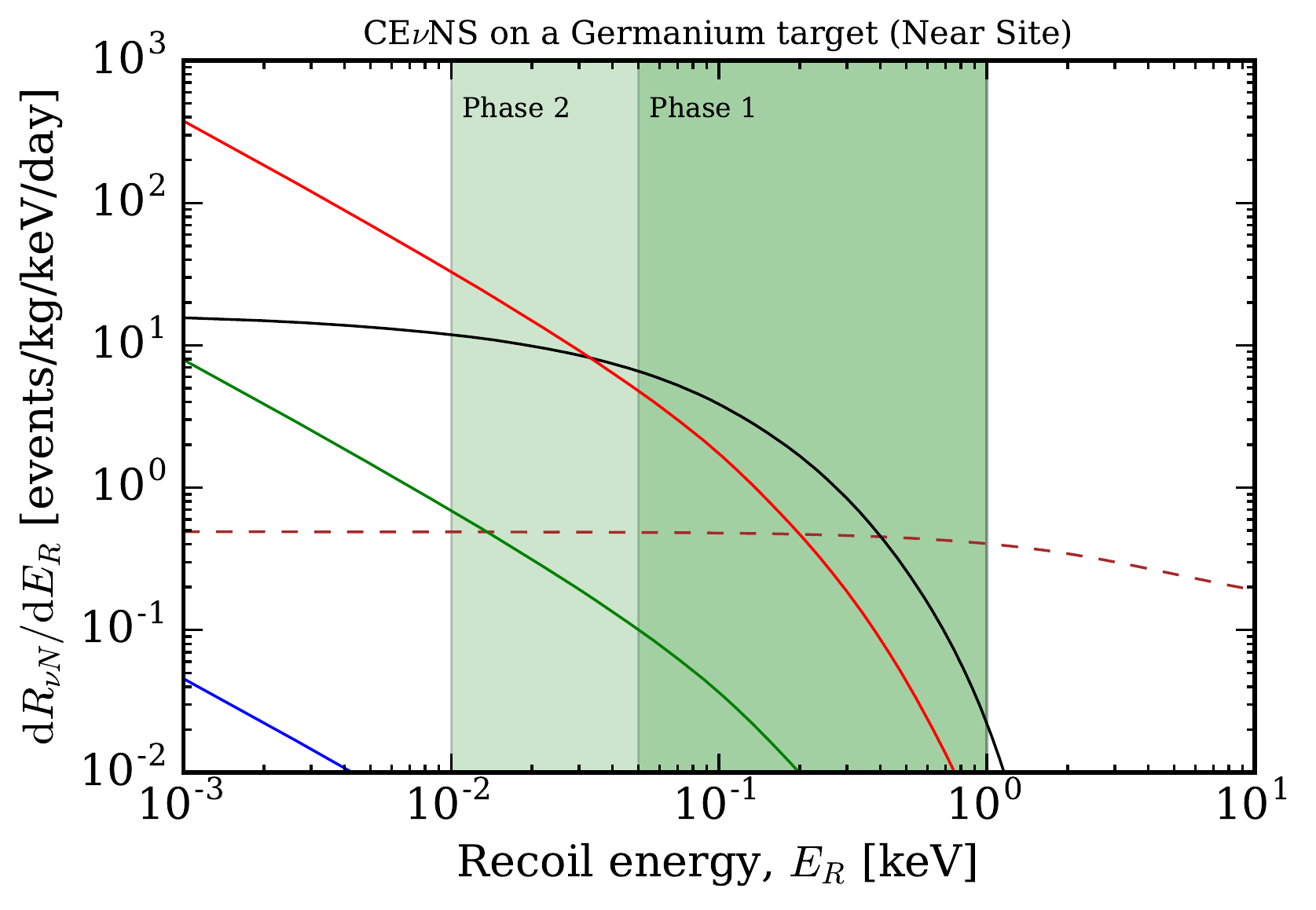}
\includegraphics[width=0.49\textwidth]{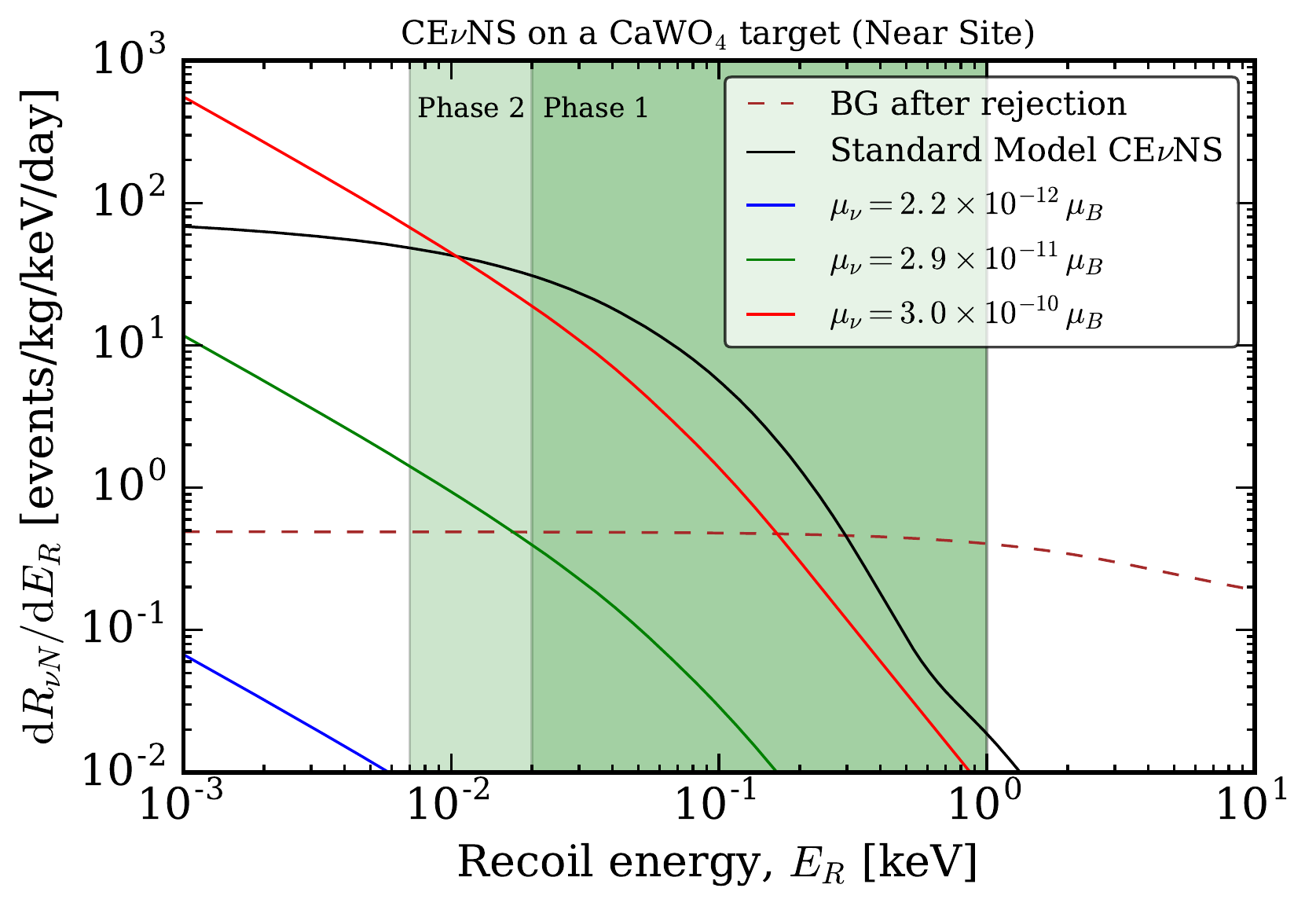}
\caption{\textbf{Nuclear recoil rate due to neutrino magnetic moment scattering.} We show the recoil rate off a Germanium target (\textbf{left}) and CaWO$_4$ target (\textbf{right}) due to a flux of Chooz Reactor neutrinos for Standard Model \CEvNS (solid black) as well as the contribution for a range of values of the neutrino magnetic moment $\mu_\nu$ (solid color). The shaded green regions show the Phase 1 and Phase 2 regions of interest (with thresholds given in Tab.~\ref{tab:experiments}). The dashed brown line shows the background after rejection (see Fig.~\ref{fig:backgrounds}).}
\label{fig:rate_magnetic}
\end{figure*}

The cross section for nuclear scattering from the neutrino magnetic moment $\mu_\nu$ is given by \cite{Vogel:1989iv}:
\begin{align}
\frac{\mathrm{d}\sigma_{\nu-N}^\mathrm{mag.}}{\mathrm{d}E_R} = \frac{\pi \alpha^2 \mu_\nu^2 Z^2}{m_e^2} \left(\frac{1}{E_R} - \frac{1}{E_\nu} + \frac{E_R}{4E_\nu^2}\right) F^2(E_R)\,.
\end{align}
This is the charge-dipole interaction, which receives a coherence enhancement from the charge of the nucleus $Z^2$ but which does not interfere with the Standard Model \CEvNS interaction. We neglect dipole-dipole interactions, which are chirality-flipping \cite{Vogel:1989iv} and therefore incoherent. The resulting cross section is proportional to the nuclear magnetic moment $\mu_N$ and is typically sub-dominant. In Fig.~\ref{fig:rate_magnetic}, we show the nuclear recoil rate for a Germanium target (left) and a $\CaWO$ target (right), comparing the Standard Model \CEvNS rate with that expected for a range of values of the neutrino magnetic moment $\mu_\nu$.

\begin{figure}[t!]
\centering
\includegraphics[width=\textwidth]{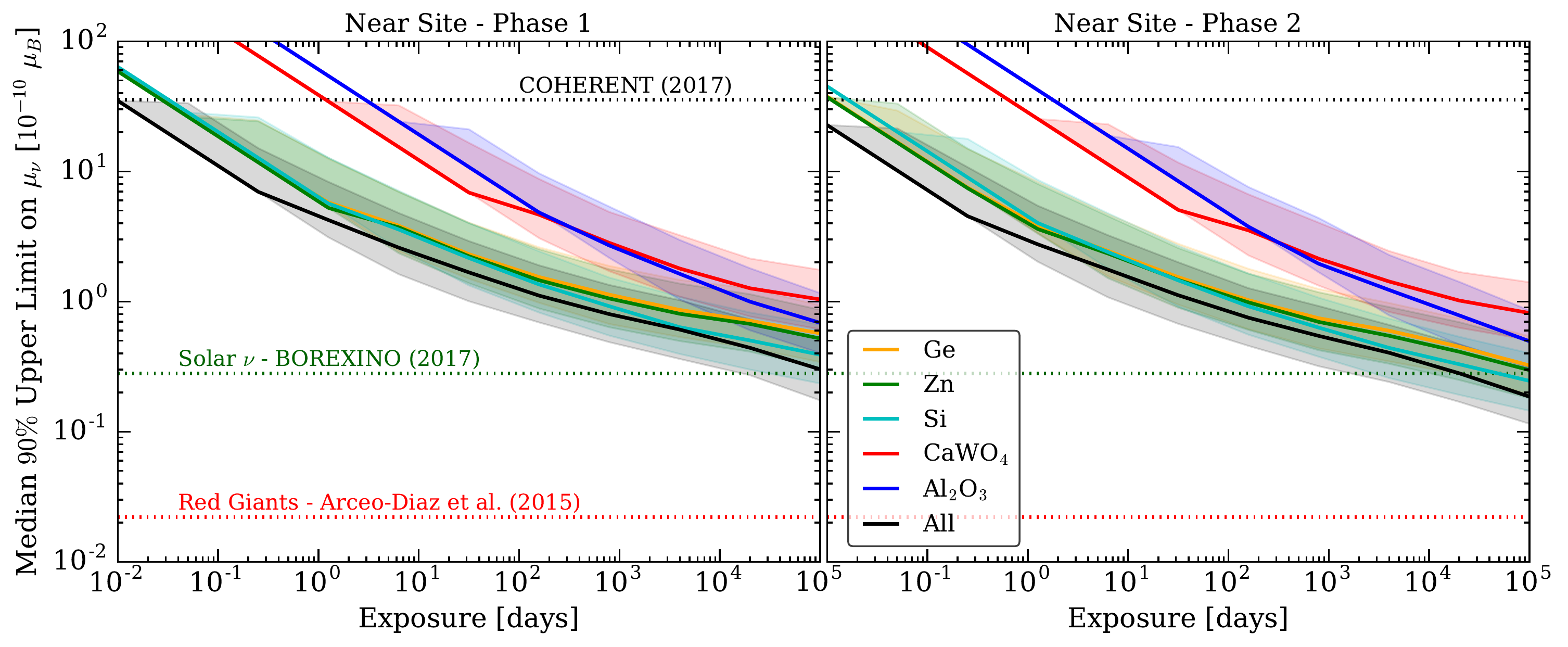}
\includegraphics[width=\textwidth]{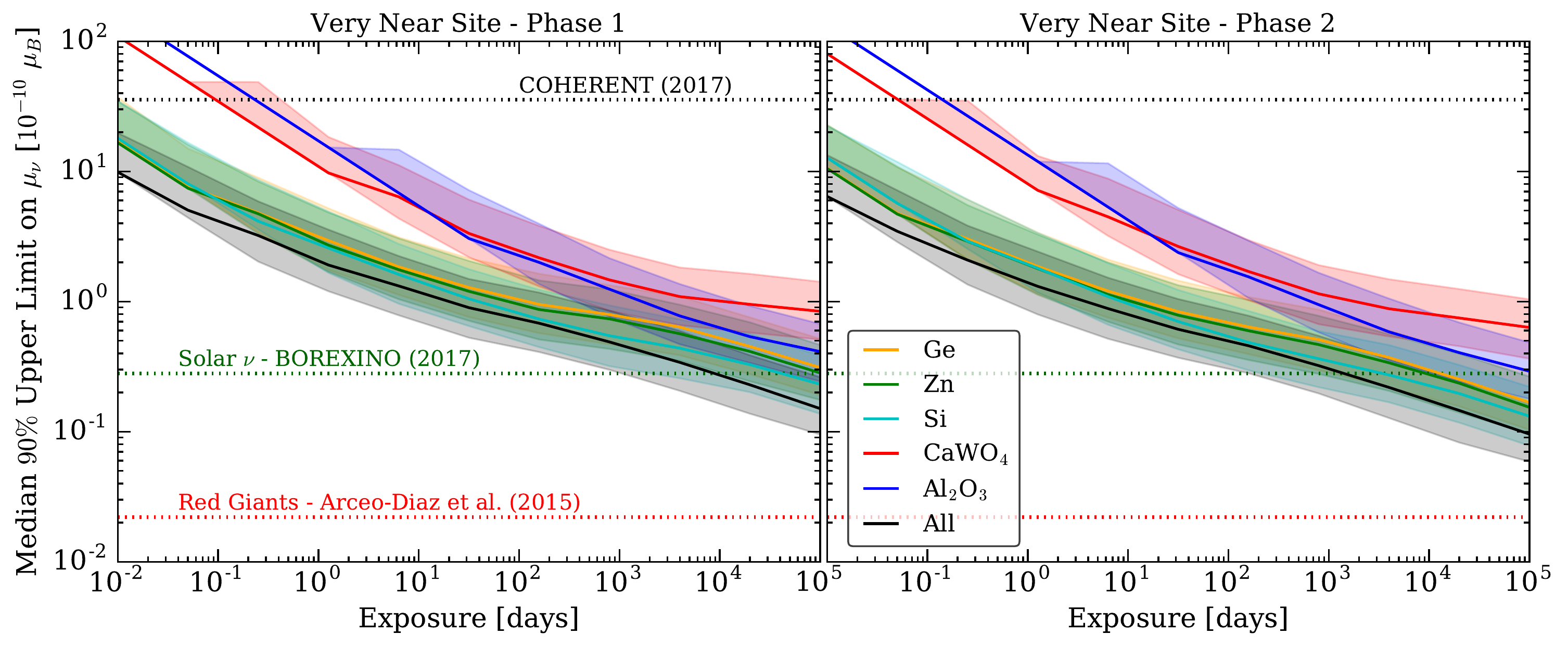}
\caption{\textbf{Projected 90\% Upper Limits on the neutrino magnetic moment as a function of exposure.} We show projections for the Near Site (top) and the Very Near Site (bottom), as well as for Phase 1 (left panels) and the lower-threshold Phase 2 (right panels). See Tab.~\ref{tab:experiments} for more details. We also show selected complementary constraints \cite{Borexino:2017fbd,Arceo-Diaz:2015pva} which are detailed in the text. The GEMMA (2013) limit \cite{Beda:2013mta} from observations of reactor $\overline{\nu}_e$ is at roughly the same level as BOREXINO (2017).} 
\label{fig:MuNuLimits}
\end{figure}

In Fig.~\ref{fig:MuNuLimits}, we show the projected constraints on the neutrino magnetic moment as a function of the detector exposure. On the top row we show results for the Near Site (NS) while on the bottom row we show results for the Very Near Site (VNS). The left and right columns show Phase 1 and Phase 2 respectively (though we remind the reader than we fix the payload mass to the Phase 1 mass and vary only the energy threshold in this case). 

Even in Phase 1, the larger Ge, Zn and Si detectors will become competitive with current constraints from COHERENT with just a few hours exposure, with $\CaWO$ and $\sapphire$ detectors surpassing COHERENT on timescales of a few days.  Reducing the energy threshold of the detectors (moving from the left to the right column of Fig.~\ref{fig:MuNuLimits}) strengthens the constraints by roughly a factor 2 at fixed exposure. The reason for this is clear from Fig.~\ref{fig:rate_magnetic}: the neutrino magnetic moment produces a spectrum which rises rapidly as $E_R^{-1}$ towards lower recoil energies meaning that lowering the energy threshold enhances the NMM contribution relative to the standard \CEvNS signal.

We note that for exposures around $10^2$ days, the upper curves for Ge, Zn and Si begin to flatten, as the experiments become dominated by systematic uncertainties on the backgrounds. At larger exposures ($>10^3$ days), the limits begin to strengthen more rapidly once again as differences in the spectra between signal and background allow them to be discriminated (c.f.~Ref.~\cite{Billard:2013qya}).  For experiments with smaller payloads and lower thresholds ($\CaWO$ and $\sapphire$), this systematics-dominated regime sets in only at much smaller values of the magnetic moment.


Depending on the detector setup and location, constraints on the neutrino magnetic moment from coherent nuclear scattering could become competitive with other terrestrial constraints \cite{Beda:2013mta,Borexino:2017fbd} at exposures of $10^4$-$10^5$ days. Constraining below the level of $\mu_\nu \sim 10^{-11}\,\mu_B$ (in competition with astrophysical constraints \cite{Arceo-Diaz:2015pva}) would require even larger exposure, though as we have discussed lower threshold experiments are more effective in this regime.

\begin{figure}[t!]
\centering
\includegraphics[width=0.7\textwidth]{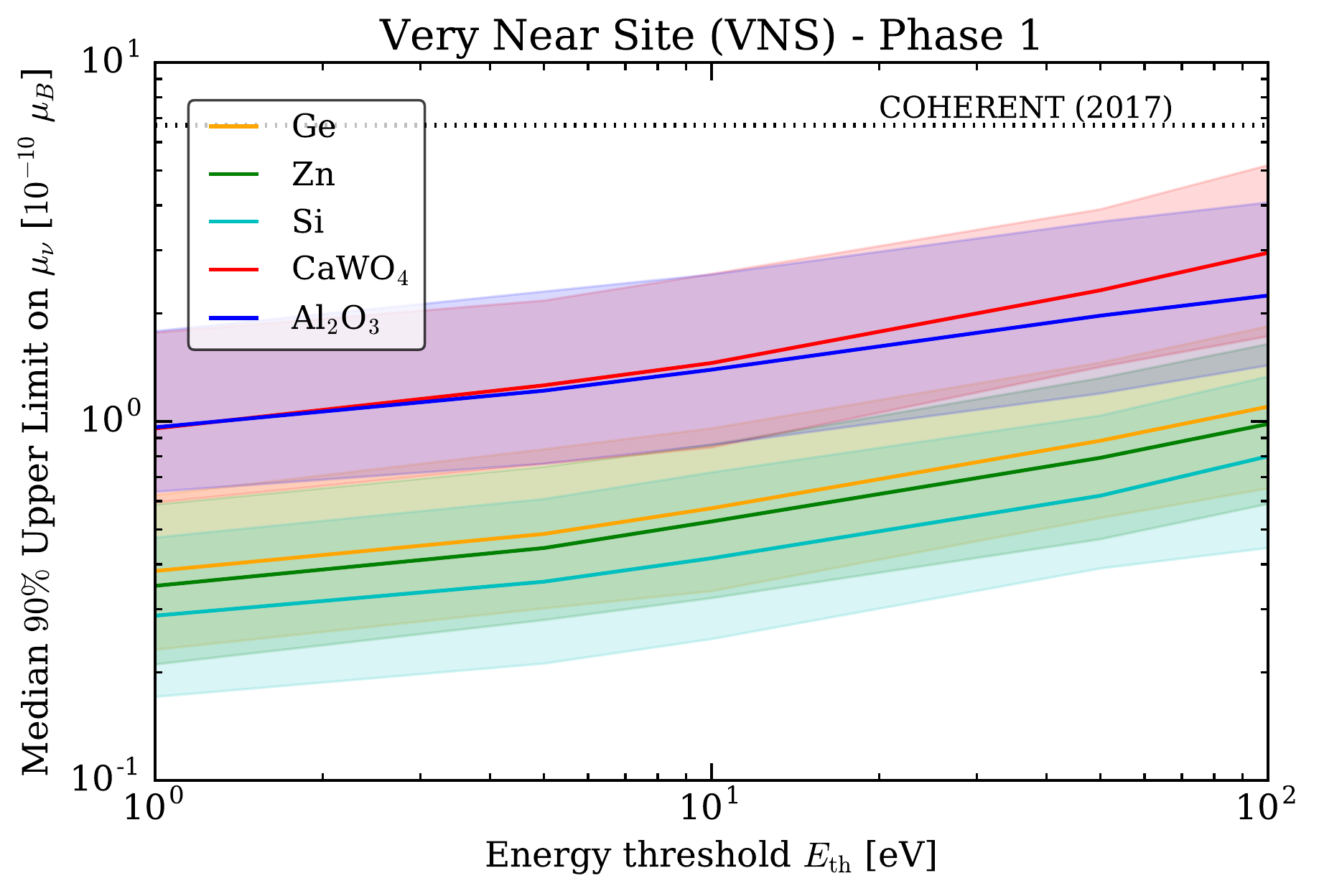}
\caption{\textbf{Projected 90\% Upper Limits on the neutrino magnetic moment as a function of energy threshold.} We show projections for different detector targets for Phase 1 at the Very Near Site, assuming a 1-year exposure. For Zn, Ge and Si, we assume a 500g payload, while for $\CaWO$ and $\sapphire$, we assume 6.84g and 4.41g respectively. See Tab.~\ref{tab:experiments} for more details.} 
\label{fig:MuNuLimits_threshold}
\end{figure}

In Fig.~\ref{fig:MuNuLimits_threshold}, we explore the impact of energy thresholds in more detail, showing the projected limits on the neutrino magnetic moment as a function of $E_\mathrm{th}$. As we have seen in Fig.~\ref{fig:MuNuLimits}, limits from  $\CaWO$ and $\sapphire$ targets are in general weaker owing to the smaller payload masses. However, this could be compensated by a lower energy threshold: $\CaWO$ and $\sapphire$ detectors with a $1 \,\mathrm{eV}$ threshold give limits which are as strong as Zn, Ge and Si detectors with a $100 \,\mathrm{eV}$ threshold. If such low thresholds can be more easily achieved with $\CaWO$ and $\sapphire$, these detectors can give competitive constraints on New Physics despite a factor of 100 smaller payload. 

\section{Non-Standard Interactions (NSI)}
\label{sec:NSI}

Going beyond anomalously large neutrino magnetic moments, as we explored in the previous section, we can consider more generic modifications of the neutrino-nucleus interaction. We parametrise new Non-Standard Interactions (NSI) by introducing new 4-fermion interactions into the Standard Model \cite{Wolfenstein:1977ue,Guzzo:1991cp,Davidson:2003ha,Barranco:2005yy}:
\begin{align}
\mathcal{L}^{\mathrm{NSI}} &= -\epsilon_{\alpha \beta}^{fP} 2 \sqrt{2} G_F (\overline{\nu}_\alpha \gamma_\rho L \nu_\beta)(\overline{f} \gamma^\rho P f)\,,
\end{align}
where $f = e,\, u,\, d$ is a first generation SM fermion, $\alpha,\, \beta = e,\, \mu,\,\tau$ are lepton flavours and $P = L, \, R$ are the left and right handed projection operators. The coefficients $\epsilon_{\alpha \beta}^{fP}$ parametrise the strength of the new NSI operators. For example, $\epsilon_{ee}^{uL}$ parametrises the (modification of the) coupling of the $\nu_e$ current to the left-handed current of $u$ quarks. This is a so called "non-universal" coupling as it concerns only the $e$-flavour neutrinos. Instead, "flavour-changing" couplings such as $\epsilon_{e\mu}^{uL}$ allow transformations between different neutrino flavours. We note that the NSI couplings simply change the overall strength of the neutrino-nucleus interaction. Thus, for a given neutrino flavour, the shape of the \CEvNS nuclear recoil spectrum (illustrated in Fig.~\ref{fig:rate_magnetic} as a solid black line) will not be affected by NSI operators, though the overall normalisation will.

A useful summary of constraints on NSI interactions can be found Table 1 of Ref.~\cite{Scholberg:2005qs}, many of which are compiled in Ref.~\cite{Davidson:2003ha} (see also Ref.~\cite{Biggio:2009nt}). In particular, the $\epsilon_{\mu\mu}$ couplings are strongly constrained by the NuTeV experiment \cite{Zeller:2001hh}, which measured the ratios of neutral current to charged current neutrino-nucleon cross sections. In addition, tight constraints can be placed on the flavour-changing couplings $\epsilon_{e \mu}$ from constraints on $\mu \rightarrow e$ conversion in nuclei \cite{deGouvea:2000cf}.\footnote{The experimental upper bound gives a limit $|\eps_{e\mu}^{qV}| \lesssim 1.5\times 10^{-3}$ \cite{deGouvea:2000cf}.}

We focus instead on the non-universal $\epsilon_{ee}$  and the flavor-changing $\epsilon_{e\tau}$ couplings, which can be probed using \CEvNS from a beam of reactor $\overline{\nu}_e$ neutrinos. We focus on modifications of the vector-current couplings to quarks, since the vector current interactions receive a coherent nuclear enhancement and therefore typically dominant. To this end, we define:
\begin{align}
\begin{split}
\epsilon_{ee}^{qV} &= \epsilon_{ee}^{qL} + \epsilon_{ee}^{qR}\,,\\
\epsilon_{e\tau}^{qV} &= \epsilon_{e\tau}^{qL} + \epsilon_{e\tau}^{qR}\,,
\end{split}
\end{align}
where $q = u, \,d$. These NSI couplings lead to a modification of the weak nuclear charge (see e.g.~Ref.~\cite{Lindner:2016wff}) for a nucleus of $N$ neutrons and $Z$ protons:
\begin{align}
\begin{split}
\label{eq:Q_NSI}
Q_W^2 \rightarrow Q_\mathrm{NSI}^2 &= 4 \left[ N\left(-\frac{1}{2} + \epsilon_{ee}^{uV} + 2 \epsilon_{ee}^{dV}\right) + Z \left(\frac{1}{2} - 2 \sin^2\theta_W + 2 \epsilon_{ee}^{uV} + \epsilon_{ee}^{dV}\right)\right]^2\\
&\qquad+ 4  \left[ N(\epsilon_{e \tau}^{uV} + 2 \epsilon_{e \tau}^{dV}) + Z(2 \epsilon_{e \tau}^{uV} + \epsilon_{e \tau}^{dV})\right]^2\,.
\end{split}
\end{align}

\begin{figure}[t!]
\centering
\includegraphics[width=0.49\textwidth]{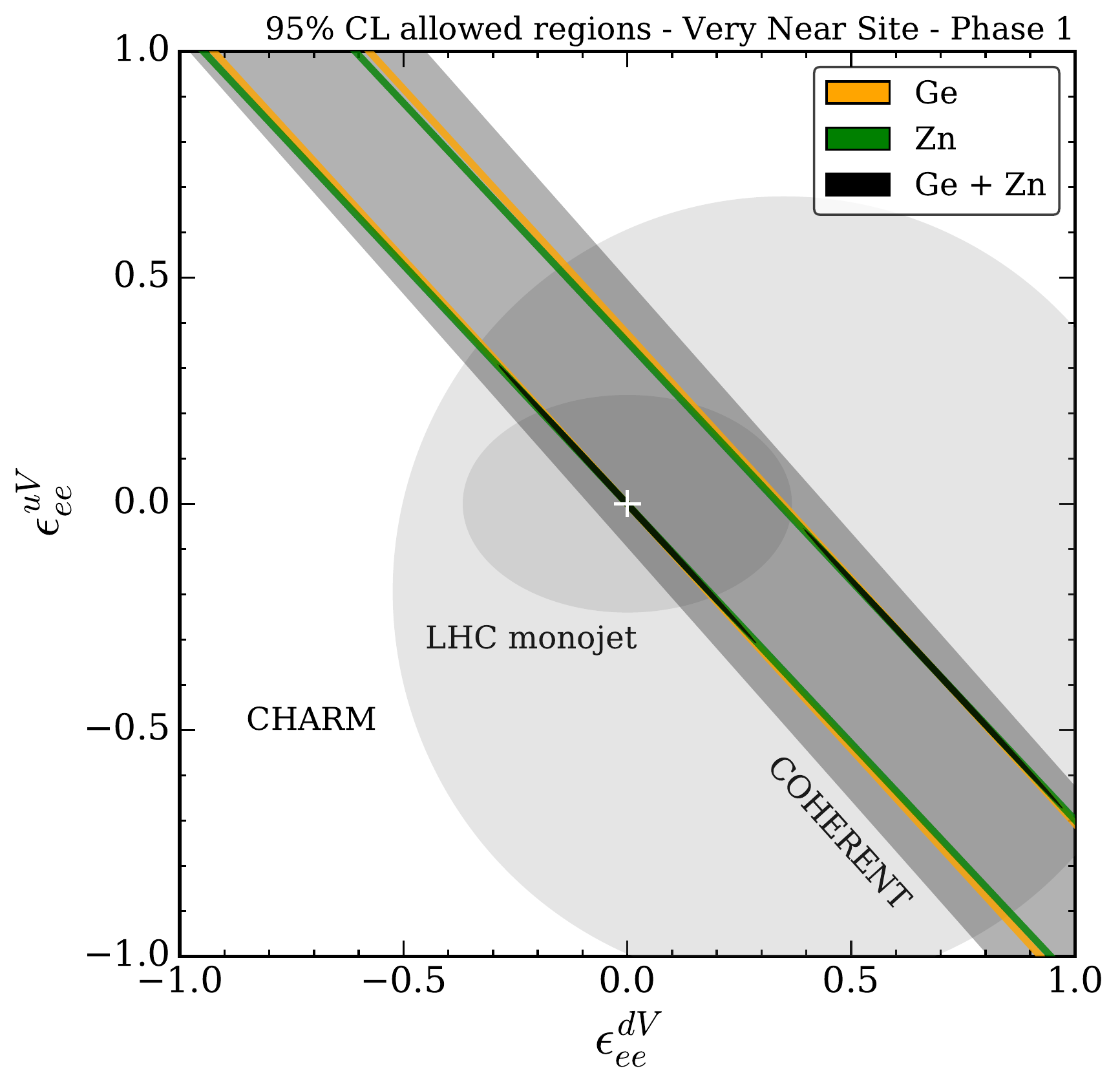}
\includegraphics[width=0.49\textwidth]{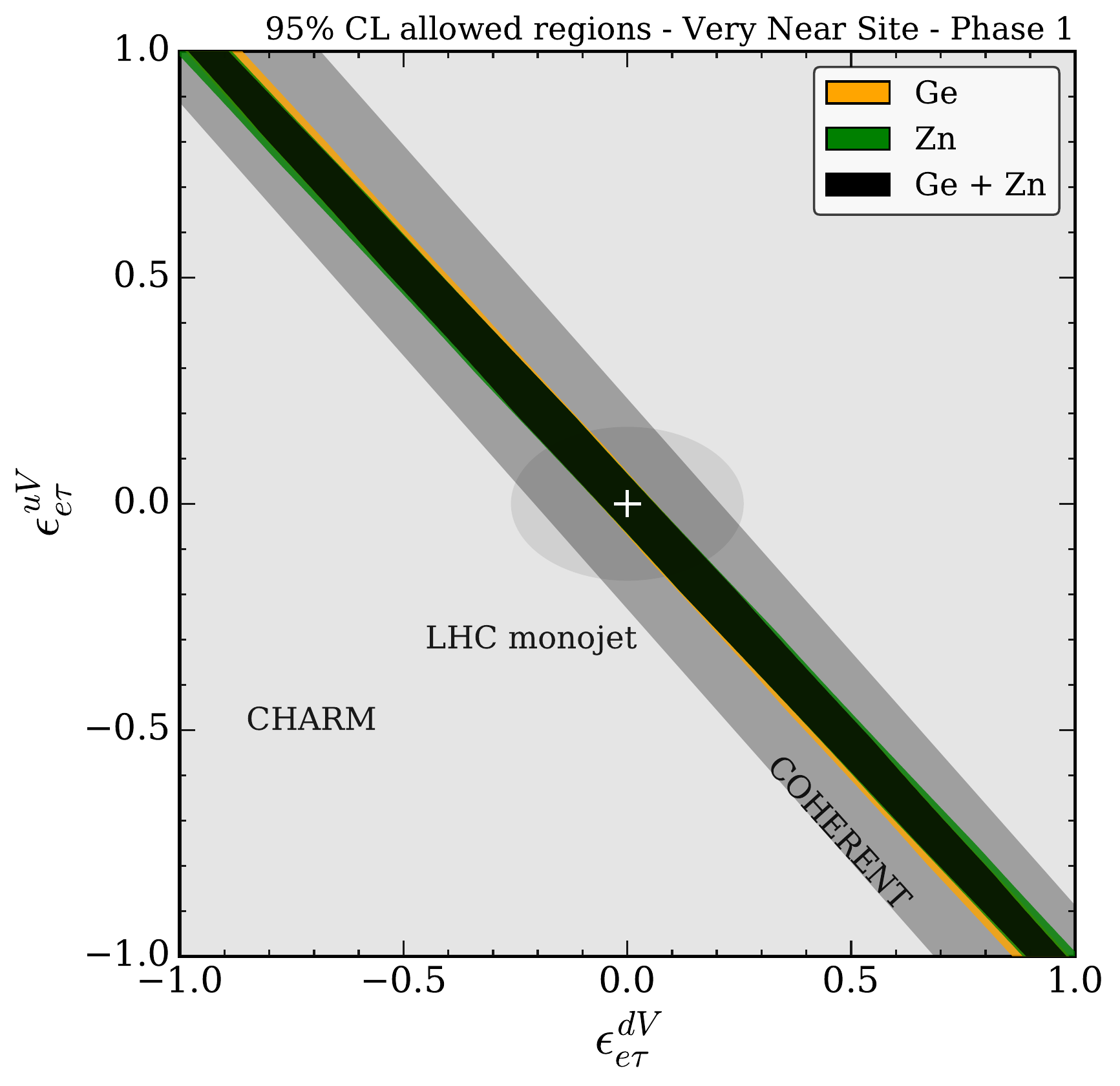}
\includegraphics[width=0.49\textwidth]{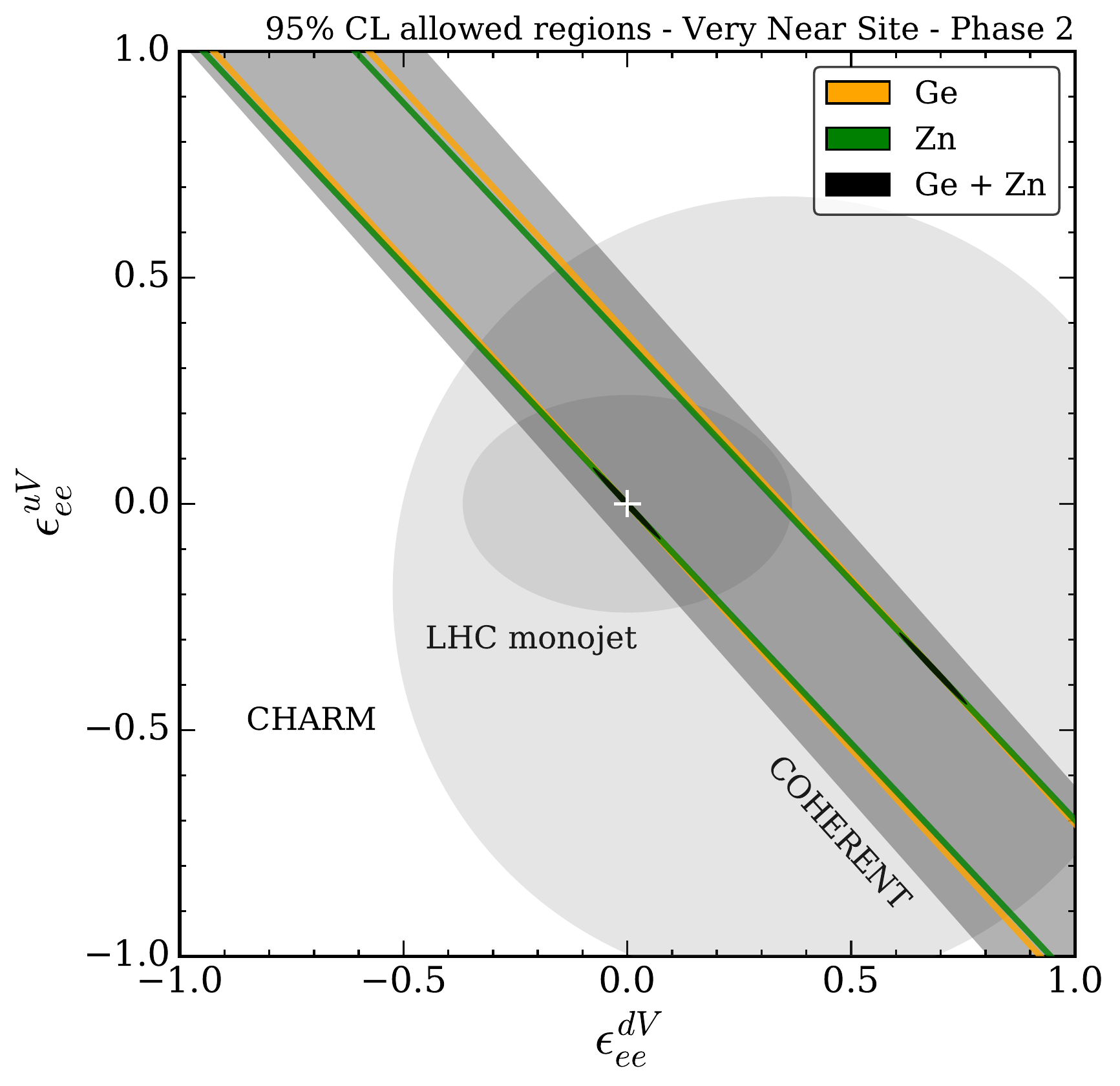}
\includegraphics[width=0.49\textwidth]{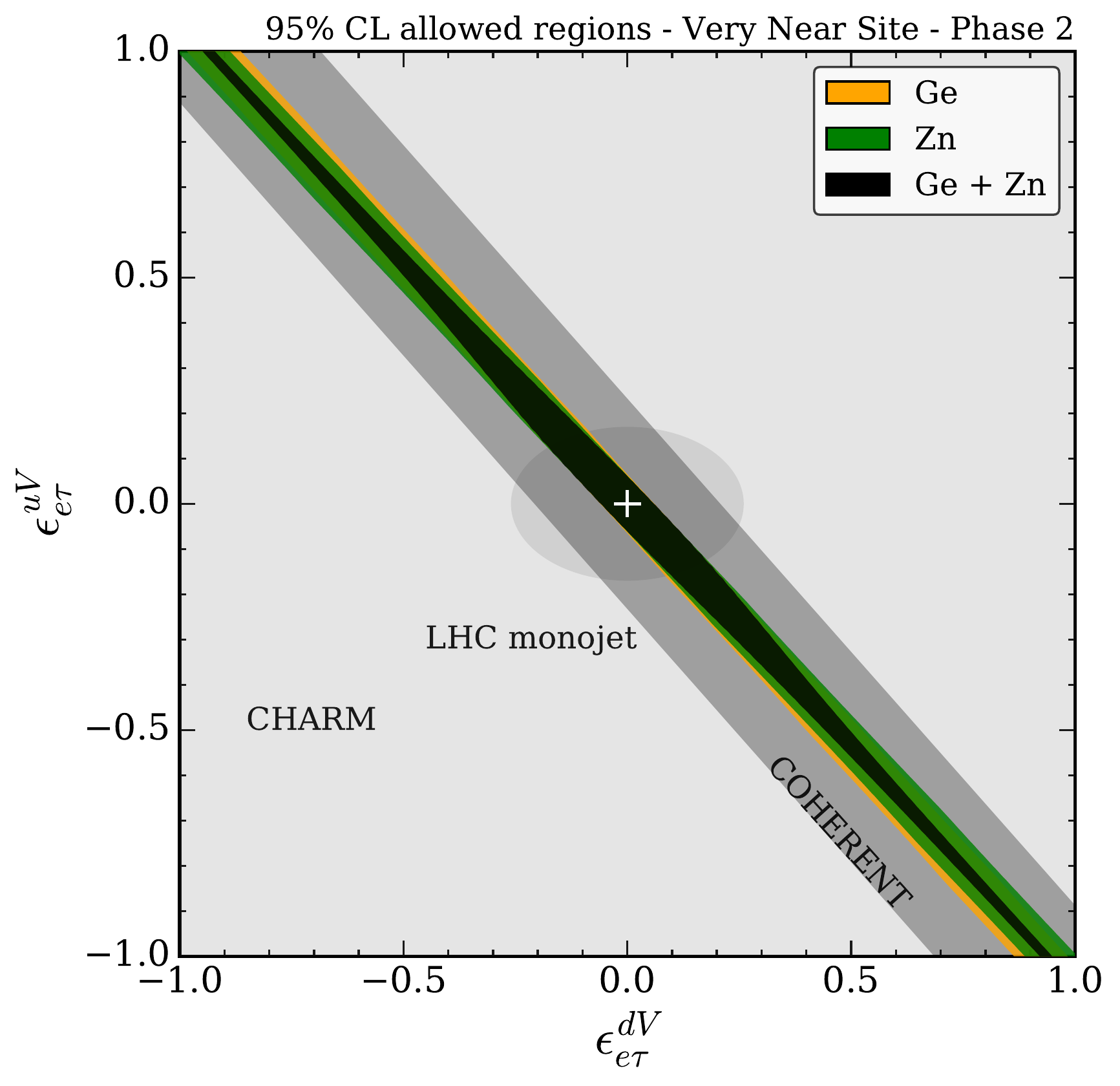}
\caption{\textbf{Projected 95\% CL allowed regions for non-standard interactions (NSI) at the Very Near Site, using Ge and Zn targets.} The colored shaded regions show projected constraints for Phase 1 (top row) and Phase 2 (bottom row) detectors at the Very Near Site. We show constraints on flavor-conserving ($ee$) interactions (left column) and flavor-changing ($e\tau$) interactions (right column). The grey shaded regions come from the CHARM experiment \cite{Dorenbosch:1986tb}, LHC monojet searches \cite{Friedland:2011za} and the COHERENT experiment \cite{Akimov:2017ade}.}
\label{fig:NSI_VNS_GeZn}
\end{figure}

\begin{figure}[t!]
\centering
\includegraphics[width=0.49\textwidth]{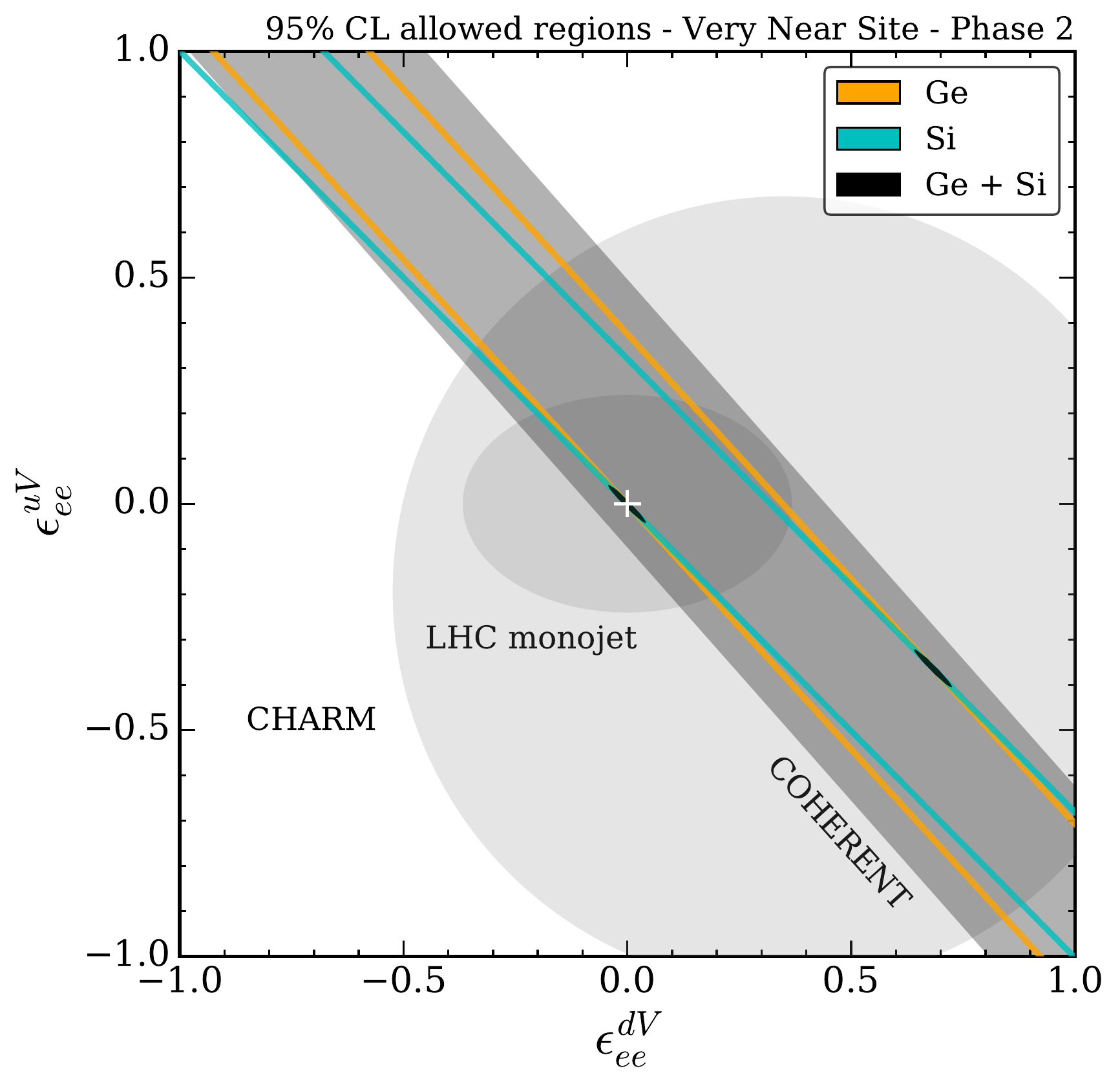}
\includegraphics[width=0.49\textwidth]{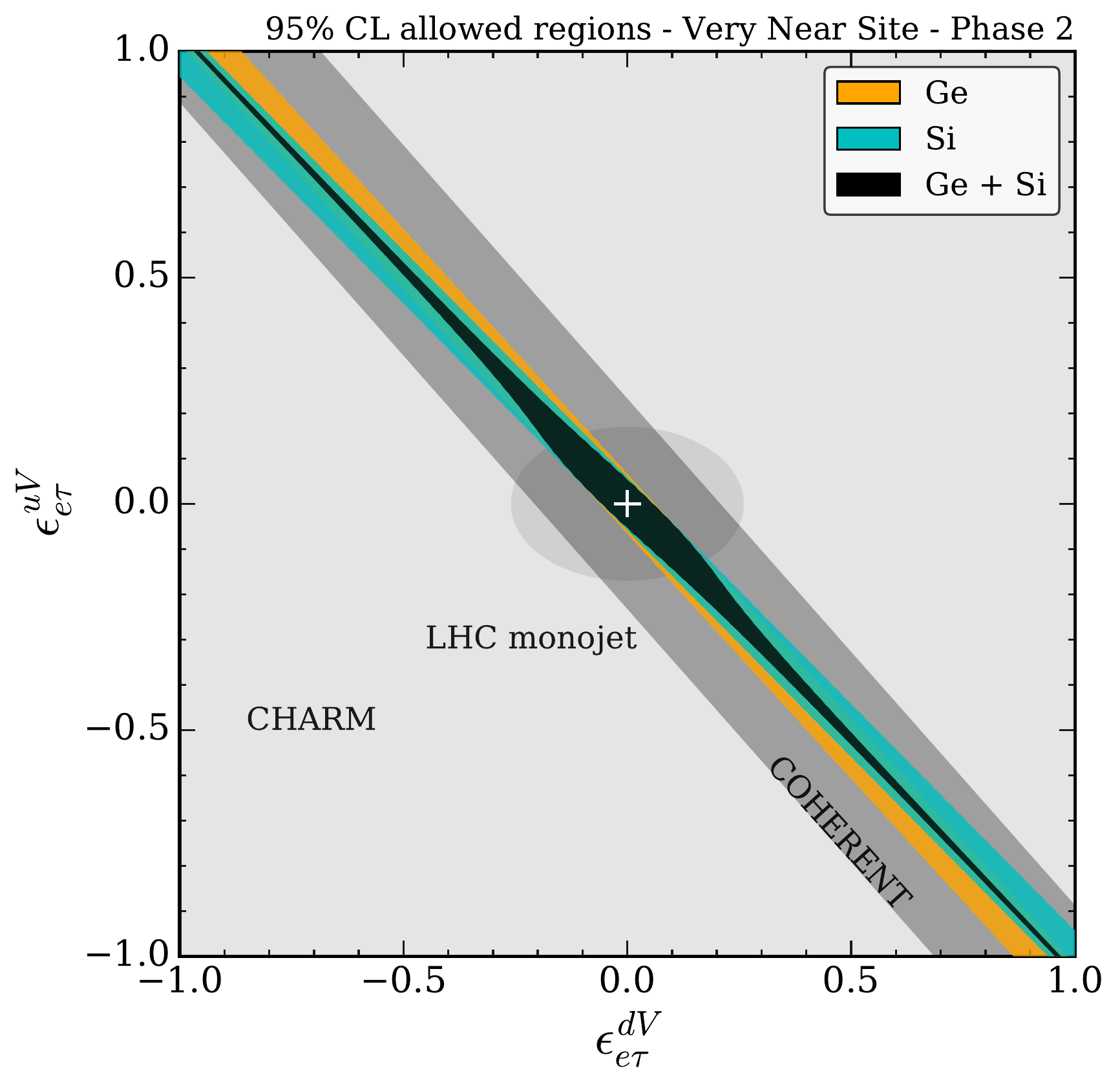}
\caption{\textbf{Projected 95\% CL allowed regions for non-standard interactions (NSI) at the Very Near Site, using Ge and Si targets.} Same as Fig.~\ref{fig:NSI_VNS_GeZn}, but for Ge and Si targets in Phase 2 only.}
\label{fig:NSI_VNS_GeSi}
\end{figure}

\begin{figure}[t!]
\centering
\includegraphics[width=0.49\textwidth]{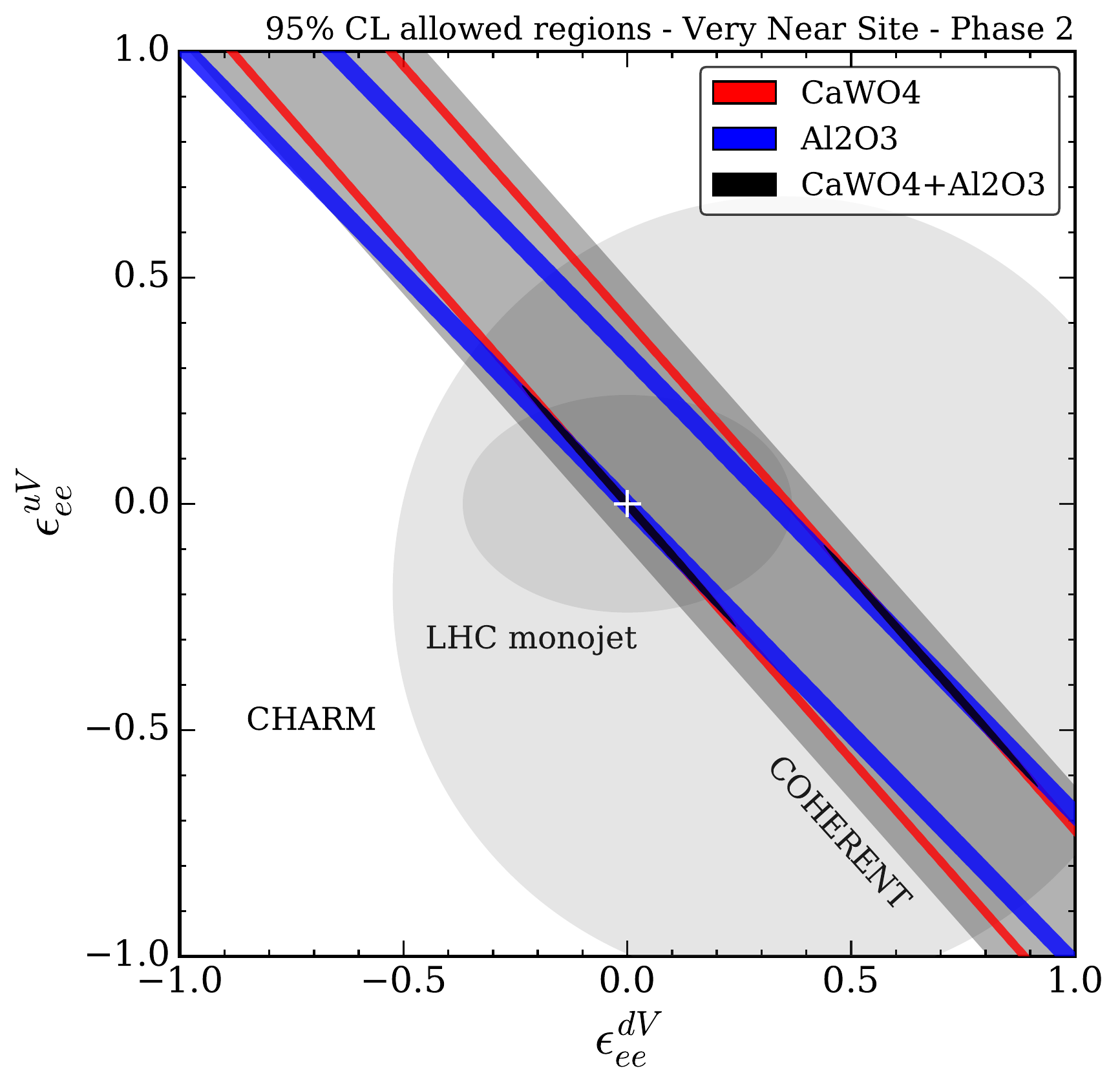}
\includegraphics[width=0.49\textwidth]{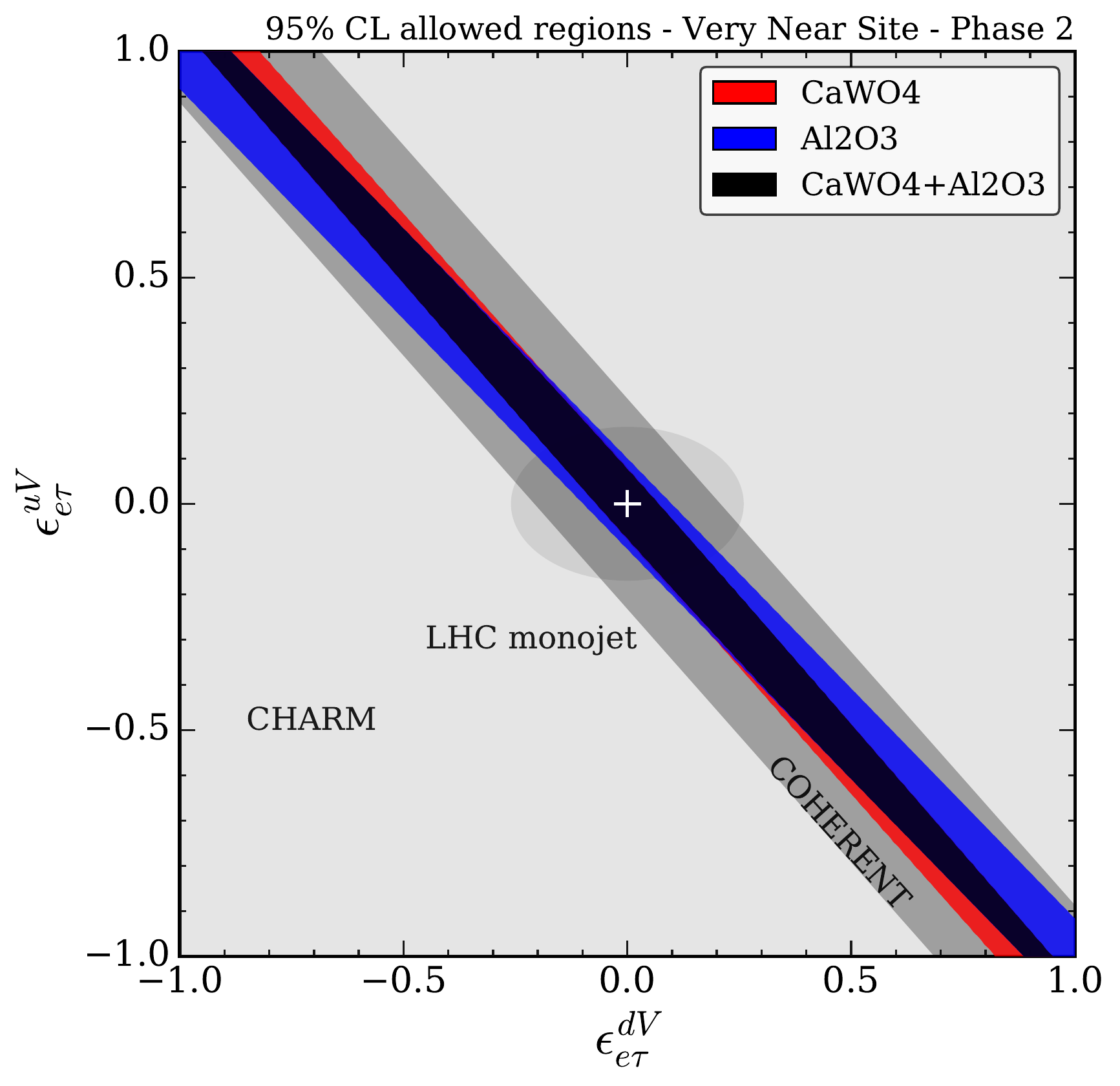}
\caption{\textbf{Projected 95\% CL allowed regions for non-standard interactions (NSI) at the Very Near Site, using CaWO$_4$ and Al$_2$O$_3$ targets.} Same as Fig.~\ref{fig:NSI_VNS_GeZn}, but for CaWO$_4$ and Al$_2$O$_3$ targets in Phase 2 only.}
\label{fig:NSI_VNS_CaWO4Al2O3}
\end{figure}

\begin{figure}[t!]
\centering
\includegraphics[width=0.49\textwidth]{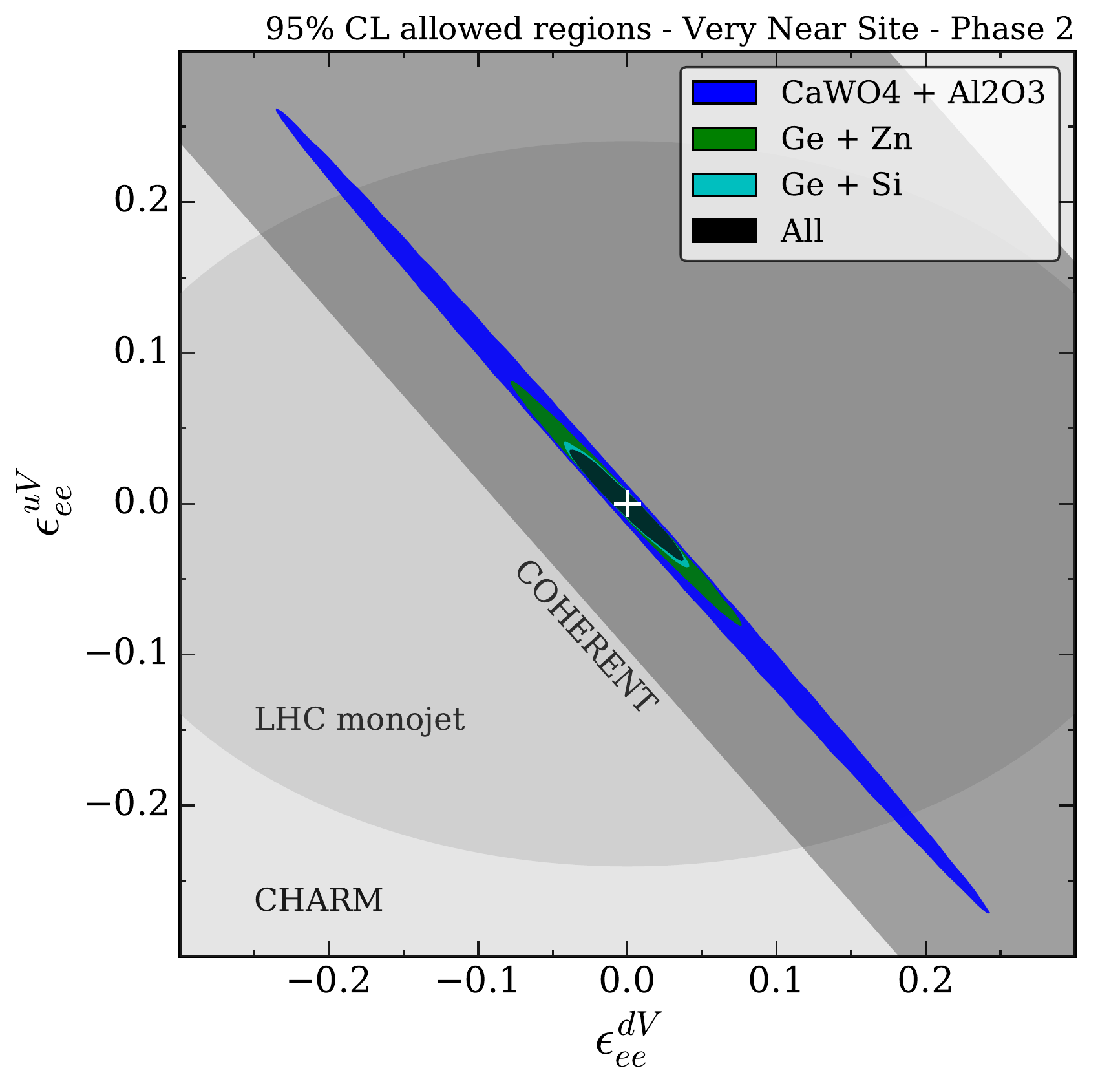}
\includegraphics[width=0.49\textwidth]{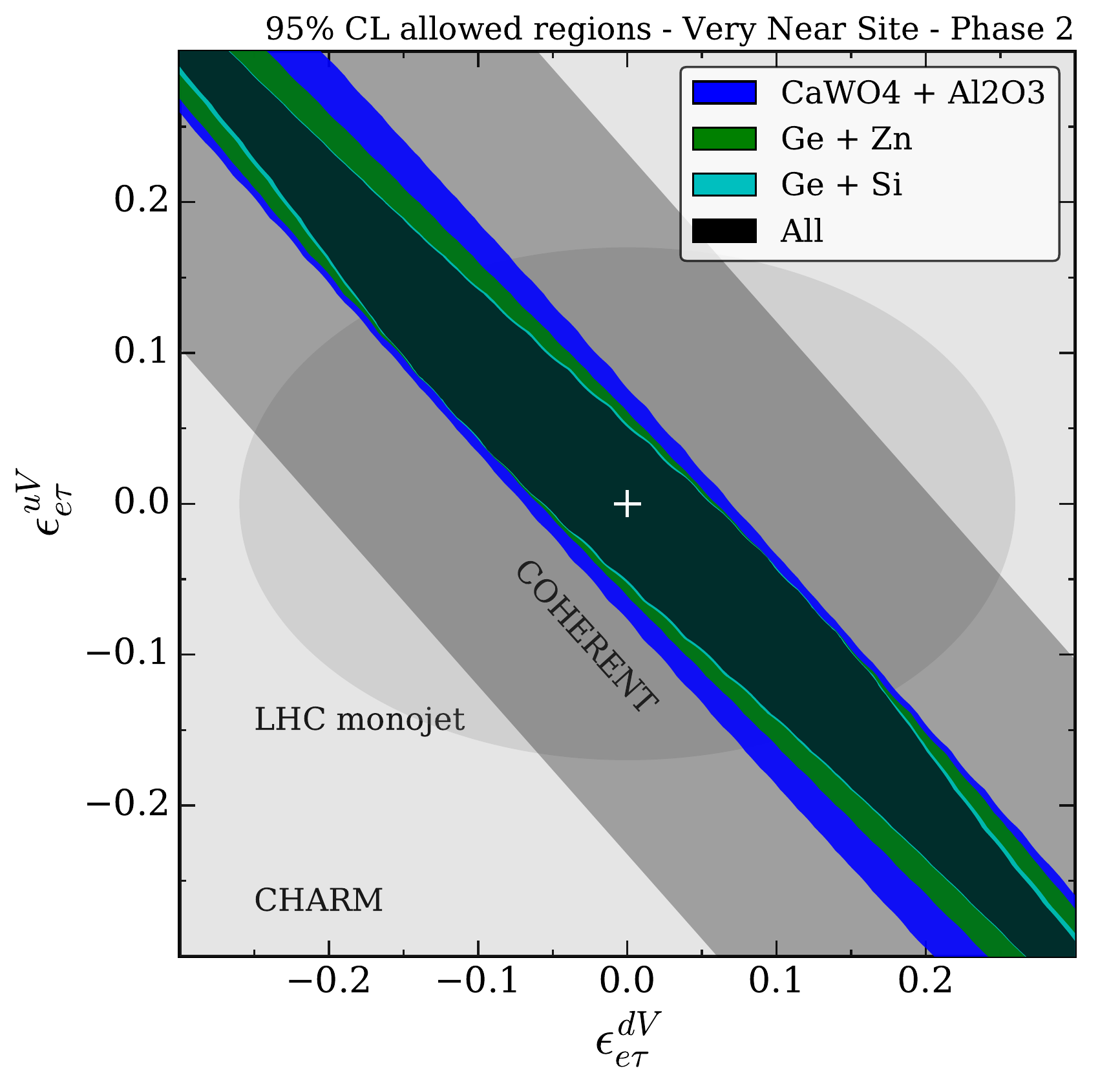}
\caption{\textbf{Projected 95\% CL allowed regions for non-standard interactions (NSI) at the Very Near Site, using different target combinations.} Same as Fig.~\ref{fig:NSI_VNS_GeZn}, but for a range of target combinations (in Phase 2 only). Here, we zoom in on the Standard Model point, shown as a white cross.}
\label{fig:NSI_VNS_combined}
\end{figure}

In \Cref{fig:NSI_VNS_GeZn,fig:NSI_VNS_GeSi,fig:NSI_VNS_CaWO4Al2O3,fig:NSI_VNS_combined}, we show the projected 95\% CL allowed regions for $(\epsilon_{ee}^{uV}, \, \epsilon_{ee}^{dV})$ (left panels) and $(\epsilon_{e\tau}^{uV}, \, \epsilon_{e\tau}^{dV})$ (right panels) assuming a Standard Model-only signal. Each figure shows results for a different combination of experimental targets, in all cases located at the Very Near Site. Results for the Near Site are qualitatively similar, but typically the allowed regions in that case are larger owing to the smaller neutrino flux. For comparison, complementary constraints are shown in grey. CHARM constraints \cite{Dorenbosch:1986tb} come from measurements of $\nu_e-N$ inelastic scattering. Also shown are LHC constraints from monojet + missing energy searches \cite{Friedland:2011za}. In this case, NSI may alter the production cross section for neutrinos, which manifest as missing energy in the detector. Finally, the limits reported by the COHERENT experiment \cite{Akimov:2017ade} are also shown in grey.

In Fig.~\ref{fig:NSI_VNS_GeZn}, we show results for Ge and Zn targets separately (yellow and green regions) as well as combined (black region), corresponding to the planned targets of the MINER experiment. The slope of the lines which form the allowed regions are fixed by the ratio of $N/Z$ which appears in Eq.~\eqref{eq:Q_NSI}. In the case of the non-universal couplings $(\epsilon_{ee}^{uV}, \, \epsilon_{ee}^{dV})$ (left column), there are two distinct allowed regions. The first is centred on the SM point (white cross), while the second is offset and represents the region where the contribution from NSI leads to $Q_\mathrm{NSI} \approx -Q_W$, mimicking a SM signal. In between the lines which form these allowed regions, the NSI contribution leads to a cancellation with the SM contribution, suppressing the \CEvNS rate. Note that the COHERENT observation leads to a single, connected allowed region. In that case both electron- and muon-flavour neutrinos contribute to the scattering, meaning that even when the $\nu_e$-$N$ scattering is suppressed, $\nu_\mu$-$N$ scattering can still provide a reasonable fit to the data. This highlights the complementarity between reactor and accelerator neutrino sources. 

Combining two targets with different ratios of $N/Z$ allows us to break degeneracies between $\epsilon_{ee}^{uV}$ and $\epsilon_{ee}^{dV}$ and reduce the allowed regions to closed areas of the parameter space. Though there is still some remaining degeneracy along the direction $\epsilon^{uV} + \epsilon^{dV} \sim 0$, the combination of Ge + Zn targets limits the couplings to two distinct regions (black regions in left panels of Fig.~\ref{fig:NSI_VNS_GeZn}).

The allowed regions for flavour-changing couplings $(\epsilon_{e\tau}^{uV}, \, \epsilon_{e\tau}^{dV})$ (right panels) are larger than in the flavor-conserving case. In this case, there can be no cancellation between the SM and NSI contributions (see Eq.~\eqref{eq:Q_NSI}) and a wider range of values can therefore reproduce the observed data. Even so, combining different targets can still improve constraints in this case; we see from the lower right panel of Fig.~\ref{fig:NSI_VNS_GeZn} that the Ge + Zn combined allowed region is beginning to shrink to a closed region in Phase 2.

We can also consider other combinations of targets. In Appendix.~\ref{app:NSI} we plot results for all targets, detector locations and phases; we show only selected results here. In Fig.~\ref{fig:NSI_VNS_GeSi} we show the projected allowed regions for Ge + Si (planned targets for the RICOCHET experiment) in Phase 2. In this case, the difference in $N/Z$ between the two experiments is larger than for Ge + Zn (the lines forming the allowed regions have noticeably different slopes for Ge and Si). This results in much smaller allowed regions when the two experiments are combined. 

In Fig.~\ref{fig:NSI_VNS_CaWO4Al2O3}, we show the allowed regions for a combination of $\CaWO$ and $\sapphire$ (the NUCLEUS experiment). In this case, the smaller payloads lead to wider allowed regions (especially in the case of $\sapphire$). For Zn, Ge and Si, the shape of the allowed regions was set by $N/Z$. For $\CaWO$ and $\sapphire$, there is no method of determining which nucleus the neutrino is scattering off, so we might expect wider, more complicated regions. In fact, for $\sapphire$, $\mathrm{Al}$ and $\mathrm{O}$ have very similar $N/Z$ ratios, so that both nuclei receive roughly the same rate enhancement for a given set of NSI couplings. For $\CaWO$, the rate is in fact dominated by scattering off $\mathrm{W}$\footnote{This can be seen in the solid black line in the right panel of Fig.~\ref{fig:rate_magnetic}; the `kink' in the rate above $\sim 0.5 \,\mathrm{keV}$ corresponds to the point where scattering on $\mathrm{W}$ begins to be kinematically disfavoured and scattering from lighter nuclei begins to dominate.}. This means that for low enough thresholds $\CaWO$ is effectively a single-nucleus target. Furthermore, the large neutron content of $\mathrm{W}$ means that relatively tight constraints can be obtained in spite of the smaller payloads. 


Finally, in Fig.~\ref{fig:NSI_VNS_combined}, we show the combined NSI constraints for different combinations of targets in Phase 2. These appear as black contours in \Cref{fig:NSI_VNS_GeZn,fig:NSI_VNS_GeSi,fig:NSI_VNS_CaWO4Al2O3}; we overlay them here and zoom in on the Standard Model point (white cross) for comparison. Focusing on flavor-conserving couplings (left panel), we see that the combined $\CaWO$+$\sapphire$ allowed regions bound the NSI couplings at the 25\% level, competitive with LHC constraints. This is in spite of the much smaller payloads compared with the Ge, Zn and Si targets. For combined Ge + Zn and Ge + Si analyses, the non-universal NSI couplings are constrained down to the level of 10\% and 5\% respectively, substantially improving over LHC constraints. We note that adding more targets beyond the combined Ge + Si analysis leads to little improvement. The size of the allowed regions is set ultimately by the targets with the most varied $A/Z$ values (in this case Ge and Si).

\subsection{Leptoquarks}

The constraints on model-independent NSI can also be recast as constraints on the masses and couplings of new, heavy leptoquarks \cite{Barranco:2007tz}. Leptoquarks are particles arising in theories beyond the Standard Model which carry the quantum numbers of both quarks and leptons \cite{Pati:1973uk}. Leptoquarks coupling to third-generation leptons and quarks have received much attention recently \cite{Tanaka:2012nw,Sakaki:2013bfa,Dorsner:2013tla,Gripaios:2014tna,Bauer:2015knc,Buttazzo:2017ixm,Dorsner:2018ynv} as a possible explanation for anomalies in flavour physics \cite{Lees:2012xj,Lees:2013uzd,Matyja:2007kt,Bozek:2010xy,Huschle:2015rga,Hirose:2016wfn,Aaij:2013qta,Aaij:2014ora,Aaij:2015yra,Aaij:2015oid,Aaij:2017vbb}. 

\CEvNS searches are instead sensitive to first-generation leptoquarks which would contribute to the process $q\nu \rightarrow q\nu$ via $s$-channel and $u$-channel exchange. Collider searches constrain possible leptoquark masses to be $> \mathcal{O}(1\,\mathrm{TeV})$ (e.g.~\cite{Abramowicz:2012tg,Khachatryan:2015vaa,Patrignani:2016xqp,DiLuzio:2017chi,Sirunyan:2018kzh}), meaning that leptoquark exchange gives rise to a contact interaction, whose contribution to the NSI couplings is of the form \cite{Davidson:1993qk}:
\begin{equation}
\label{eq:leptoquark}
\epsilon^{qV} = \frac{\lambda_q^2}{m_{LQ}^2} \frac{\sqrt{2}}{4 G_F}\,,
\end{equation}
where $\lambda_q$ is the leptoquark coupling to quark flavour $q$. Here, we will focus on neutrino flavour-conserving interactions and models in which either $\epsilon^{uV}$ or $\epsilon^{dV}$ is generated but not both (as is the case, for example, for specific couplings of the $\mathrm{SU}(2)_L$-singlet vector leptoquark \cite{Davidson:1993qk,Barbieri:2015yvd,Buttazzo:2017ixm}).

We recast some of the limits on $\epsilon^{uV}_{ee}$ presented in Sec.~\ref{sec:NSI} using Eq.~\eqref{eq:leptoquark}, fixing $\epsilon_{ee}^{dV} = 0$. The resulting upper limits on $\lambda_u^2/m_{LQ}^2$ are shown in Tab.~\ref{tab:leptoquark}. We compare with the corresponding NSI limits from COHERENT and CHARM. We also give the limit from Atomic Parity Violation (APV) constraints (see Sec.~8.3 of Ref.~\cite{Davidson:1993qk}) as well as the limit (at $m_{LQ} = 1\,\mathrm{TeV}$) from first-generation leptoquark searches as the HERA collider \cite{Abramowicz:2012tg}.

\begin{table}[t]
\begin{center}
  \begin{tabulary}{0.9\textwidth}{Lp{3.5cm}}
	Experiment	& 90\% CL upper limit on $\lambda_u^2/(m_{LQ}/\mathrm{TeV})^2$\\
    \hline
    \hline
    Ge + Si (NS, Phase 1) & 0.50 \\
    Ge + Si (VNS, Phase 1) & 0.30 \\
    Ge + Si (NS, Phase 2) & 0.30 \\
    Ge + Si (VNS, Phase 2) & 0.23 \\
    \hline
    CHARM & 28.4\\
    COHERENT & 14.5\\
    HERA & 0.9 \\
    APV & 0.2 \\
  \end{tabulary}  
\end{center}
\caption{ \textbf{Limits on vector leptoquark couplings.} Projected limits on the coupling of a vector leptoquark of mass $m_{LQ}$ to $u$-quarks. These limits are recast from generic NSI limits presented in Sec.~\ref{sec:NSI}, assuming $\eps_{ee}^{dV} = 0$. We also compare with constraints from CHARM, COHERENT, Atomic Parity Violation (APV) constraints and first-generation leptoquark searches at HERA (see the text for more details). For this choice of couplings, limits from other combinations of targets are similar.}
\label{tab:leptoquark}
\end{table}

Future \CEvNS searches should be able to constrain $\lambda_u^2/(m_{LQ}/\mathrm{TeV})^2 < 0.23$, exceeding the sensitivity of direct searches at HERA and approaching the sensitivity of APV constraints. We note however that the APV constraints are valid only if $\epsilon^{uL} \neq \epsilon^{uR}$ (otherwise parity is not violated). The constraints we have presented here are not intended as a complete exploration of leptoquark models or parameters but instead give an indication of the typical sensitivity of future \CEvNS searches. Whether or not \CEvNS can give relevant constraints on models of leptoquarks invoked to solve the recent flavour anomalies depends on the detailed flavour structure and symmetry-breaking properties of the theory \cite{Barbieri:2015yvd,Buttazzo:2017ixm}. We leave this question to future studies.



\section{Simplified models}
\label{sec:simplifiedmodels}

In the previous section, we discussed projected constraints on Non-Standard Interactions (NSI), which allow us to explore future experimental sensitivity while remaining agnostic about the nature of the New Physics which may appear in the neutrino sector. In particular, the NSI couplings induce only a change in the normalisation of the \CEvNS signal, valid for an interaction mediated by some new mediator which is much heavier than the $Z$-boson. If the new force mediator has a mass comparable with the typical momentum transfer,$< \mathcal{O}(\mathrm{GeV})$, however, it could induce changes in the recoil energy spectrum, leading to stronger constraints.

In this section, we explore projected constraints in the `Simplified Model' framework, which has become popular in Dark Matter phenomenology \cite{Abdallah:2014hon,Abercrombie:2015wmb,Abdallah:2015ter,Boveia:2016mrp} and has also been used to study New Physics signals in the neutrino sector \cite{Cerdeno:2016sfi,Bertuzzo:2017tuf}. Here, we specify the new mediator and aim to constrain its couplings to leptons and quarks. This new mediator could be a vector (Sec.~\ref{sec:vector}) or a scalar (Sec.~\ref{sec:scalar}). While such Simplified Models are necessarily incomplete and should be used with care (see e.g.~Refs.~\cite{Kahlhoefer:2015bea,DEramo:2016gos,Ellis:2017tkh}), they allow us to explore broad classes of New Physics signals without specifying a full high-energy theory.

\subsection{Massive scalar mediators}
\label{sec:scalar}

We begin by considering a new scalar mediator $\phi$, which mediates an interaction between neutrinos and quarks. We consider the interaction Lagrangian:
\begin{align}
\label{eq:L_scalar}
\mathcal{L}_{\phi} =  \phi \left[ g_\nu \overline{\nu_{R}}\nu_{L} + g_\nu^* \overline{\nu_{L}}\nu_{R}  + g_\ell \overline{\ell} \ell + g_u \overline{u} u + g_d \overline{d} d\right]\,,
\end{align}
where $\nu_{L,R}$ are left- and right-handed neutrinos, $\ell = e,\,\mu,\,\tau$ are charged leptons and $u$, $d$ are up- and down-type quarks\footnote{The keen-eyed reader may have noticed that the Lagrangian in Eq.~\eqref{eq:L_scalar} is defined after electro-weak (EW) symmetry breaking. Restoring the full Standard Model gauge-invariance and defining the scalar-mediator simplified model above the EW-scale requires a more complicated New Physics sector and typically leads to a richer phenomenology \cite{Bell:2016ekl}. However, we neglect these complications for the moment.}. The exchange of this new scalar mediator does not interfere with Standard Model $Z$-exchange, so we simply add the cross section from $\phi$ exchange to the SM \CEvNS cross section \cite{Bertuzzo:2017tuf,Farzan:2018gtr}:
\begin{align}
\frac{\mathrm{d}\sigma_{\phi}}{\mathrm{d}E_R} = \frac{(g_\nu)^2 Q_{\phi}^2}{4\pi} \frac{E_R m_N^2}{E_\nu^2 (q^2 + m_\phi^2)^2} F^2(E_R)\,.
\end{align}
Here, $m_\phi$ is the mass of the scalar mediator, $Q_\phi$ is the nuclear charge under $\phi$ exchange and $q = \sqrt{2 m_N E_R}$ is the momentum transfer. This `scalar charge' of the nucleus is computed in terms of the quark couplings $g_q$ and the contributions of the quarks to the nucleon mass $f_{Tq}^{(N)}$ (see e.g.~Appendix B.1 in Ref.~\cite{DelNobile:2013sia}). An approximate expression (for universal couplings to all quarks) was given in Ref.~\cite{Cerdeno:2016sfi}:
\begin{align}
Q_{\phi} \approx (15.1 \,Z + 14\, N)g_q\,.
\end{align}

\begin{figure*}[t!]
\centering
\includegraphics[width=0.49\textwidth]{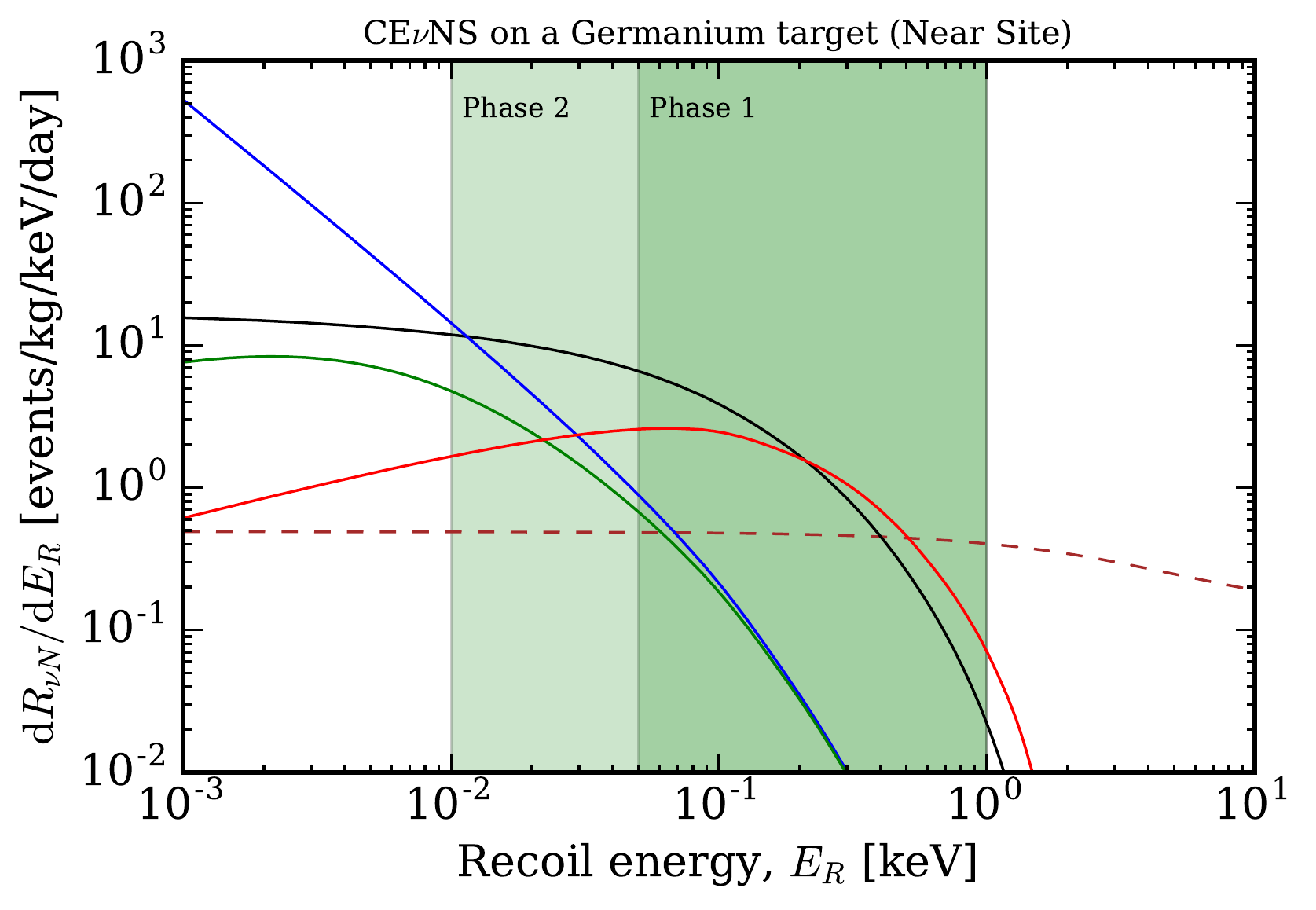}
\includegraphics[width=0.49\textwidth]{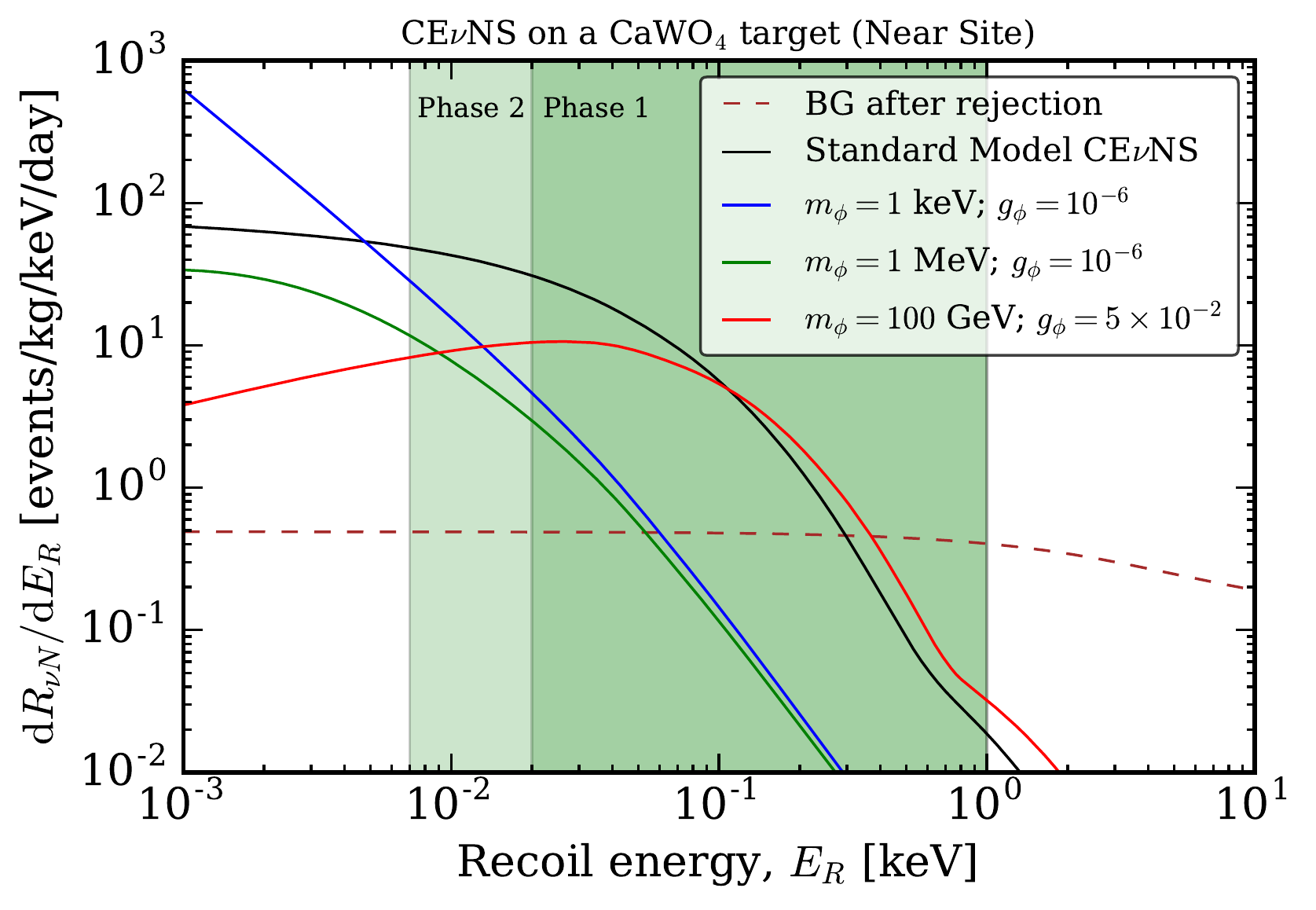}
\caption{\textbf{\CEvNS rate including a new scalar mediator.} We show the recoil rate off a Germanium target (left) and CaWO$_4$ target (right) due to a flux of Chooz Reactor neutrinos for Standard Model-only \CEvNS (solid black) as well as the contribution from a new scalar mediator $\phi$ with a range of masses and couplings (solid color). The shaded green regions show the Phase 1 and Phase 2 regions of interest (with thresholds given in Tab.~\ref{tab:experiments}). The dashed brown line shows the background after rejection (see Fig.~\ref{fig:backgrounds}).}
\label{fig:rate_scalar}
\end{figure*}

In Fig.~\ref{fig:rate_scalar}, we show the contribution to the nuclear recoil rate from a new scalar mediator, with a range of masses and couplings. The total rate is the sum of the Standard Model \CEvNS rate (solid black) and the scalar exchange contribution (one of the solid colored lines). For light mediators, $m_\phi < \mathcal{O}(\mathrm{MeV})$, the interaction is effectively long range, leading to a rapidly falling recoil spectrum proportional to $q^{-4} \sim E_R^{-2}$. For heavier mediators, the spectrum becomes peaked; the cross section in this case scales as $E_R$ for small recoil energies, before cutting off at high energy due to a loss of coherence. As was pointed out in Ref.~\cite{Farzan:2018gtr}, if the mass of the new scalar is similar to the neutrino energy, it should be possible also to measure $m_\phi$. 

\begin{figure*}[t!]
\centering
\includegraphics[width=0.49\textwidth]{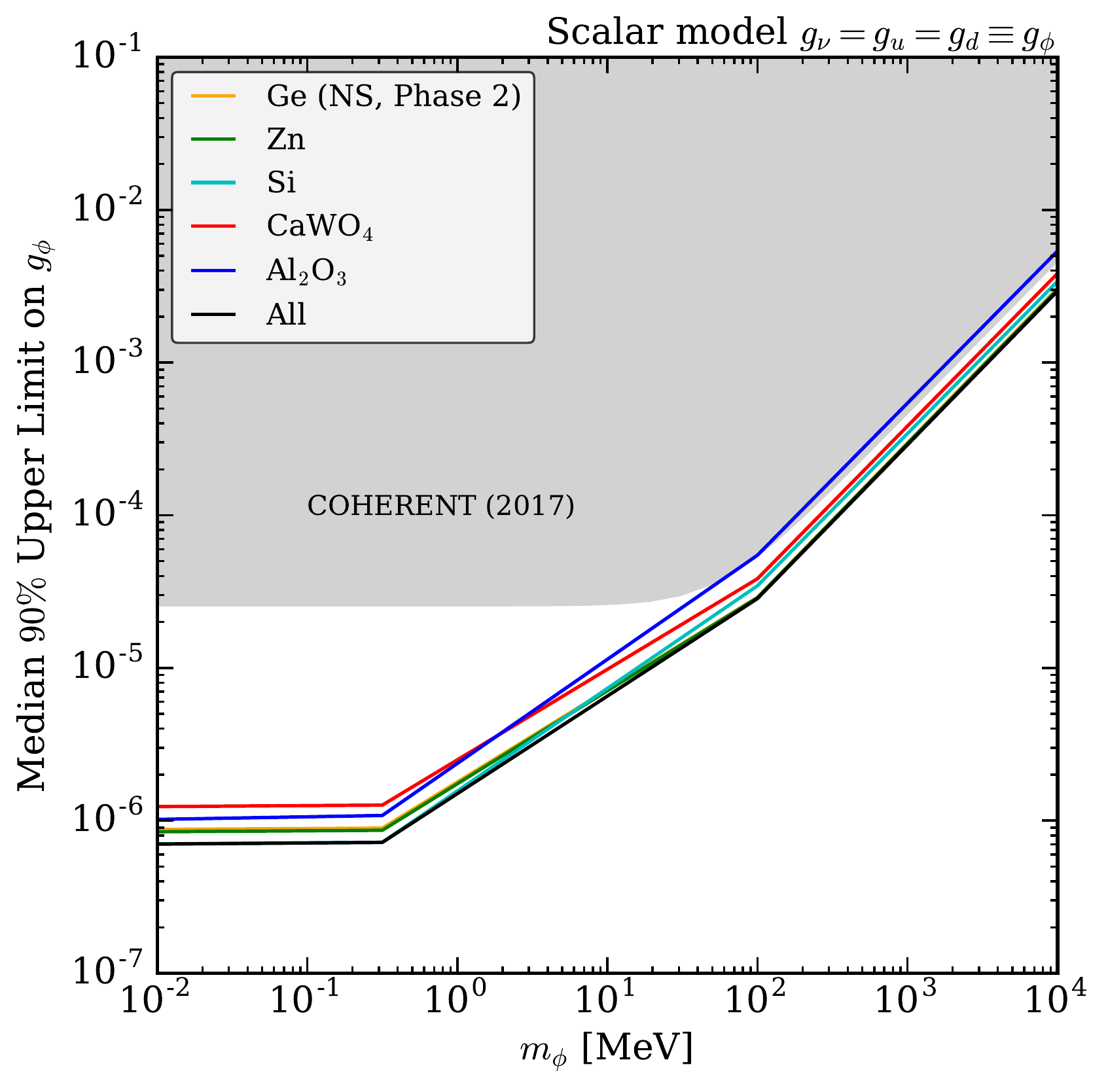}
\includegraphics[width=0.49\textwidth]{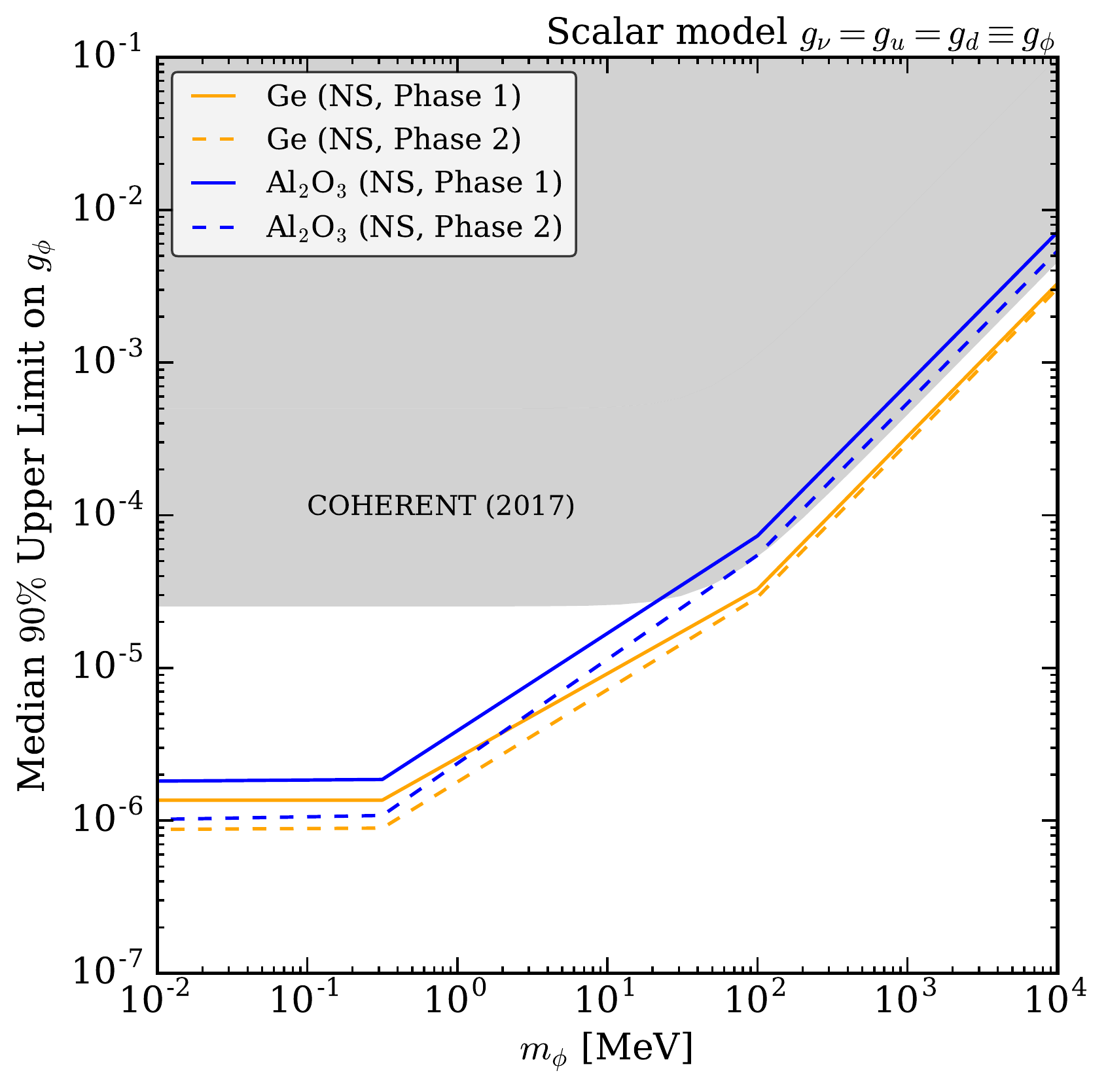}
\caption{\textbf{Projected 90\% upper limits on a new scalar mediator.} We assume that the new mediator $\phi$ couples equally to $u$ and $d$ type quarks and to neutrinos, with coupling $g_{\phi}$. The left panel shows the effect of using different detector targets, while the right panel shows the impact of varying the detector threshold. In both cases, we assume the detectors are situated at the Near Site. The grey shaded region is the area excluded by results from the COHERENT experiment \cite{Akimov:2017ade}.}
\label{fig:scalar}
\end{figure*}

In Fig.~\ref{fig:scalar}, we show the projected limits on the coupling of the new scalar mediator to fermions, $g_\phi$. We assume that the scalar couples universally to all Standard Model fermions\footnote{In general, to maintain minimal flavor violation (MFV), we would require the quark-scalar couplings to scale with the quark masses \cite{DAmbrosio:2002vsn,Buras:2003jf,Bell:2016ekl}. Given the small ratio between the up and down quark masses, we do not expect the picture to change substantially in a more realistic scenario.}. The left panel shows projections for all of the target materials we consider (for Phase 2 at the Near Site). Detectors using Ge, Zn and Si give projected limits which are very similar, improving on constraints from COHERENT by a factor of around $1.5$ at high mediator mass (the $g_\phi^4$ scaling of the rate means that this corresponds to a factor of around 5 in the signal strength). Constraints using $\CaWO$ and $\sapphire$ are competitive with COHERENT at high mass, but are substantially stronger for scalar mediators lighter than $\sim 10\,\mathrm{MeV}$, owing to the lower energy thresholds which capture the sharply rising recoil spectrum towards low energy.

In the right panel of Fig.~\ref{fig:scalar}, we explore the impact of energy thresholds more closely. We consider a Germanium detector (50 eV threshold in Phase 1 and 10 eV threshold in Phase 2), as well as an Al$_2$O$_3$ detector (20 eV threshold in Phase 1 and 4 eV threshold in Phase 2). Going from Phase 1 to Phase 2, there is a reduction in threshold, as well as an increase in target mass, resulting in a larger signal and therefore tighter constraints on the $\phi$ coupling. At high mediator mass, we note that there is little improvement in the projected constraints from Ge when going from Phase 1 to Phase 2. In this case, the recoil spectrum for scalar-mediated scattering is similar to the \CEvNS spectrum and so, despite the increased exposure, the scalar contribution can easily be absorbed by uncertainties in the signal and background rates. Furthermore, the reduced threshold has little impact, as the scalar contribution appears only at high recoil energies. For light mediators, however, there is a noticeable improvement from Phase 1 to Phase 2. We further note that in Phase 2, constraints on light mediators from Ge and $\sapphire$ are comparable, in spite of the factor $\sim 70$ smaller target mass for $\sapphire$. Again, this highlights that a reduction in threshold (10 eV for Ge compared to 4 eV for $\sapphire$) can yield a large increase in sensitivity for light mediators (see e.g.~Ref.~\cite{Dent:2016wcr}).

\subsection{Massive vector mediators}
\label{sec:vector}

We now consider adding a generic vector mediator $Z'$, which couples both to the SM neutrinos and quarks. As in the case of NSI interactions, we focus on vector couplings to quarks, which receive a coherent nuclear enhancement and therefore typically dominate.  The interaction Lagrangian is then:\footnote{As in the case of a new scalar mediator, care must be taken to respect Standard Model gauge-invariance and unitarity when introducing couplings to new vector mediators. Spin-1 mediators with only vector couplings (as in our model) are typically less problematic than those with axial-vector couplings \cite{Kahlhoefer:2015bea}. However, one should bear in mind that even in that case, a richer New Physics sector is required, for example, to cancel gauge anomalies \cite{Ellis:2017tkh}.}
\begin{align}
\mathcal{L}_{Z'} =  Z'_\mu \left[ g_\nu \overline{\nu} \gamma^\mu \nu + g_\ell \overline{\ell} \gamma^\mu \ell + g_u \overline{u} \gamma^\mu u + g_d \overline{d} \gamma^\mu d\right]\,.
\end{align}
The charge of the nucleus under the exchange of this new $Z'$ is given by:
\begin{align}
\label{eq:Qzpa}
Q_{Z'} = (2Z+N)g_u g_\nu + (2N + Z)g_d g_\nu\,,
\end{align}
where we have absorbed the coupling of the $Z'$ to neutrinos into the definition of $Q_{Z'}$. The full cross section comes from coherently summing the contributions from $Z$ and $Z'$ exchange, which may interfere. Thus, the $Z'$ effectively induces an \textit{energy-dependent} modification of the weak nuclear charge (see e.g.~\cite{Bertuzzo:2017tuf}). The differential cross section is then given by Eq.~\eqref{eq:CEvNS}, with the replacement:
\begin{align}
\label{eq:Qzp}
Q_W \rightarrow Q_\mathrm{SM+Z'} = Q_W - \frac{\sqrt{2}}{G_F} \frac{Q_{Z'}}{q^2 + m_{Z'}^2}\,.
\end{align}
Depending on the sign and magnitude of $g_\nu g_q$ (equivalently, $Q_{Z'}$), the addition of the $Z'$ can lead to either constructive or destructive interference. 

\begin{figure*}[t!]
\centering
\includegraphics[width=0.49\textwidth]{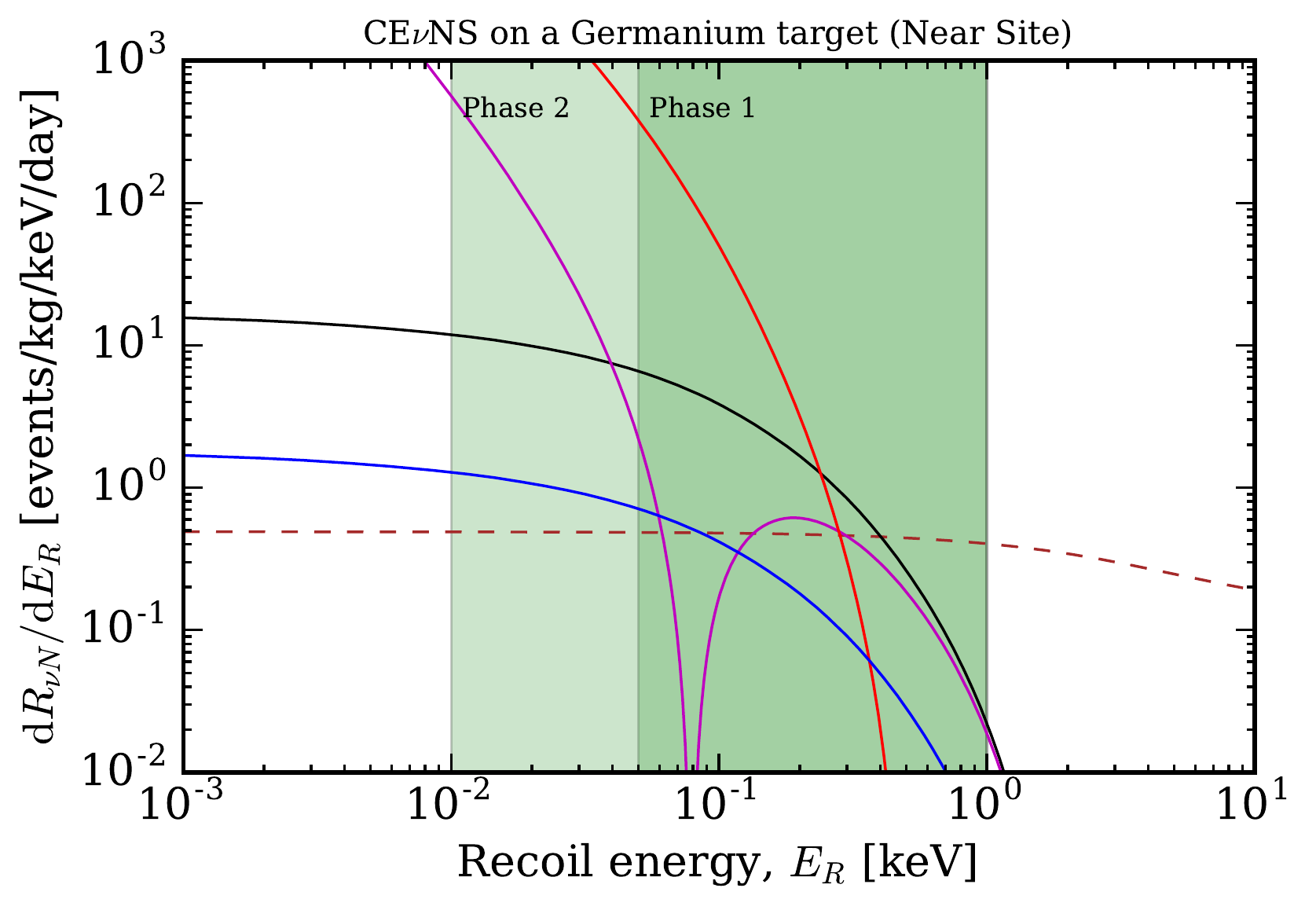}
\includegraphics[width=0.49\textwidth]{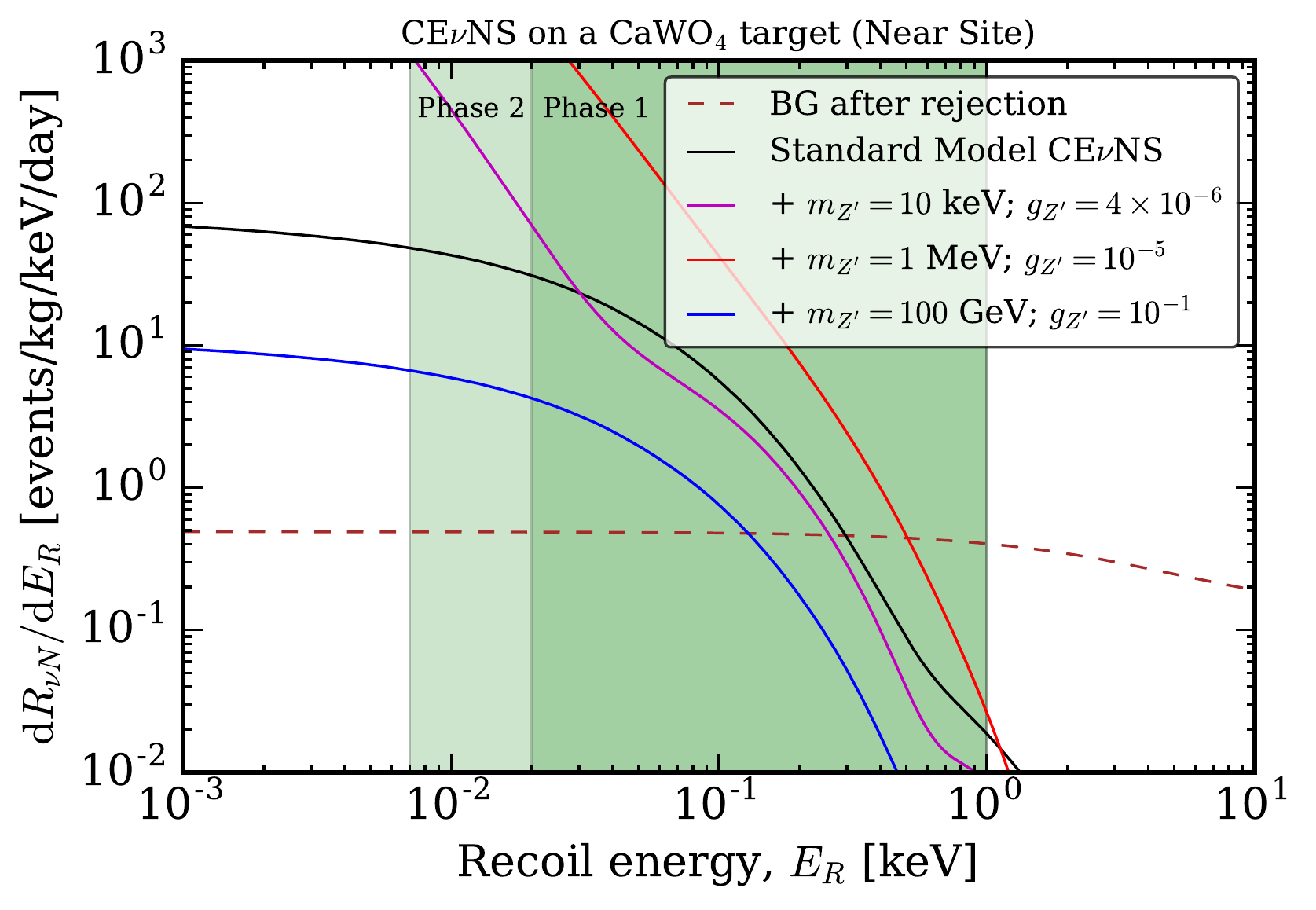}
\caption{\textbf{\CEvNS rate including a new vector mediator.} We show the recoil rate off a Germanium target (left) and CaWO$_4$ target (right) due to a flux of Chooz Reactor neutrinos for Standard Model-only \CEvNS (solid black) as well as the rate when including a new vector mediator $Z'$ with a range of masses and couplings (solid color). In all cases, due to $Z-Z'$ interference, the total rate is given by one of the solid lines (\textit{not} by the sum of the colored and black lines). The shaded green regions show the Phase 1 and Phase 2 regions of interest (with thresholds given in Tab.~\ref{tab:experiments}). The dashed brown line shows the background after rejection (see Fig.~\ref{fig:backgrounds}).}
\label{fig:Zprime_rate}
\end{figure*}

In Fig.~\ref{fig:Zprime_rate}, we show the \CEvNS recoil rate off a Germanium target (left panel) and a $\CaWO$ target (right panel), due to a flux of Chooz Reactor neutrinos. The solid black line shows the Standard Model-only contribution (from $Z$ exchange). Each of the solid colored lines shows the total recoil rate (SM + $Z'$) for a different vector mediator mass. 
The addition of a heavy vector mediator (blue line) leads to an overall rescaling of the \CEvNS rate, in which case the model reduces to the NSI framework presented in Sec.~\ref{sec:NSI}. Decreasing the mediator mass begins to distort the \CEvNS spectrum. In the left panel of Fig.~\ref{fig:Zprime_rate}, we note a dip in the rate for certain recoil energies (purple line), arising from a cancellation in the $Z'$ couplings (where $Q_\mathrm{SM+Z'}^2 = 0$ in Eq.~\eqref{eq:Qzp}). 
For multi-nucleus targets (such as $\CaWO$ in the right panel Fig.~\ref{fig:Zprime_rate}), this dip is far less pronounced, because the cancellation is only possible for one target nucleus at a time\footnote{In general, even for single-nucleus targets, the dip in the rate will be smeared out by the presence of multiple isotopes (as is that case for Ge). For simplicity, in this work we have assumed only a single isotope and we do not expect the projections to change substantially.}.


\begin{figure*}[t!]
\centering
\includegraphics[width=0.49\textwidth]{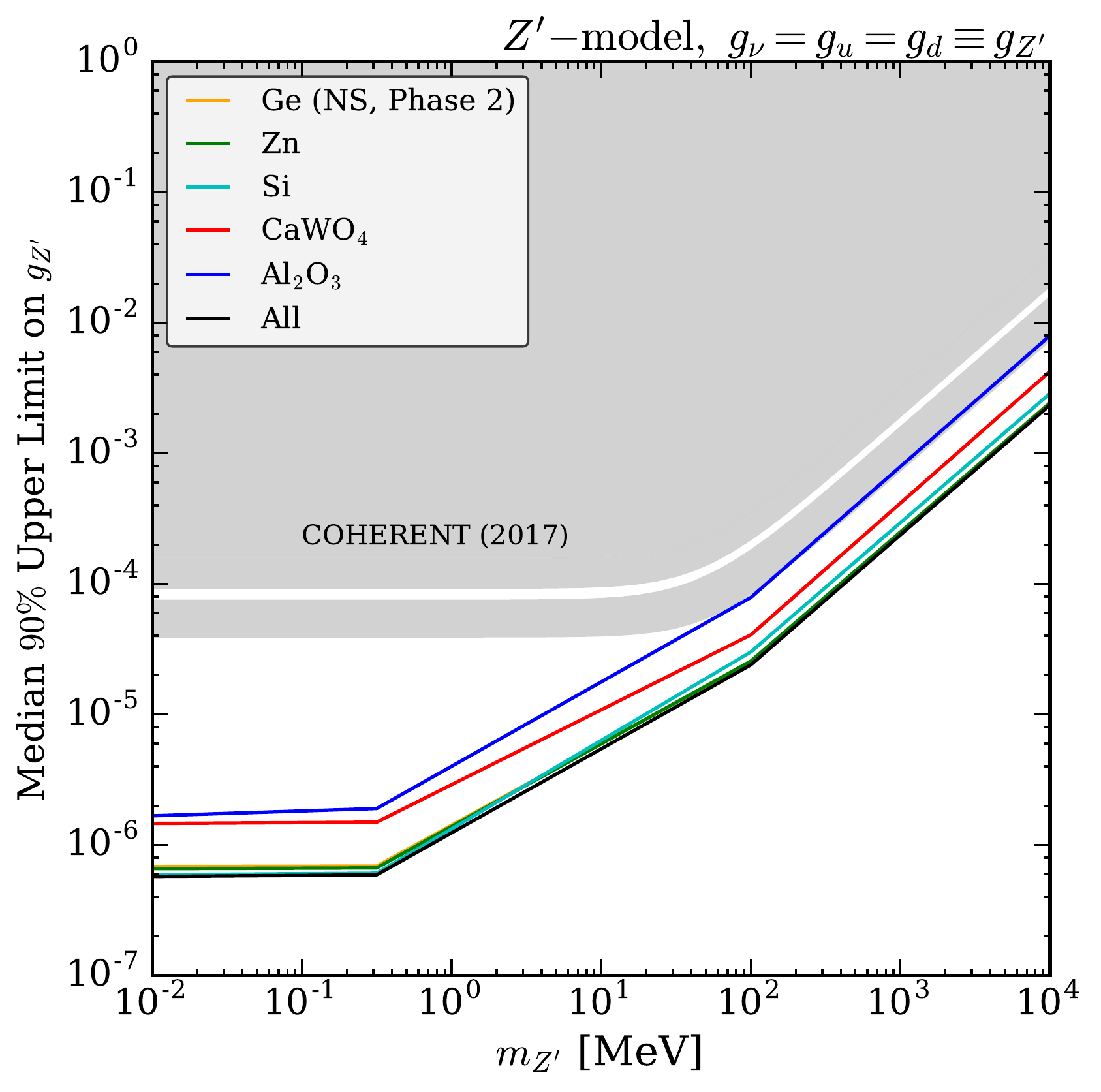}
\includegraphics[width=0.49\textwidth]{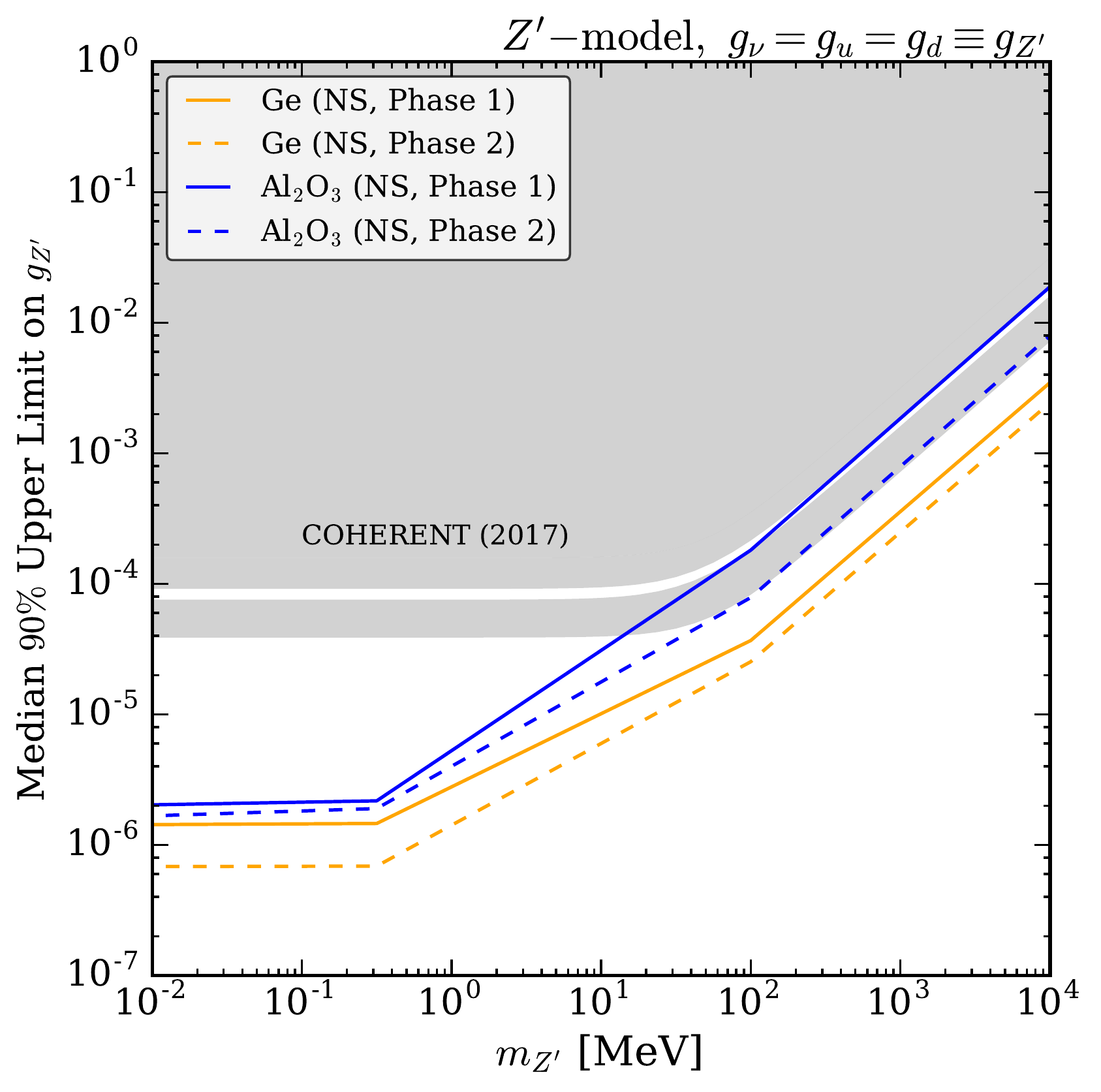}
\caption{\textbf{Projected 90\% upper limits on a new vector mediator.} We assume that the new mediator $Z^\prime$ couples equally to $u$ and $d$ type quarks and to neutrinos, with coupling $g_{Z^\prime} > 0$. The left panel shows the impact of varying the detector threshold while the right panel shows the effect of using different detector targets. In both cases, we assume the detectors are situated at the Near Site. The grey shaded region is the area excluded by results from the COHERENT experiment \cite{Akimov:2017ade}. In the left panel, the limits from Ge, Zn and Si are almost indistinguishable.}
\label{fig:Zprime}
\end{figure*}

In Fig.~\ref{fig:Zprime}, we present projected upper limits on the coupling of the $Z'$ to SM fermions $g_{Z'} = g_u = g_d = g_\nu$. Converting these limits on universal $Z'$ couplings into constraints assuming different couplings to $u$, $d$ and $\nu$ is straightforward using Eq.~\eqref{eq:Qzp}. The grey region in Fig.~\ref{fig:Zprime} is the area excluded by the COHERENT experiment \cite{Akimov:2017ade}. The thin white band which is not excluded by COHERENT corresponds to $Z^\prime$ couplings for which (see Eq.~\eqref{eq:Qzp}):
\begin{equation}
\label{eq:degeneracy}
\left(Q_W - \frac{\sqrt{2}}{G_F} \frac{Q_{Z'}}{q^2 + m_{Z'}^2}\right)^2 = Q_W^2\,.
\end{equation}
For these values of the couplings, the interference of the Standard Model $Z$ and the new $Z^\prime$ is such that the cross section (in the recoil energy range probed by the experiment) is indistinguishable from the SM-only case. Note that for $m_{Z'} \lesssim 100 \,\mathrm{MeV}$, the shape of the recoil spectrum becomes distorted. Here, we use a simple, single-bin analysis for the COHERENT data, but in principle this white degeneracy band (for light vector mediators) can be excluded if spectral information is included in the analysis (see e.g.~Refs.~\cite{Liao:2017uzy,Denton:2018xmq}).

In the left panel Fig.~\ref{fig:Zprime}, we examine the effect of using different target nuclei. For Phase 2 experiments at the Near Site, we see that all targets should be able to achieve a sensitivity competitive with COHERENT. For Ge, Si and Zn targets, constraints on the $Z'$ coupling could be improved by a factor of 4 at high mediator mass, and by more than 2 orders of magnitude at low mediator mass. As for scalar mediators, $\CaWO$ and $\sapphire$ targets are most useful in constraining the low mass region, where constraints on $g_{Z'}$ can be improved by a factor of around 15 compared to COHERENT. 

In the right panel of Fig.~\ref{fig:Zprime}, we explore the impact of varying the experimental threshold. We note in particular that for $\sapphire$ reducing the threshold (going from Phase 1 to Phase 2) has only a small impact at low mediator masses, in contrast with the scalar case. The reason for this is evident when looking at the spectra in the right panel of Fig.~\ref{fig:Zprime_rate}, where for light mediators (purple and red curves) the \CEvNS spectrum is distorted for all recoil energies and does not simply appear as a low energy excess as in the case of the scalar. Thus, reducing the energy threshold for $\sapphire$ does not substantially improve constraints.


We now focus on projected constraints on a specific model - the $U(1)_{B-L}$ model \cite{Mohapatra:1980qe,Khalil:2006yi} - in order to explore the complementarity between \CEvNS New Physics searches and other constraints. Here, $B$ and $L$ are Baryon and Lepton number respectively. The $B-L$ symmetry of the Standard Model is elevated to a fundamental $U(1)$ gauge symmetry, with quarks carrying a $B-L$ charge of $Q_q = 1/3$ and leptons a charge of $Q_\ell = -1$. The gauge boson mediating this interaction can be described as a new vector mediator $Z'$, with $g_q = -g_\nu/3$. Such a model has been examined in the context of \CEvNS (from Solar neutrinos) in Refs.~\cite{Cerdeno:2016sfi,Dent:2016wcr}.

\begin{figure*}[t!]
\centering
\includegraphics[width=0.75\textwidth]{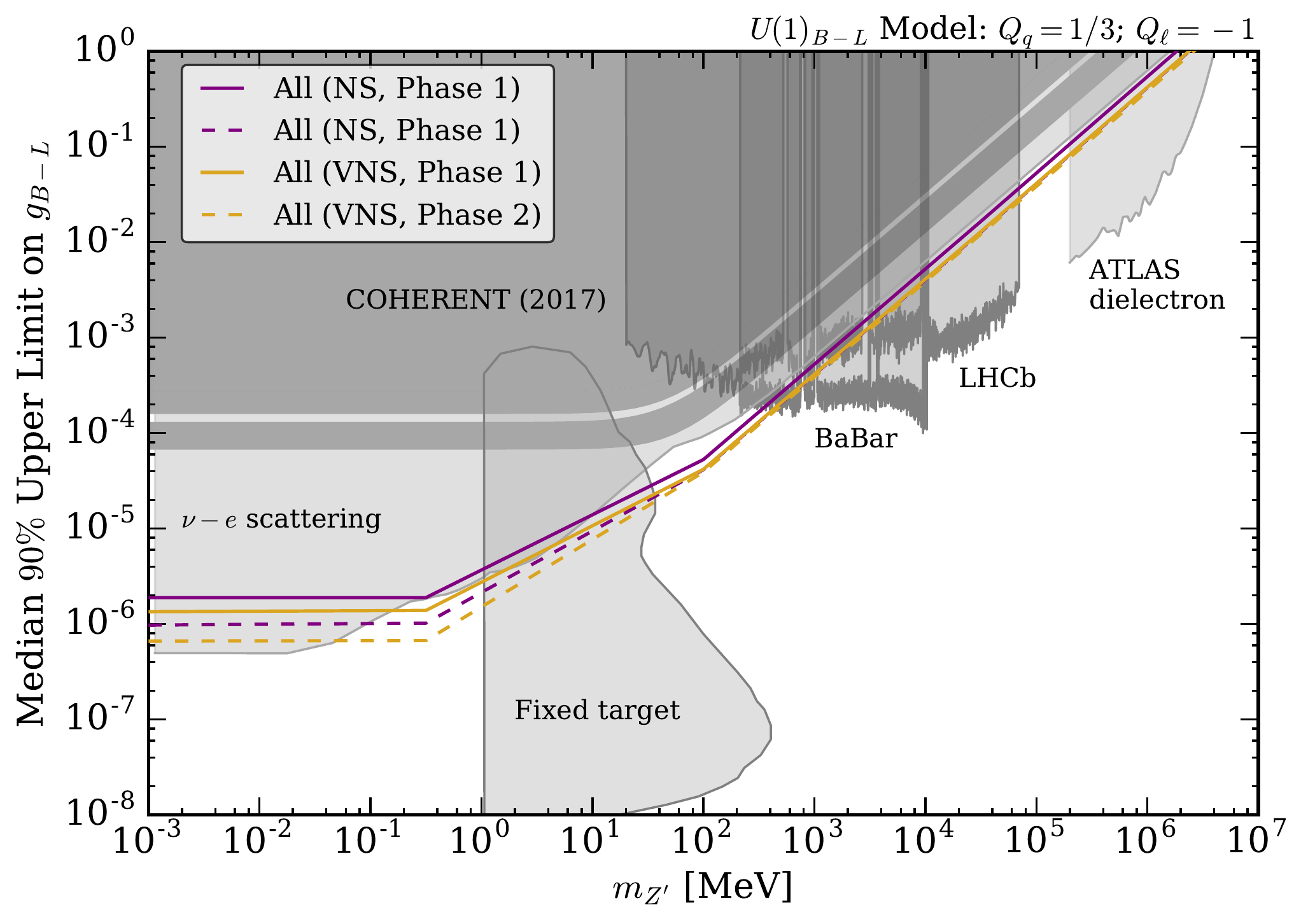}
\caption{\textbf{Projected 90\% CL upper limits on the $U(1)_{B-L}$ gauge coupling $g_{B-L}$.} We consider constraints from \CEvNS using a combined analysis of all targets (Zn, Ge, Si, $\CaWO$ and $\sapphire$. We also show a selection of other terrestrial limits (see the text for details).}
\label{fig:B-L}
\end{figure*}

Recasting the results of the previous section, we plot in Fig.~\ref{fig:B-L} the projected limits on the gauge coupling $g_{B-L}$. In the same plot, we also show complementary terrestrial limits on the $U(1)_{B-L}$ Model as grey shaded regions. These are: limits from the ATLAS search for dielectron resonances \cite{Aaboud:2016cth} (probing the production of the $Z'$ through its coupling to quarks, followed by its subsequent decay into leptons); constraints from electron beam-dump fixed-target experiments \cite{Harnik:2012ni,Ilten:2018crw}; constraints from measurements of $\nu$-$e$ scattering (results from TEXONO, GEMMA, BOREXINO, LSND and CHARM II compiled in Refs.~\cite{Harnik:2012ni,Bilmis:2015lja}); constraints from Dark Photon searches at BaBar \cite{Lees:2014xha,Lees:2017lec} and LHCb \cite{Aaij:2017rft}, and constraints from the COHERENT experiment \cite{Akimov:2017ade}.  Note that in many cases there are a number of overlapping constraints and we do not show them all. For a more detailed discussion of all the relevant limits on $B-L$ models, see Refs.~\cite{Ilten:2018crw,Bauer:2018onh}.

We note that current constraints from COHERENT are less stringent than those coming from $\nu$-$e$ scattering experiments for all mediator masses. However, future $\nu$-$N$ scattering experiments (such as those we consider here) should be able to probe new parameter space. For mediators heavier than around 100 GeV, the dominant limits come from LHC resonance searches. Below this we find that our projections are competitive with current results from $\nu$-$e$ scattering, though dark photon searches at BaBar and LHCb are typically stronger above around 1 GeV in mediator mass. The greatest improvement in constraints will appear for mediator masses around $100 \,\mathrm{MeV}$. Below this, fixed target experiments dominate down to around $1 \,\mathrm{MeV}$ (roughly twice the electron mass). Below $1\,\mathrm{MeV}$, future \CEvNS searches may again be competitive with $\nu$-$e$ experiments, but this would require the low energy thresholds of Phase 2 experiments (dashed lines), in order to maximise sensitivity to such light mediators.

Of course, the comparisons we have presented in this section are valid only in the framework of the $U(1)_{B-L}$ Model. Looking at more general values of the couplings (i.e.~changing the ratios of $g_u$, $g_d$, $g_\ell$ and $g_\nu$) will change the relative importance of the different constraints. Thus, models such as gauged $U(1)_{L_\mu - L_\tau}$ models (see e.g.~Refs.~\cite{Baek:2008nz,Heeck:2010ub,Elahi:2015vzh,Patra:2016shz,Xing:2015fdg}), or leptophobic $Z'$ models (see e.g.~Refs.~\cite{Ko:2010at,Gondolo:2011eq,Chiang:2015ika}) will be constrained more or less strongly by different searches \cite{Abdullah:2018ykz}. However, the sensitivity of \CEvNS to both lepton and quark couplings should make it a useful complementary probe in all cases.

\section{Conclusions}
\label{sec:conclusions}

In this work, we have explored the prospects for constraining New Physics in the neutrino sector using measurements of Coherent Elastic Neutrino-Nucleus Scattering (\CEvNS) from reactor neutrino sources. We considered a number of possible targets (and combinations of targets) --- Ge, Zn, Si, $\CaWO$ and $\sapphire$ --- which have been proposed for future \CEvNS experiments, as we summarised in Tab.~\ref{tab:experiments}. The first three targets have a larger payload mass, while the last two have a lower recoil energy threshold, allowing us to assess which experimental setups provide the best prospects for constraining and detector a range of New Physics signals. 

We summarise below our key findings:
\begin{description}
\item[Discovery significance (Sec.~\ref{sec:discovery})] The first step for any detector will be to conclusively observe the Standard Model \CEvNS signal. It should be possible to exceed the COHERENT significance of $6.7\sigma$ within $\sim70$ days for Ge or Zn detectors located at the Chooz Near Site (NS), owing to their low proposed energy threshold ($E_\mathrm{th} = 50 \,\mathrm{eV}$). For the remaining targets, discovery may require $\mathcal{O}(100\mathrm{s})$ of days, though this would be substantially reduced at the Very Near Site (VNS) or in Phase 2 (with lower threshold and larger payload). 

\item[Neutrino magnetic moment (Sec.~\ref{sec:magnetic})] An anomalously large neutrino magnetic moment (NMM) would manifest in a \CEvNS experiment as an excess of low energy recoil events. Future low threshold detectors could constrain the NMM below the COHERENT limits of $\mu_\nu \lesssim 2 \times 10^{-9} \,\mu_B$ within a couple of days of exposure. We also find that with $10^{3} - 10^{4}$ days of exposure, future experiments should be competitive with the best current terrestrial constraints from Solar $\nu$ scattering ($\mu_\nu \sim 3 \times 10^{-11}\,\mu_\mathrm{B}$), though astrophysical constraints (at the level of $\mu_\nu \sim 2 \times 10^{-12}\,\mu_\mathrm{B}$) remain out of reach.

\item[Non-standard Neutrino Interactions (Sec.~\ref{sec:NSI})]  Taking an agnostic approach to New Physics in the neutrino sector, we explored projections for Non-standard Neutrino Interactions (NSI), which modify the coupling of neutrinos to quarks. Constraints on NSI depend strongly on the ratio of $A/Z$ for the nuclear target. The ideal detector combinations have $A/Z$ ratios which are as diverse as possible, with constraints on $\epsilon_{ee}^{uV}$ and  $\epsilon_{ee}^{dV}$ being the most stringent for a combined Ge+Si analysis. While multi-nucleus targets do not in general have a well defined $A/Z$ ratio, we find that those we consider, $\CaWO$ and $\sapphire$, can be treated as effectively single-nucleus. In the case of $\sapphire$, Al and O have a common value of $A/Z$, while for $\CaWO$, recoils on W dominate the rate, owing to the low energy threshold of the detector. Thus, these detectors give surprisingly tight constraints on NSI couplings, despite their target composition and small payload.

Combining $\CaWO$+$\sapphire$ targets, Ge+Zn targets and Ge+Si targets, we find that non-universal $\nu_e$-$q$ couplings can be constrained at the 25\%, 10\% and 5\% level. In all cases, this is competitive with current constraints from LHC monojet searches. Recasting these projected limits as constraints on first-generation leptoquarks, we also find that future \CEvNS searches should be competitive with direct leptoquark searches at colliders.

\item[Simplified Models (Sec.~\ref{sec:simplifiedmodels})] Going beyond NSI couplings, we also considered simplified models with new heavy mediators coupling neutrinos to quarks. A new scalar mediator will provide an enhancement over the \CEvNS rate, which drops rapidly with recoil energy in the case of a light scalar ($m_\phi \lesssim 1\,\mathrm{MeV}$). In this case, future low threshold detectors could improve constraints on the scalar coupling $g_\phi$ by around a factor of 20 (corresponding to a factor of $20^4 \approx 10^5$ in the signal rate). For heavy mediators, improvements in energy thresholds have little effect on the limit, as the scalar-mediated signal in that case peaks at a few hundred eV. Substantial gains are made for light mediators, however, with the low threshold $\sapphire$ and $\CaWO$ detectors becoming competitive with the (larger payload) Ge, Zn and Si detectors.

We also explored constraints on new vector mediators $Z'$, where future detectors should improve constraints on couplings to light (heavy) vectors by a factor of 100 (4). For vectors,  reductions in energy thresholds have a less pronounced effect on the limits. This is because the new $Z'$ can interfere with the Standard Model $Z$, leading to distortions of the \CEvNS spectrum for almost all recoil energies. Finally, we compared current and projected \CEvNS constraints on a more complete $Z'$ model, namely $U(1)_{B-L}$. In this case, future detectors should provide the most stringent constraints for mediator masses from $\sim 100 \,\,\mathrm{MeV}$ up to $\sim 100 \,\,\mathrm{GeV}$. For mediators lighter than $1\,\,\mathrm{MeV}$, \CEvNS searches could be competitive with current $\nu$-$e$ scattering experiments, depending on energy threshold. 
\end{description}

In this work, we have emphasised the importance of target complementarity, showing that employing Silicon detectors along with other targets provides the best prospects for pinning down New Physics couplings to up and down quarks. We have also emphasised the importance of a low detector threshold. As we showed explicitly in Fig.~\ref{fig:MuNuLimits_threshold}, smaller $\CaWO$ and $\sapphire$ detectors can be competitive with more massive Ge, Zn and Si detectors in constraining neutrino magnetic moments, provided that a low enough energy threshold is reached. We expect similar results to hold for any New Physics signature which is expected to appear at low energy, such as for light vector and scalar mediators. 

We hope that the projections we have presented here will prove a useful tool to the experimental community in guiding future \CEvNS searches for New Physics. 

\acknowledgments

We thank Joe Formaggio and Lindley Winslow for checking a draft of this manuscript. We also thank the authors of Ref.~\cite{Ilten:2018crw} for pointing out (and sharing data files for) a number of the limits appearing in Fig.~\ref{fig:B-L}. 

BJK acknowledges funding from the European Research Council ({\sc Erc}) under the EU Seventh Framework Programme (FP7/2007-2013)/{\sc Erc} Starting Grant (agreement n.\ 278234 --- `{\sc NewDark}' project) and from the NWO through the VIDI research program "Probing the Genesis of Dark Matter" (680-47-532). We wish to thank the Heising-Simons Foundation and the MIT MISTI-France Program for their support of this work.

\appendix

\section{COHERENT constraints}
\label{app:COHERENT}

In this appendix, we describe in more detail the recent observations of \CEvNS by the COHERENT experiment \cite{Akimov:2017ade} and describe how we derive limits on New Physics from these observations. The COHERENT collaboration detected the coherent elastic scattering of neutrinos from the Spallation Neutron Source (SNS) off nuclei in a 14.6-kg CsI[Na] scintillating detector. The detection was made at 6.7$\sigma$ significance, with a rate consistent with the Standard Model expectation at  the 1$\sigma$ level.

\begin{figure}[t]
\centering
\includegraphics[width=0.7\textwidth]{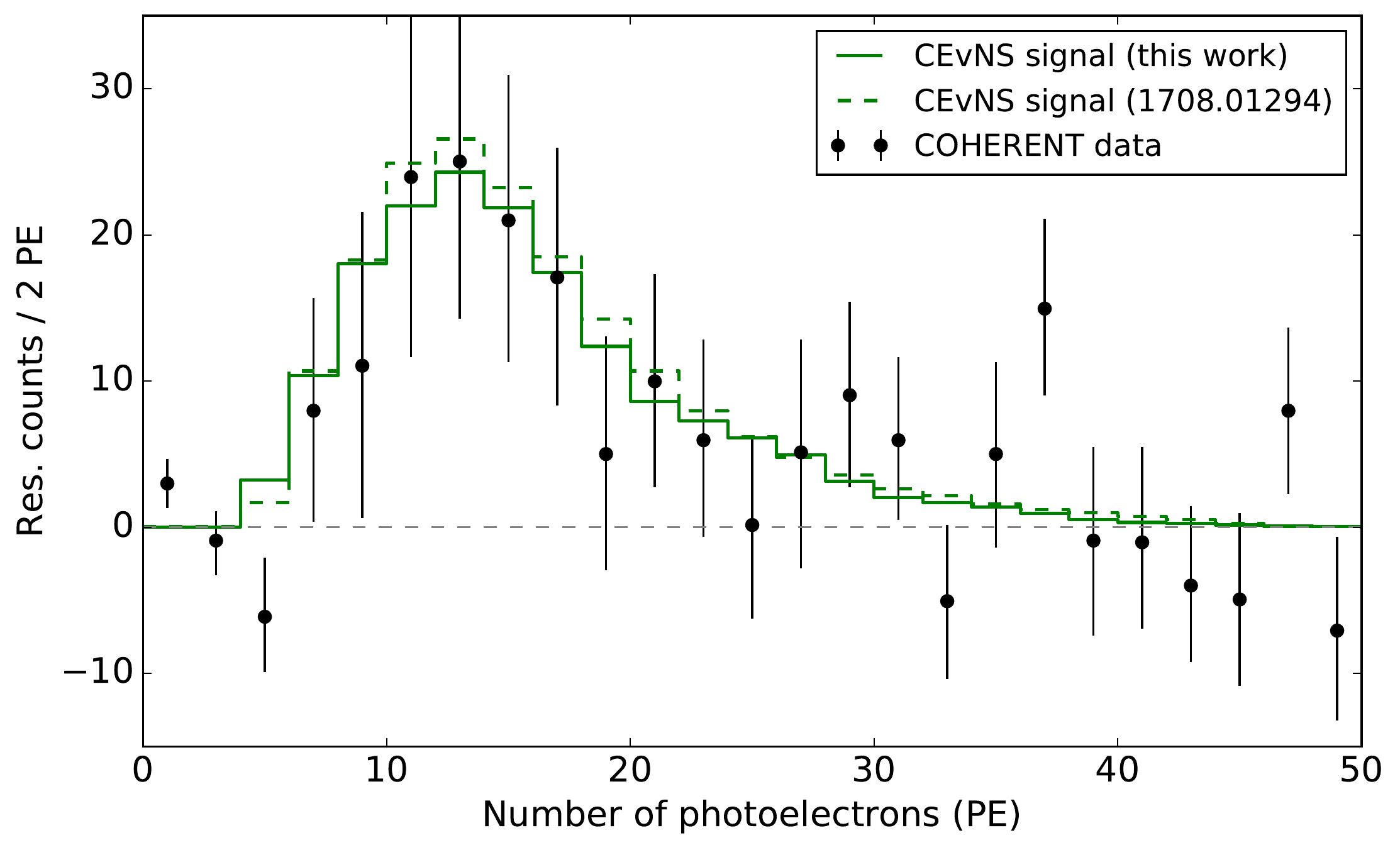}
\caption{Expected and observed \CEvNS signal at the COHERENT experiment. The solid green line shows the expected signal we obtain following the description of Ref.~\cite{Akimov:2017ade} (see text for more details), while the dashed green line shows the signal reported in Ref.~\cite{Akimov:2017ade} (see the upper-right panel of Fig.~3 therein).}
\label{fig:COHERENT_data}
\end{figure}

The flux of $\nu_\mu$, $\overline{\nu}_\mu$ and $\nu_e$ neutrinos is obtained by digitising\footnote{Where it is necessary to digitise data from plots, we use the publicly available WebPlotDigitizer \cite{WebPlotDigitizer}.} the upper panel of Fig.~S2 in Ref.~\cite{Akimov:2017ade}, which gives the energy distribution of SNS neutrinos obtained from Geant4 simulations. In addition to the continuum spectra shown in Fig.~S2 \cite{Akimov:2017ade}, we add in the contribution of 29.65 MeV prompt $\nu_\mu$ from pion decay at rest as a $\delta$-function in the flux (this appears as a sharp peak in Fig.~S2). Such decay-at-rest (DAR) neutrinos contribute to this mono-chromatic flux of $\nu_\mu$ as well as the continuum flux of $\overline{\nu}_\mu$ and $\nu_e$ up to 53.5 MeV. To fix the overall normalisation of the flux, we assume 0.08 decay-at-rest neutrinos per proton, $5 \times 10^{20}$ protons per day and a detector placed 19.3m from the neutrino production point. In Fig.~\ref{fig:COHERENT_data} we show the expected COHERENT signal, taking into account the detector efficiency given in Fig.~S9 of Ref.~\cite{Akimov:2017ade}. There is good agreement between the expected signal we calculate (solid green) and that reported by the COHERENT collaboration (dashed green). The observed data points are shown in black for comparison.

In order to derive New Physics constraints from the COHERENT data, we follow the single-bin $\chi^2$ approach given in Ref.~\cite{Akimov:2017ade}. In this case, the $\chi^2$ test-statistic is defined as:

\begin{equation}
\label{eq:chi2}
\chi^2 = \frac{\left(N_\mathrm{obs} - N_\mathrm{sig} [1+\alpha] - B_\mathrm{on}[1+\beta]\right)^2}{\sigma_\mathrm{stat}^2} + \left( \frac{\alpha}{\sigma_\alpha}\right)^2 + \left( \frac{\beta}{\sigma_\beta}\right)^2\,,
\end{equation}
where $N_\mathrm{obs} = 142$ is the number of background-subtracted counts, $N_\mathrm{sig}$ is the number of expected signal counts, $B_\mathrm{on} = 6$ is the estimated beam-on background and $\sigma_\mathrm{stat} = 30.95$ is the statistical uncertainty on the number of counts. Systematic uncertainties on the signal and background are accounted for by the nuisance parameters $\alpha$ and $\beta$, with corresponding uncertainties $\sigma_\alpha = 0.28$ and $\sigma_\beta = 0.25$. 

In practice, we calculate the expected number of signal events for a given model and set of model parameters $\boldsymbol{\theta}$ then minimise $\chi^2$ (given in Eq.~\ref{eq:chi2}) with respect to $\alpha$ and $\beta$, to obtain $\hat{\chi}^2$. We then use the $\chi^2$ difference, $\Delta \chi^2 (\boldsymbol{\theta}) = \hat{\chi}^2(\boldsymbol{\theta}) - \hat{\chi}^2_\mathrm{min}$ to calculate limits or allowed regions for the model parameters. We have checked explicitly that we recover the same limits on NSI interactions as reported in Fig.~4 of Ref.~\cite{Akimov:2017ade}. 

Code for calculating \CEvNS spectra and signal rates and for obtaining New Physics limits from the COHERENT data is publicly available at \href{https://github.com/bradkav/CEvNS/}{https://github.com/bradkav/CEvNS/} \cite{CEvNS-code}.

During the preparation of this work, the COHERENT collaboration released data from the first \CEvNS observation with CsI \cite{Akimov:2018vzs}. Given the good agreement shown in Fig.~\ref{fig:COHERENT_data}, we do not expect our constraints from a single-bin analysis to change substantially using the officially released data.

However, Ref.~\cite{Akimov:2018vzs} also includes more information about background distributions as well as full energy and timing information about the observed events, which would allow for an improve analysis in future. Performing a full bin-by-bin analysis including energy information would allow different \textit{shapes} of recoil spectra to be discriminated. This would likely allow us to break the degeneracy between  mass and coupling of new vector mediators, for example (see the discussion below Eq.~\eqref{eq:degeneracy} in Sec.~\ref{sec:vector}). Instead, timing information can act as a proxy for neutrino flavor ($\nu_\mu$ are produced in the prompt decay of stopped pions, while $\overline{\nu}_\mu$ and $\nu_e$ are produced in the subsequent muon decays). Better discrimination between neutrino flavors would allow us to disentangle the rates for $\nu_e$ and $\nu_\mu$ and thus improve constraints on NSI for different flavors. We leave these more detailed analyses to future work.

\section{NSI Constraints}
\label{app:NSI}

\begin{figure}[tp!]
\centering
\includegraphics[width=0.49\textwidth]{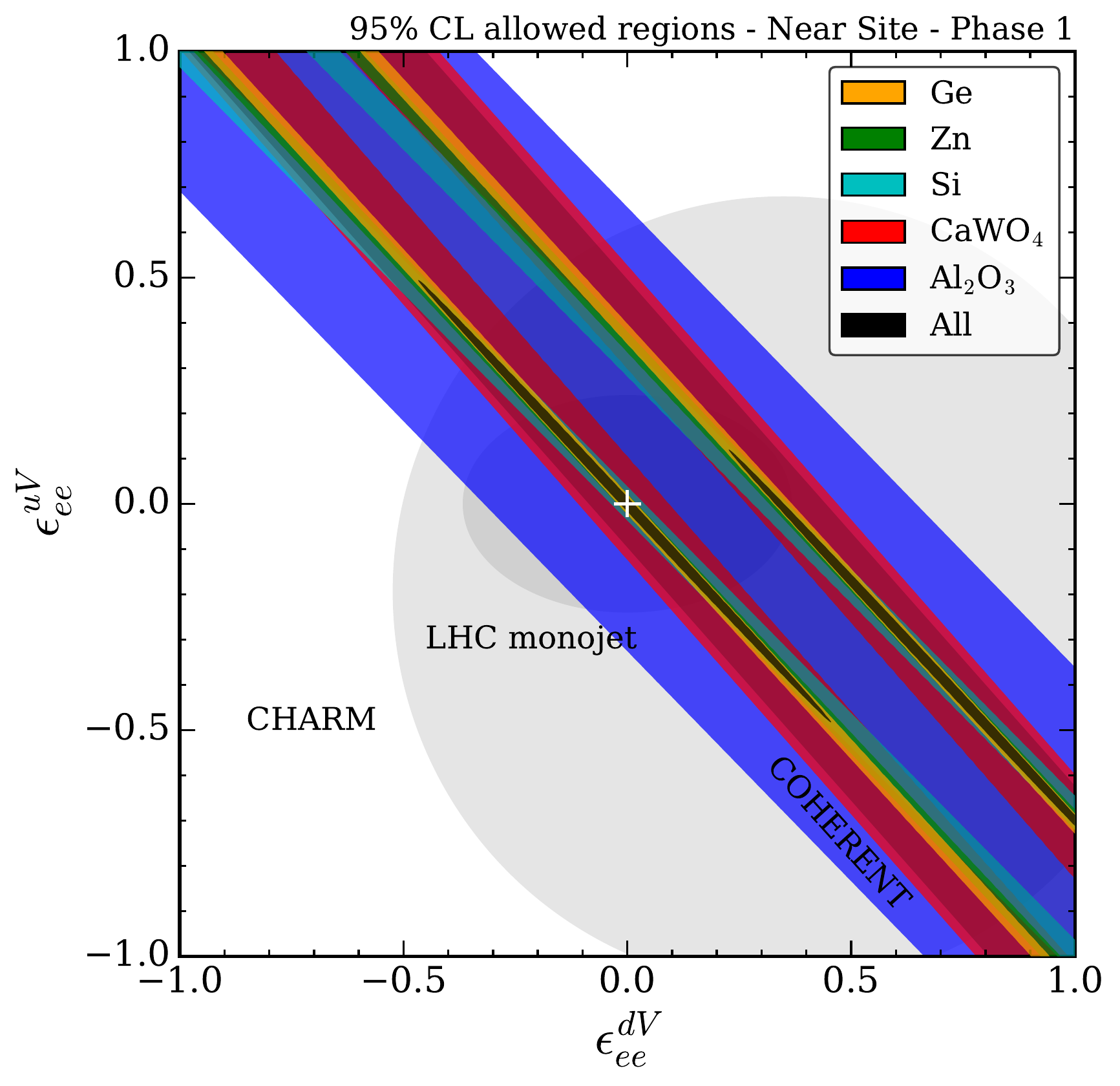}
\includegraphics[width=0.49\textwidth]{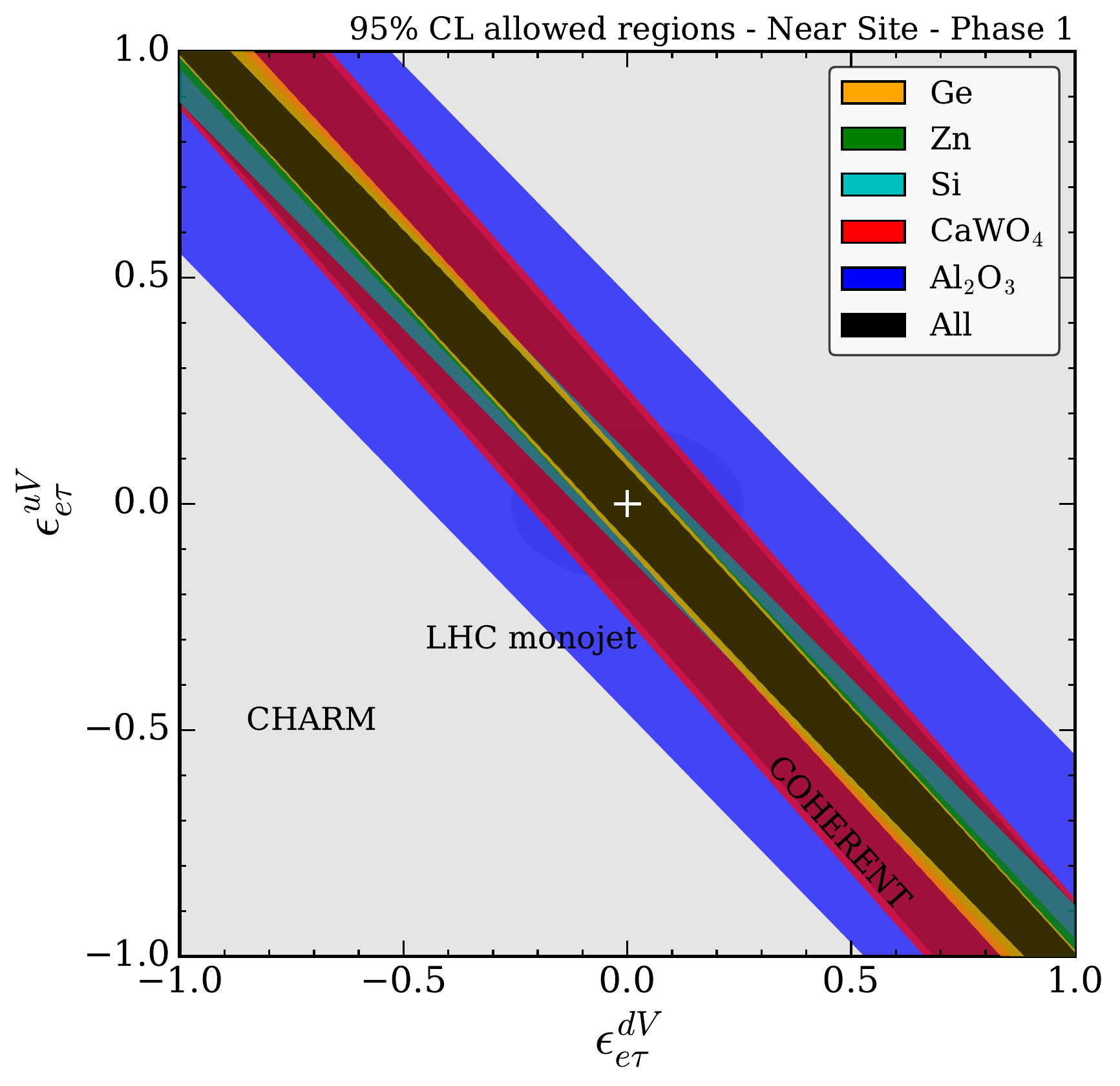}
\includegraphics[width=0.49\textwidth]{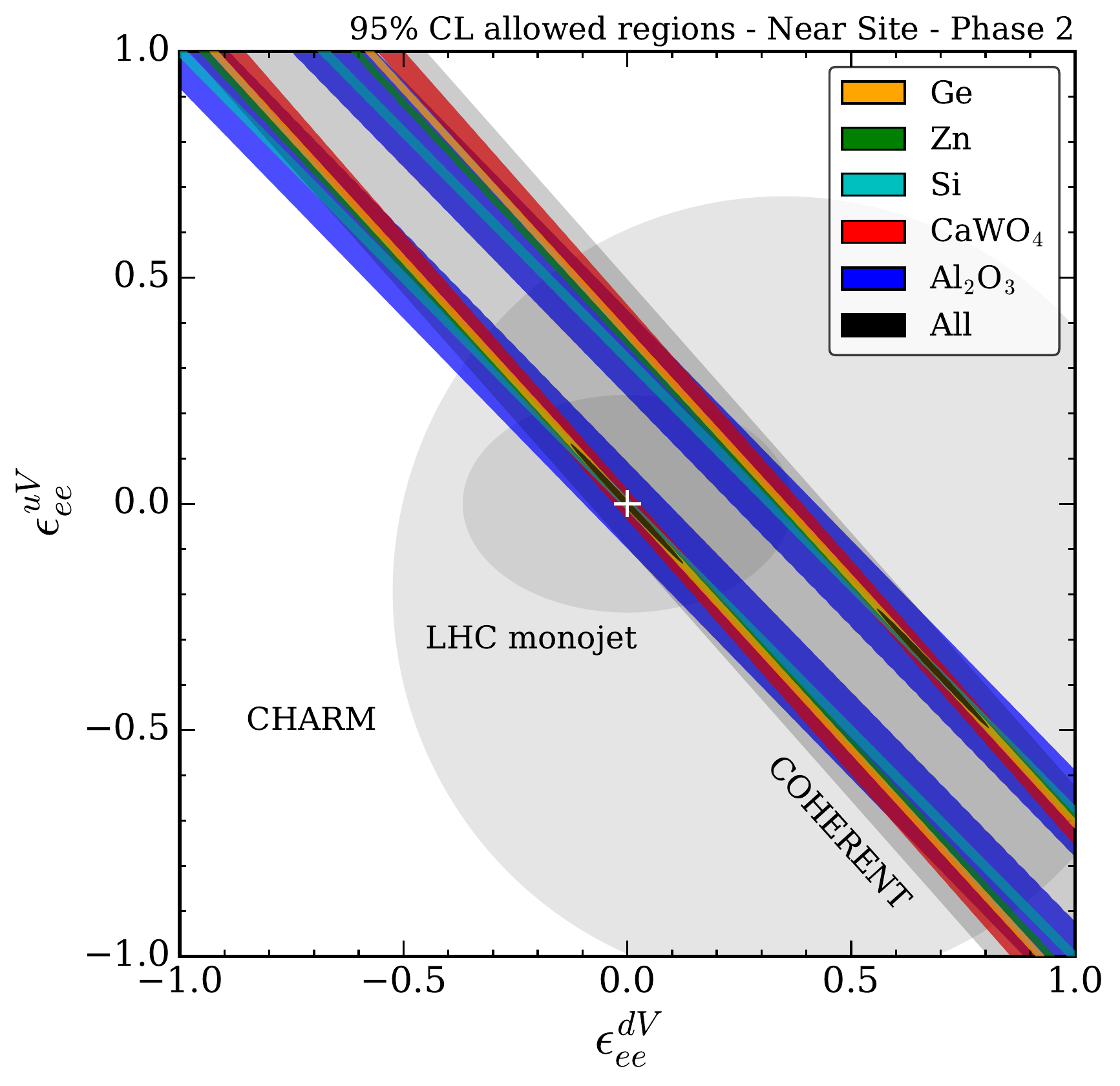}
\includegraphics[width=0.49\textwidth]{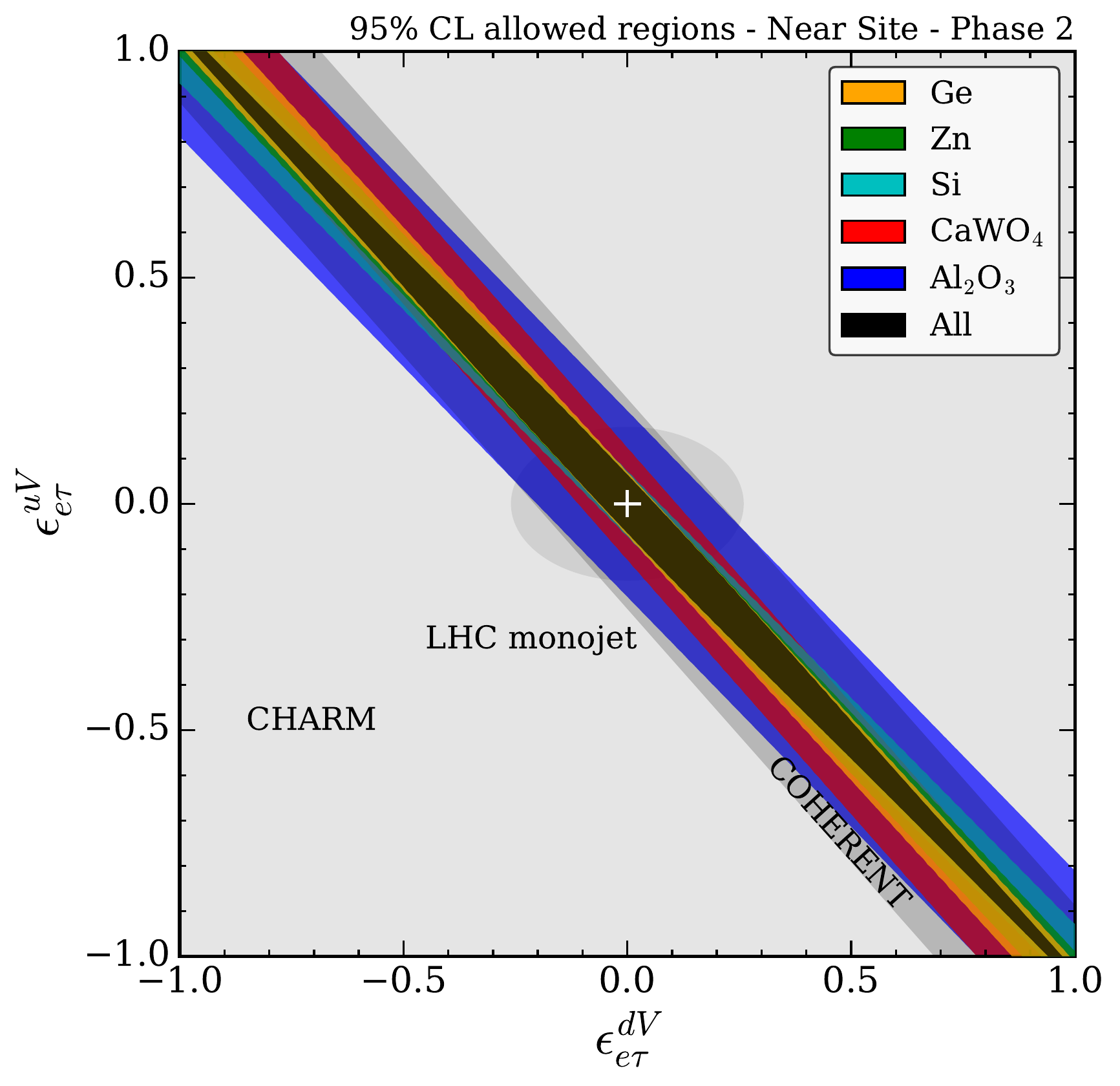}
\caption{\textbf{Projected 95\% CL allowed regions for non-standard interactions (NSI) at the Near Site, using all targets.} Same as Fig.~\ref{fig:NSI_VNS_GeZn}, but for all targets.}
\label{fig:NSI_NS_all}
\end{figure}

\begin{figure}[tp!]
\centering
\includegraphics[width=0.49\textwidth]{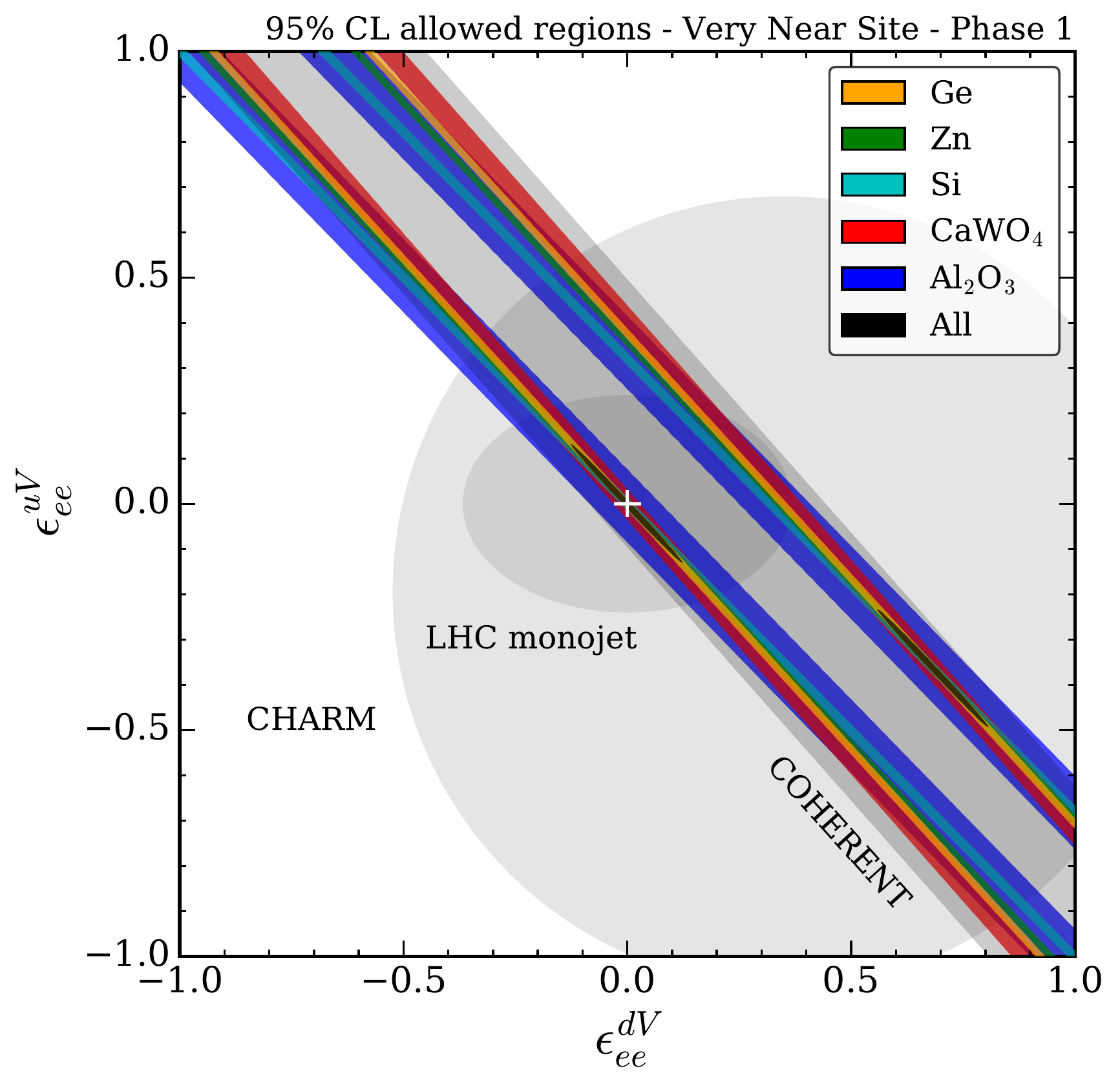}
\includegraphics[width=0.49\textwidth]{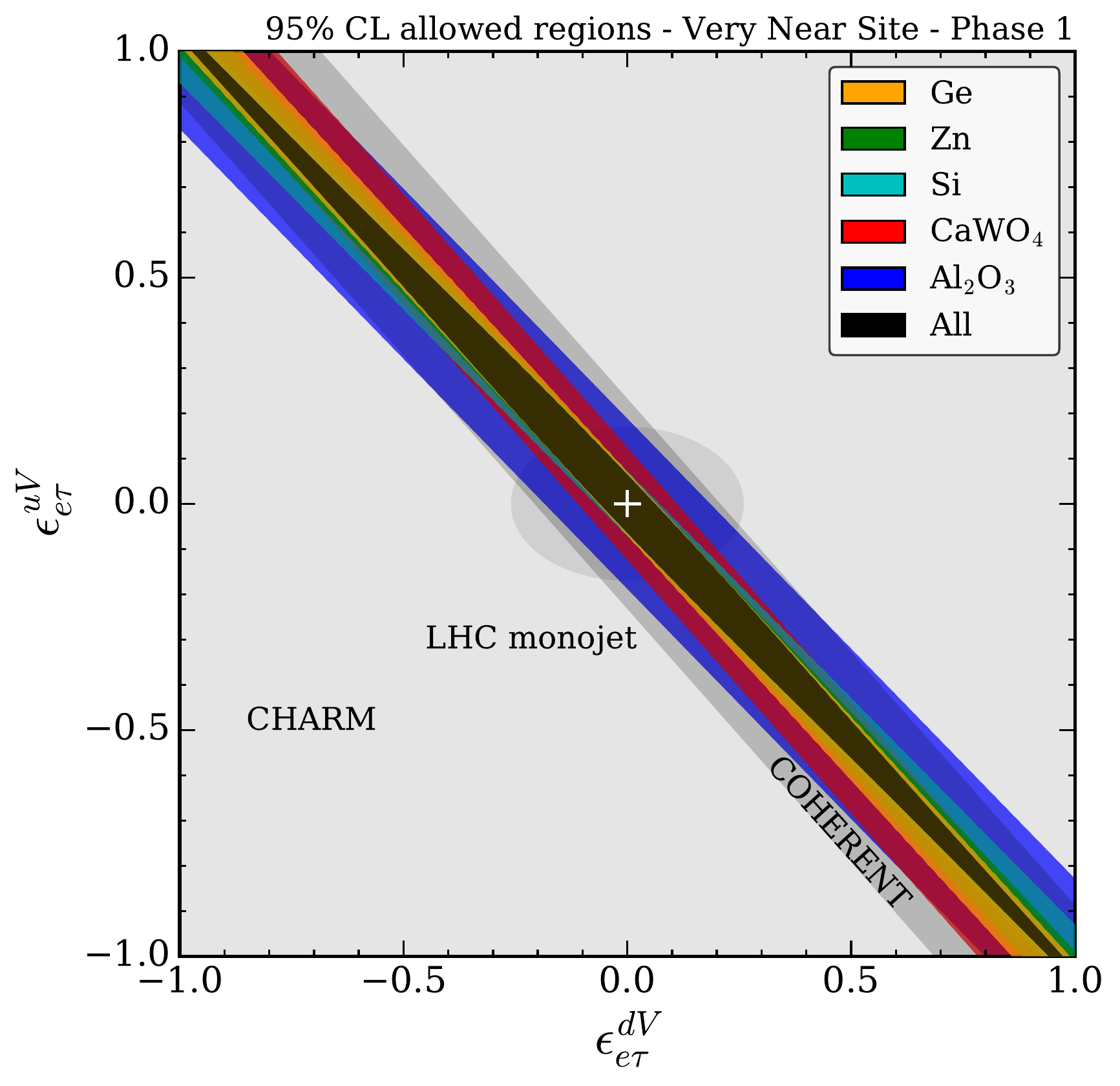}
\includegraphics[width=0.49\textwidth]{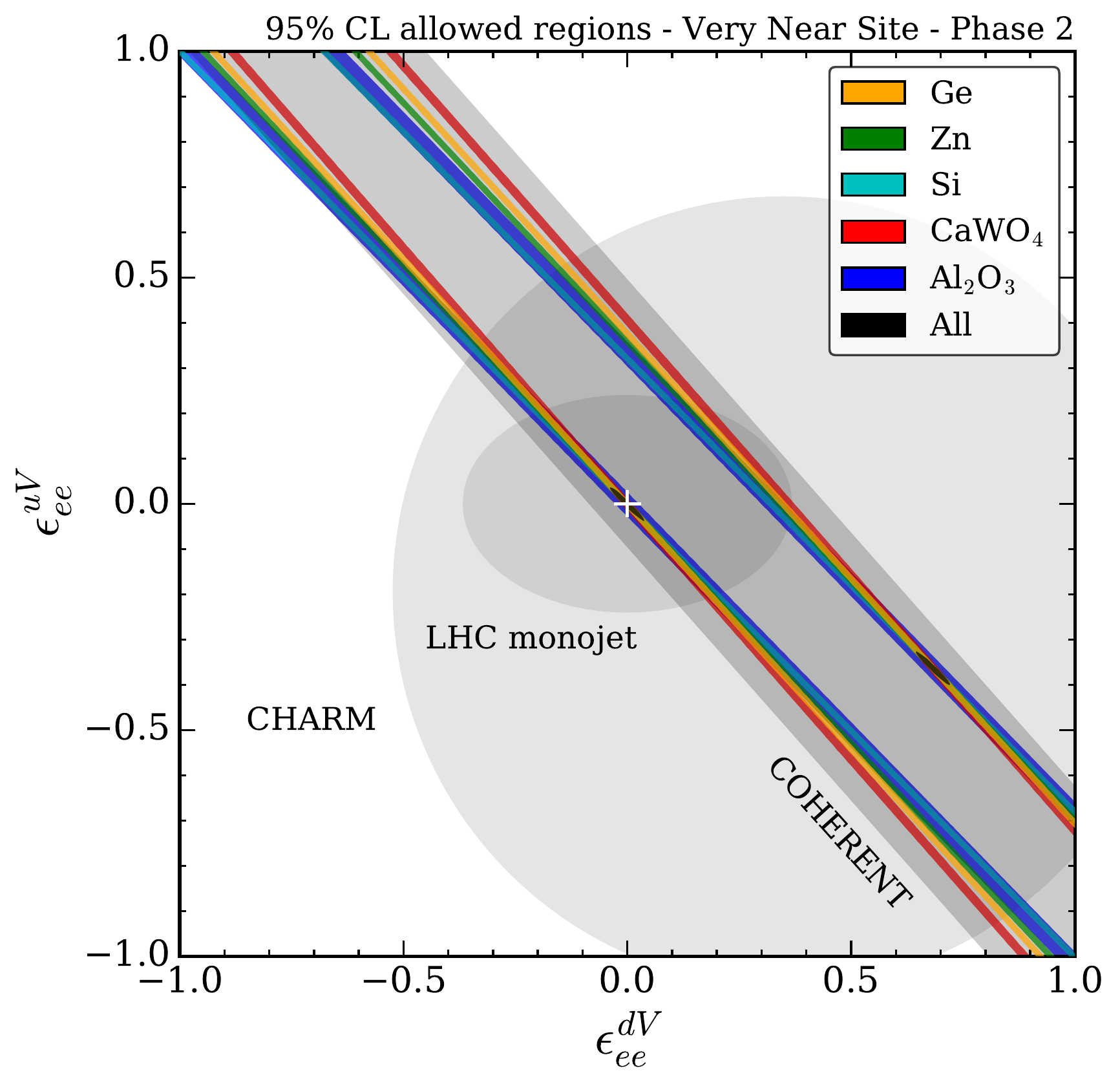}
\includegraphics[width=0.49\textwidth]{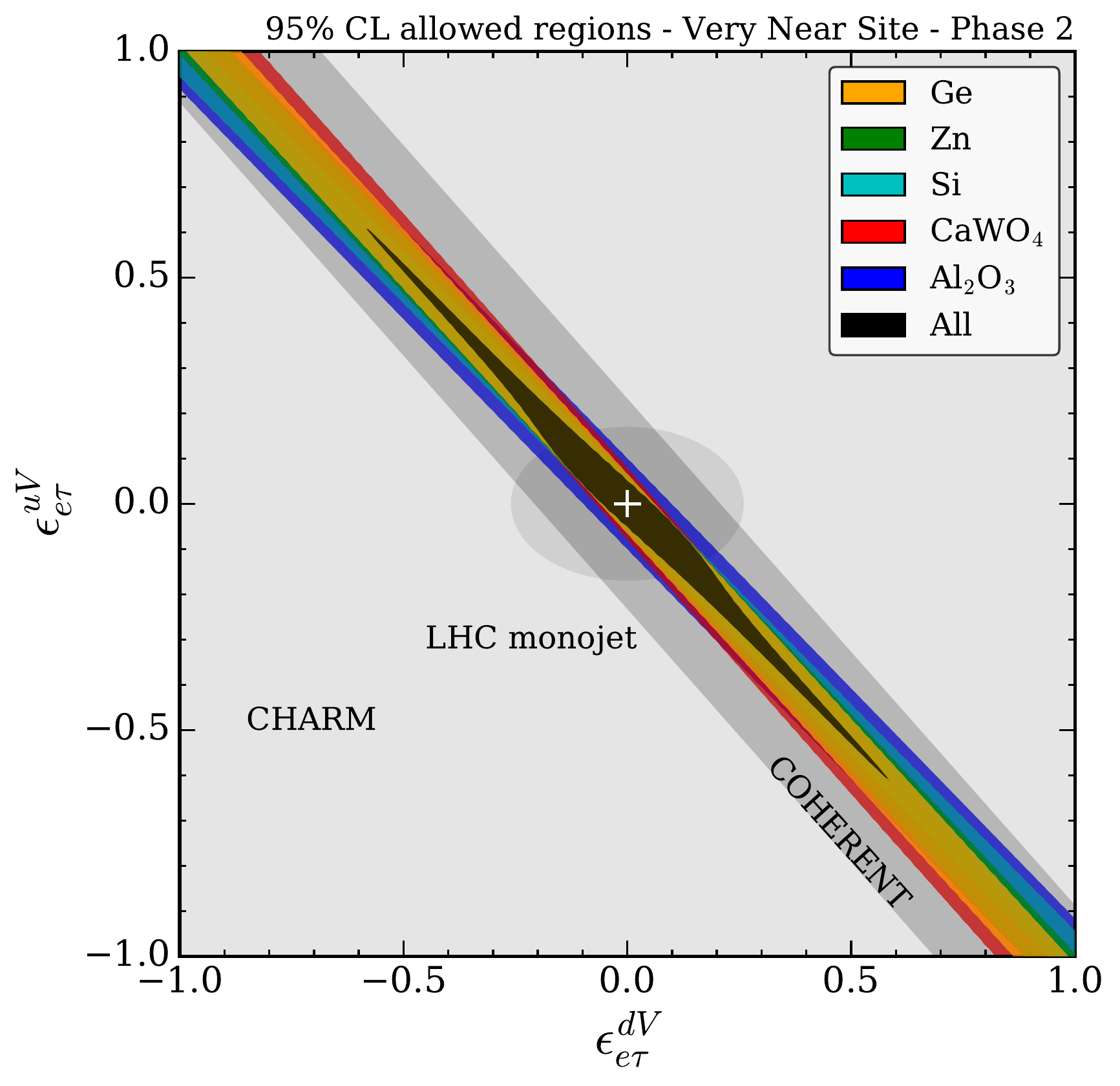}
\caption{\textbf{Projected 95\% CL allowed regions for non-standard interactions (NSI) at the Very Near Site, using all targets.} Same as Fig.~\ref{fig:NSI_VNS_GeZn}, but for all targets.}
\label{fig:NSI_VNS_all}
\end{figure}

We include here plots showing the allowed regions of the NSI parameters, as discussed in Sec.~\ref{sec:NSI}, for each of the different experimental targets. These are shown in Fig.~\ref{fig:NSI_NS_all} for the Near Site in Phase 1 (top row) and Phase 2 (bottom row). The corresponding results for the Very Near Site are shown in Fig.~\ref{fig:NSI_VNS_all}.

\bibliographystyle{JHEP}
\bibliography{RICOCHET.bib}

\end{document}